\documentclass[acmlarge]{acmart}
\usepackage{subfigure}
\usepackage{enumitem}
\usepackage{multirow}
\usepackage{makecell}
\usepackage{tikz}
\usepackage{xcolor} 
\usetikzlibrary{shapes.geometric, arrows} 
\usepackage{booktabs} 
\usepackage{tabularx} 
\usepackage{caption}  %
\definecolor{wifiblue}{HTML}{1F77B4}
\renewcommand{\textcolor}[2]{#2}
\AtBeginDocument{%
  }

\setcopyright{cc}
\setcctype{by}
\copyrightyear{2026}
\acmYear{2026}
\acmDOI{10.1145/3832029}

\acmJournal{IMWUT}
\acmVolume{10}
\acmNumber{3}
\acmArticle{179}
\acmMonth{9}

\begin{document}

\title{XGait: A Multi-Modality Wireless Sensing Dataset for Indoor Human Tracking and Identification}


\author{Wei Xu}
\email{xuwei9610@mail.nwpu.edu.cn}
\orcid{0000-0003-3079-7893}

\author{Zhu Wang}
\authornote{Corresponding Author.}
\email{wangzhu@nwpu.edu.cn(corresponding-author)}
\orcid{0000-0003-2368-8947}

\author{Yifan Guo}
\email{guoyf@mail.nwpu.edu.cn}
\orcid{0009-0009-0284-8965}

\author{Changlong Cheng}
\email{changlong_cheng@mail.nwpu.edu.cn}
\orcid{0009-0005-1527-6661}

\author{Yin Zhang}
\email{zy1494337725@outlook.com}
\orcid{0009-0000-4667-7170}

\author{Zhihui Ren}
\email{renzhihui@mail.nwpu.edu.cn}
\orcid{0000-0002-2929-9822}

\author{Bin Guo}
\email{guob@nwpu.edu.cn}
\orcid{0000-0001-6097-2467}

\affiliation{%
 \institution{Northwestern Polytechnical University}
 \city{Xi'an}
 \state{Shaanxi}
 \country{China}
 \postcode{710129}
}

\author{Zhiwen yu}
\email{zhiwenyu@nwpu.edu.cn}
\orcid{0000-0002-9905-3238}
\affiliation{%
 \institution{Harbin Engineering University}
 \city{Harbin}
 \state{Heilongjiang}
 \country{China}
 \postcode{150001}
}
\affiliation{%
 \institution{Northwestern Polytechnical University}
 \city{Xi'an}
 \state{Shaanxi}
 \country{China}
 \postcode{710129}
}

\renewcommand{\shortauthors}{Xu et al.}

\begin{abstract}
  Wireless sensing has emerged as a promising approach for tracking and identification using commodity Internet of Things devices. However, the features derived from a single wireless modality are often fragile to variations in environmental layouts and walking trajectories. Furthermore, most existing studies are based on datasets collected in specific scenarios with limited trajectory diversity and sensing modalities, preventing a robust evaluation of system generalization. \textcolor{blue}{To address this gap, we introduce \textbf{XGait}, a multi-modality wireless sensing dataset that synchronously captures human walking using Wi-Fi and acoustic transceivers across three indoor scenarios, with vision-based measurements serving as ground truth. Specifically, XGait contains more than 22K walking samples from 27 participants, covering diverse directions and trajectories to support both indoor tracking and identity recognition. To bridge the heterogeneity of wireless sensing modalities, we propose a unified Doppler spectrogram representation that maps Wi-Fi and acoustic signals into a shared time--frequency space, along with a standardized benchmark pipeline for pre-processing, temporal alignment, and feature construction, enabling reproducible evaluation and systematic cross-modal analysis. Extensive evaluations demonstrate that Wi-Fi and acoustic sensing exhibit complementary strengths, particularly under complex trajectories and challenging propagation conditions, thereby paving the way for novel research in the field of multi-modality wireless sensing.} The dataset and code are available at \url{https://github.com/warrior-087/XGait}. 
  
\end{abstract}

\ccsdesc[500]{Human-centered computing~Ubiquitous and mobile computing systems and tools}

\keywords{dataset, Wi-Fi, acoustic, multi-modal, wireless sensing, indoor tracking, human identification}

\received{1 February 2026}
\received[revised]{1 May 2026}
\received[accepted]{1 July 2026}

\maketitle

\section{INTRODUCTION}
Walking pattern is a distinctive and relatively stable biometric that reflects physiological and habitual characteristics~\cite{braune2012human}. Owing to its unobtrusive nature, wireless sensing of walking has attracted growing interest for applications ranging from indoor tracking~\cite{Wu2021GaitWay} to identity recognition~\cite{topham2022human} and health monitoring~\cite{mirelman2019gait,wang2016recognizing}. Compared to vision and wearable solutions, wireless sensing enables device-free perception, while providing better privacy and robustness to lighting conditions and visual occlusion~\cite{wang2018wi}. These advantages make wireless walking sensing well suited for everyday indoor settings such as homes and offices.

Recent studies have shown that human walking can be captured with diverse wireless modalities, including Wi-Fi~\cite{zhang2018crosssense,wang2022caution,zhang2021wi,zhang2020gaitid,liu2025csid,yan2025pushing,yang2025environment,liang2024dcs,tong2025stagr}, millimeter-wave radar~\cite{cheng2021person,li2023passive,wang2024rdgait,meng2025gr}, and acoustic signals~\cite{cai2021we,lian2021echospot,xu2019acousticid,wang2024agr}, largely leveraging motion-induced Doppler effects. Despite this progress, wireless walking sensing remains challenging. Specifically, wireless observations are indirect and often entangled with deployment geometry~\cite{niu2022rethinking,qian2017widar}, multipath propagation, and time-varying walking trajectories~\cite{wang2016human,niu2022rethinking,zhang2021wi,wang2016gait}, leading to strong viewpoint dependency and environment specificity. In practice, systems based on a single modality provide only partial evidence of human motion and may fail under unfavorable link geometry, occlusions, or trajectory changes. Meanwhile, publicly available datasets for systematic evaluation remain scarce. Most existing datasets are constrained to simple scenarios with limited trajectory and direction diversity and are predominantly single-modality datasets. These constraints hamper reproducible evaluation, fair comparison, and robustness analysis under realistic conditions. Moreover, the effectiveness of modern data-hungry representation learning models fundamentally depends on training data scale and coverage~\cite{dai2025babel,guo2025mmpencil,wang2026meco,jiang2026edge}, which are lacking in current datasets.  

This work takes a data-centric step toward addressing these barriers by introducing XGait, a multi-modality wireless sensing dataset designed for tracking and identity recognition. The dataset comprises synchronized recordings of Wi-Fi CSI and active acoustic signals from three representative indoor scenarios. It also includes camera-based ground-truth trajectories where reliable visual annotations are available.

\begin{figure}	
	\centering	
	\includegraphics[height=4.8cm]{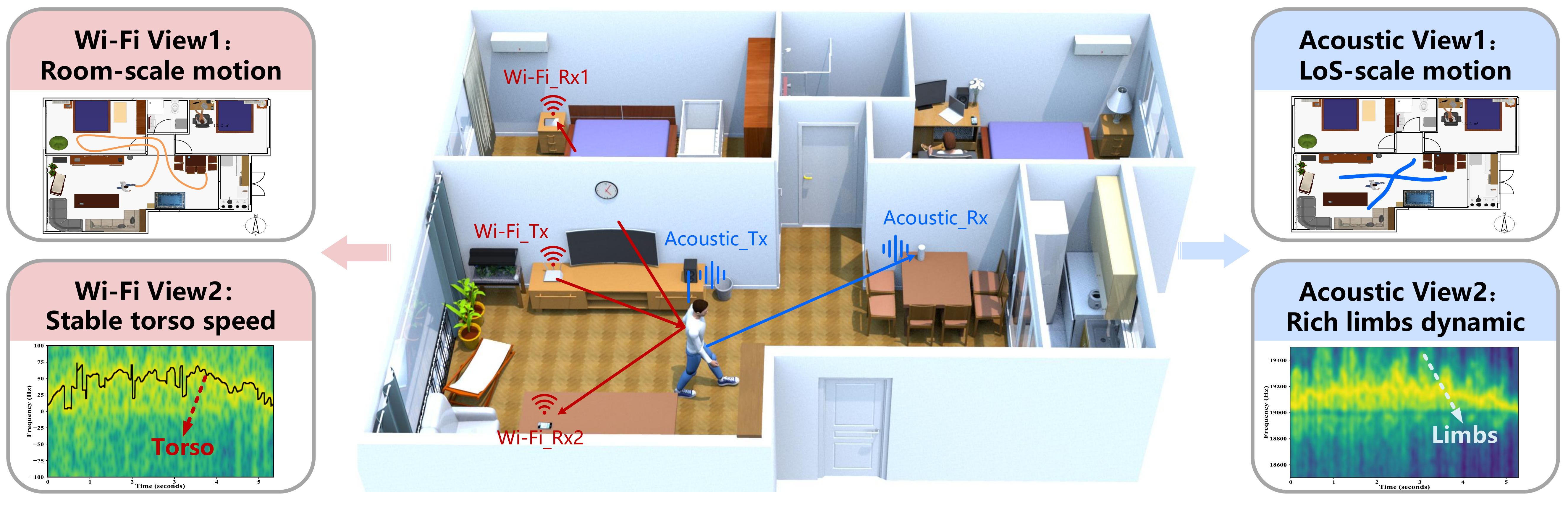}
	\caption{Complementary properties of Wi-Fi and acoustic sensing.}
    \label{Fig1}
\end{figure}

XGait is designed with three key considerations. First, it focuses on complementarity of Wi-Fi and acoustic sensing. As illustrated in Fig.~\ref{Fig1}, Wi-Fi typically offers broader coverage and stronger robustness in non-line-of-sight (NLoS) conditions, making it well suited for coarse motion tracking, whereas active acoustic signals can provide finer velocity resolution in the line-of-sight (LoS) settings, which helps capture subtle gait dynamics. Second, it emphasizes trajectory diversity under loose behavioral constraints. Instead of restricting participants to a small set of straight-line trajectories, the dataset includes diverse paths and walking directions with natural speeds and turns, thereby capturing realistic variations. Third, it spans three environments with distinct complexity levels, enabling comparative analysis of how environments and layouts affect sensing performance. 

In addition, to illustrate the utility of XGait in investigating the complementarity of heterogeneous wireless sensing modalities, we establish a unified evaluation paradigm centered on two core tasks: \textbf{human tracking} and \textbf{identity recognition}. Given the distinct physical properties and sensing mechanisms of Wi-Fi and acoustic signals, we first map both modalities to a shared time--frequency representation via signal processing. From the aligned spectrograms, we extract torso-level motion cues as a modality-invariant descriptor for continuous position tracking. Then, for identity recognition, we compare two strategies: a signal-driven approach that learns discriminative patterns directly from Doppler spectra, and a model-driven approach that employs structured motion representations for higher-level gait interpretation. Together, these task-driven signal processing and feature extraction steps establish a unified benchmarking framework. This framework reveals the complementary strengths of Wi-Fi and acoustic sensing in capturing motion and biometric traits, illustrating the distinctive insights enabled by the XGait dataset. Specifically, experimental evaluation uncovers a clear synergy between modalities: Wi-Fi provides a robust foundation for global trajectory tracking, while the finer spectral granularity of active acoustics yields more stable and discriminative signatures for identity recognition. This synergy shows that XGait facilitates not only performance benchmarking but also a deeper, physically-grounded understanding of trade-offs in multi-modality wireless sensing.

The contributions of this work can be summarized as follows:
\begin{itemize}[leftmargin=*]
\item {We release XGait, a multi-modality wireless sensing dataset that synchronously captures multi-node Wi-Fi CSI and active acoustic signals, and provides vision-based measurements as ground truth for walking trajectories. \textcolor{blue}{XGait spans three scenarios and contains 22K walking samples from 27 participants.} To our knowledge, it is the first large-scale dataset for multi-modality wireless sensing that supports both tracking and identification}.

\item {\textcolor{blue}{We unify heterogeneous Wi-Fi and acoustic signals through a shared Doppler spectrogram, providing a velocity abstraction that harmonizes disparate physical modalities. Built upon this representation, we establish a benchmark pipeline that standardizes pre-processing, temporal alignment, and feature construction, facilitating fair comparison and systematic cross-modal fusion.}}

\item {Comprehensive experiments quantify a clear synergy between Wi-Fi and acoustic sensing. Task-oriented complementarity is observed, in which Wi-Fi provides superior tracking stability, while acoustics offers stronger biometric discrimination. This finding delivers actionable insights for the building of robust multi-modality wireless sensing systems.}
\end{itemize}

The rest of this paper is organized as follows. We review related work and existing datasets in Section~\ref{sec:related work}. Section~\ref{sec:dataset acquisition and description} describes the data acquisition setup and provides an overview of the collected dataset, followed by the benchmark processing pipeline developed in Section~\ref{sec:benchmark_pipeline}. Section~\ref{sec:validation} validates the complementarity of different modalities through extensive experiments. \textcolor{blue}{Section~\ref{sec:discussion} discusses mechanisms underlying the synergy and limitations of the methods, as well as potential research opportunities enabled by XGait}. We give the dataset and the code release information in Section~\ref{sec:dataset and code availability}, and conclude the paper in Section~\ref{sec:conclusion}.

\section{RELATED WORK}
\label{sec:related work}
\subsection{Human Tracking and Identification based on Wireless Sensing}
Recent years have witnessed increasing interest in device-free wireless sensing with commodity infrastructure. Human-centered wireless sensing research aims to answer three key questions in the environment: who the person is (identity), where the person is (location), and what the person is doing (activity). Among them, identity and location are particularly challenging because they require stable, person-specific, and geometry-aware motion cues beyond coarse signal variations. Walking data naturally offers a distinctive yet repeatable biometric whose spatiotemporal dynamics encode both movement direction and walking patterns, making it an important building block for wireless indoor tracking and identity recognition. 

Prior work has developed tracking and identification systems using various wireless modalities. These systems typically exploit motion-induced Doppler shifts embedded in received signals to derive velocity-centric gait signatures for motion analysis and recognition. Wi-Fi-based approaches extract gait-related dynamics from CSI in various deployment settings, allowing device-free indoor tracking and identity recognition~\cite{zeng2016wiwho,wang2016gait,zhang2018crosssense, Wu2021GaitWay,zhang2021wi,zhang2021gaitsense,zhang2020gaitid,wang2022caution}. Millimeter-wave sensing has also been explored for gait analysis, benefiting from its higher spatial resolution to capture fine-grained motion patterns~\cite{meng2020gait,cheng2021person,li2023passive,wang2024rdgait}. In parallel, acoustic sensing takes advantage of the slower propagation speed of mechanical waves to achieve higher velocity resolution, enabling detailed characterization of gait dynamics using Doppler-based features from commodity speakers and microphones~\cite{Kalgaonkar2007Acoustic,Garreau2011Gait,xu2019acousticid,wang2024agr,xu2025acoustic}.

Despite this progress, wireless gait sensing remains inherently challenging in practice~\cite{WiFiSurvey,wang2025survey}. Human walking speeds extracted from wireless signals are often sensitive to deployment geometry, trajectories, and multipath effects, leading to viewpoint-dependent and environment-specific observations. Although recent studies~\cite{Lww,Through-wall,CSI_DIFF,freeGait,liu2025csid,yan2025pushing,yang2025environment,liang2024dcs,tong2025stagr,al2026multimodal} have proposed increasingly sophisticated signal processing and learning-based models to improve robustness, most systems still rely on a single wireless modality, which provides only partial and indirect observation of human motion. These limitations indicate that, although wireless gait sensing has demonstrated strong potential, further advances require deeper investigation into sensing complementarity and motion representation beyond what a single modality can capture.

\begin{table}[!t]
  \caption{Publicly available wireless sensing datasets for human tracking and identity recognition.}
  \label{tab:wireless_gait_datasets}
  \setlength{\tabcolsep}{3pt}
  \begin{tabular}{ccccccc}
    \toprule
    Dataset & \#Subjects & \#Samples & Constraint & Modality & Functionality & Year  \\
    \midrule
    Widar~\cite{qian2017widar}            & 5   & 54 & Strong     & Wi-Fi & Tracking       & 2017  \\
    CrossSense~\cite{zhang2018crosssense} & 100 & 600000 & Strong & Wi-Fi & Identification & 2018  \\
    Widar2.0~\cite{qian2018widar2}        & 6   & 24 & Strong     & Wi-Fi & Tracking       & 2018  \\
    GaitID~\cite{zhang2020gaitid}         & 10  & 4600 & Medium   & Wi-Fi & Identification & 2020  \\
    RISE~\cite{zhai2021rise}              & 15  & 1150 & Strong   & Wi-Fi & Identification & 2021  \\
    CAUTION~\cite{wang2022caution}        & 20  & 900  & Strong   & Wi-Fi & Identification & 2022  \\
    EIGait~\cite{yang2025environment}     & 8   & 1920 & Medium   & Wi-Fi & Identification & 2025  \\
    \textbf{XGait (Ours)}    & \textcolor{blue}{27}   & \textcolor{blue}{>22000} & Weak & Wi-Fi, acoustic, camera & Tracking, Identification & 2026  \\
    \bottomrule
  \end{tabular}
\end{table}

\subsection{Wireless Sensing Datasets for Human Tracking and Identification}
Although wireless sensing has been studied extensively, publicly available datasets that support reproducible performance evaluation remain limited. As summarized in Table~\ref{tab:wireless_gait_datasets}, early open datasets such as Widar~\cite{qian2017widar} and Widar2.0~\cite{qian2018widar2} provide CSI recordings for tracking, typically under constrained walking patterns and on a relatively small scale. Later efforts shifted toward identification, including GaitID~\cite{zhang2020gaitid}, RISE~\cite{zhai2021rise}, and CAUTION~\cite{wang2022caution}, which curate CSI samples for recognition under controlled trajectories. CrossSense~\cite{zhang2018crosssense} substantially increased both subject and sample sizes, yet its access is restricted. More recently, EIGait~\cite{yang2025environment} further considers robustness factors, but the scale and sensing modalities are still limited. Overall, existing open datasets are predominantly Wi-Fi-only, often tailored to a single task (tracking or identity recognition), and frequently collected under strict trajectory constraints.

While valuable, these datasets leave important gaps for studying generalization and modality complementarity. First, many datasets impose strong trajectory constraints (e.g., fixed directions or straight-line paths), limiting their ability to capture realistic gait variation and deployment diversity. Second, most datasets focus on a single modality and thus inherit viewpoint dependency and sensing ambiguity, which restricts the investigation of modality complementarity. Third, datasets for tracking and identification are often separated, preventing joint evaluation across the end-to-end pipeline from motion estimation to biometric recognition. These gaps motivate a dataset that combines complementary commodity modalities in realistic indoor environments and supports unified evaluation for both tracking and identification. 

Several recent efforts have released multi-modality human sensing datasets, such as XRF55~\cite{wang2024xrf55} and CUHK-X~\cite{jiang2025large}, to facilitate cross-modality learning and generalizable representations. However, these datasets are designed primarily for general human activity sensing, while XGait is specifically tailored for human tracking and identification. Concretely, it is uniquely structured to investigate cross-modality complementarity under realistic conditions by pairing Wi-Fi CSI (electromagnetic waves) with active acoustics (mechanical waves), two carriers with complementary physical sensitivities. To our knowledge, XGait is the first public dataset that jointly provides these commodity signals and vision-based ground truth for unified human tracking and identification evaluation.

\section{DATASET ACQUISITION AND DESCRIPTION}
\label{sec:dataset acquisition and description}
\subsection{Hardware setups}
To collect multi-modality walking data, we deploy a heterogeneous sensing system with commodity Wi-Fi devices, acoustic devices, and a camera. All devices are connected through a local area network (LAN) to facilitate data transmission and coordination. The recordings across all sensors are synchronized using a software trigger to initiate and terminate each trial. We assign a unique ID to each node and index nodes by modality (Wi-Fi vs. acoustic) throughout the dataset.

\begin{figure}[!t]	
	\centering	
	\subfigure[]{		
		\label{fig2a} 
		\includegraphics[height=1.3in]{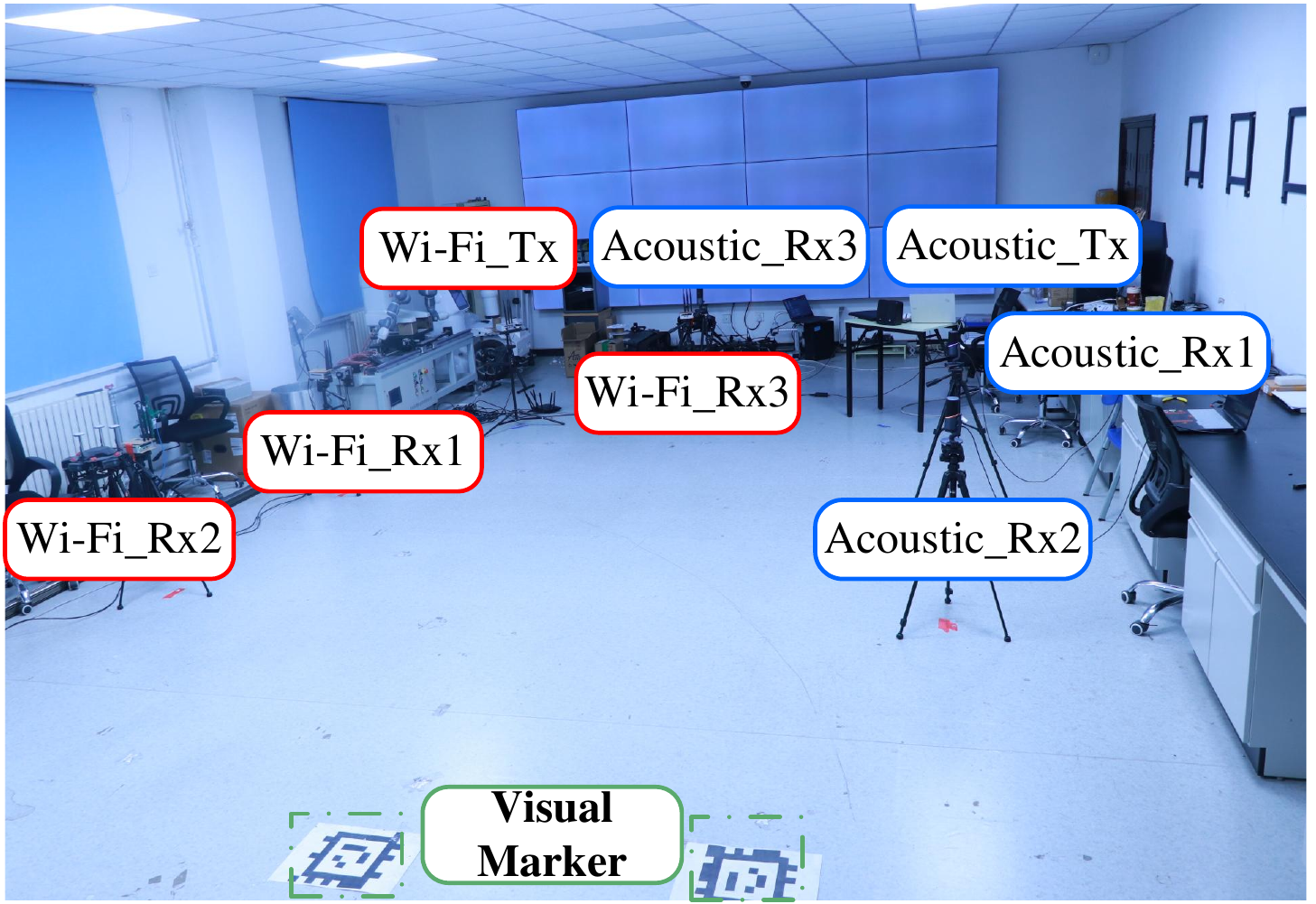}}	
    \hspace{0.05in}
	\subfigure[]{		
		\label{fig2b} 
		\includegraphics[height=1.3in]{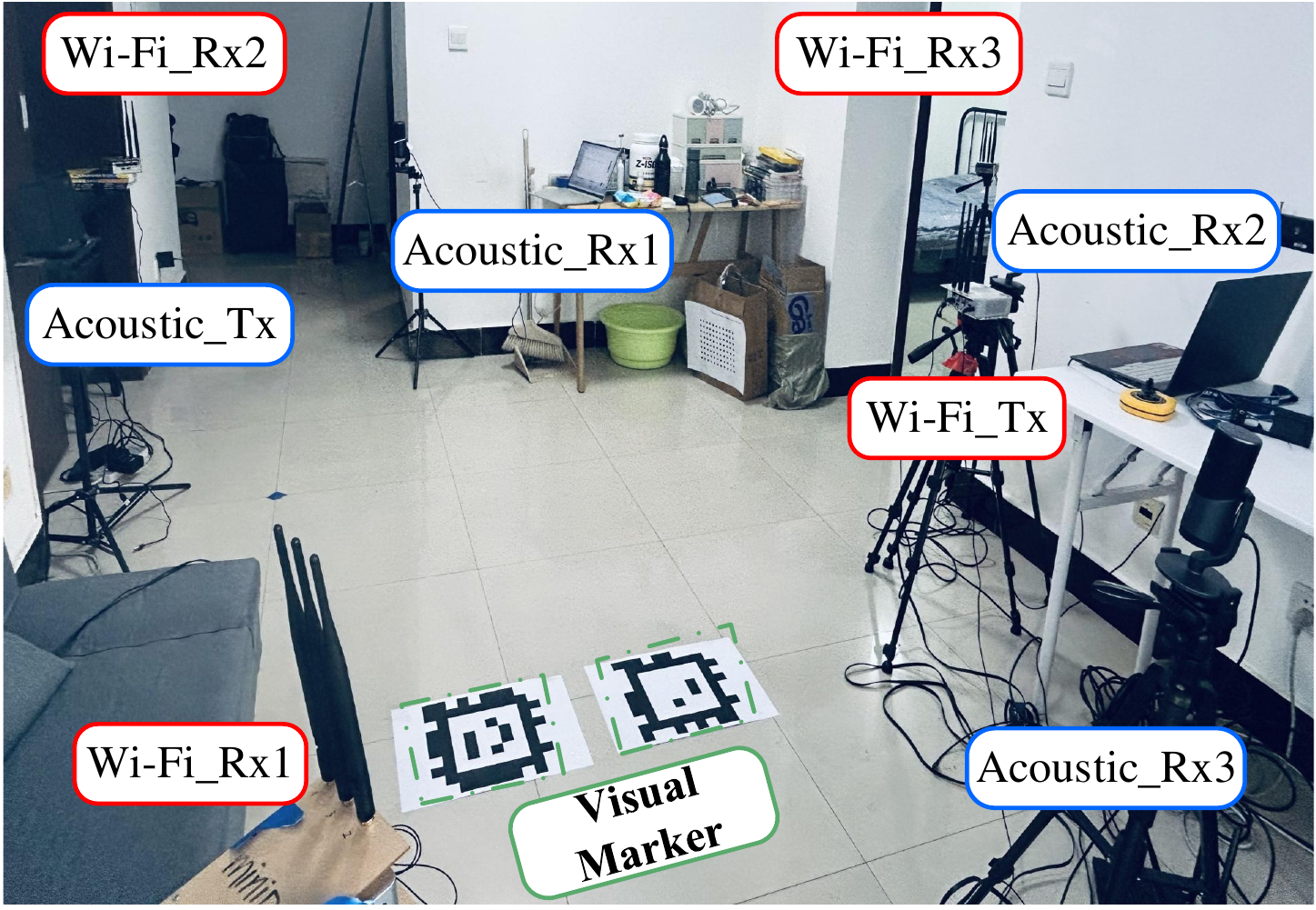}}
    \hspace{0.05in}
    \subfigure[]{		
		\label{fig2e} 
		\includegraphics[height=1.3in]{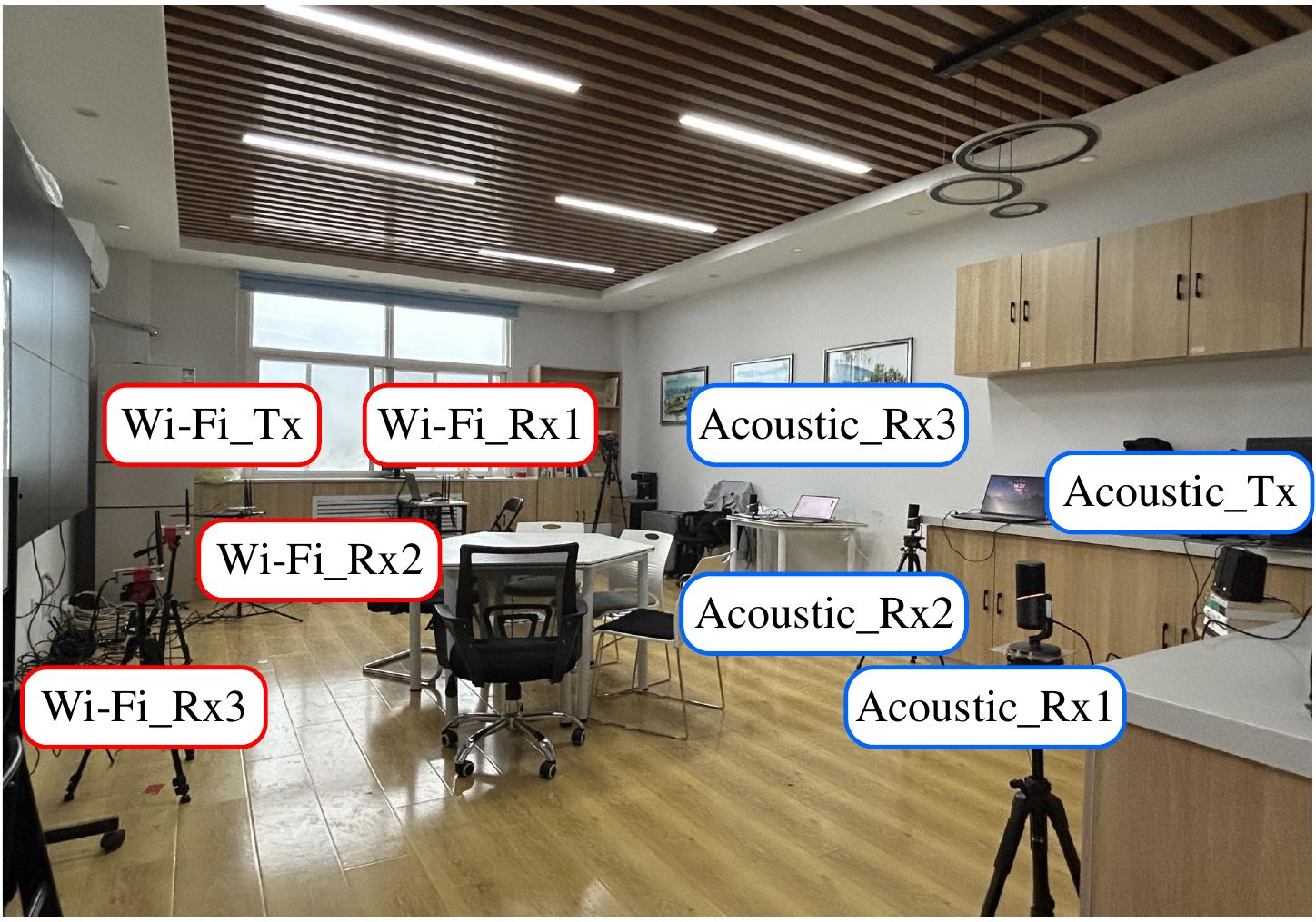}}
	\subfigure[]{		
		\label{fig2c} 
		\includegraphics[height=1.3in]{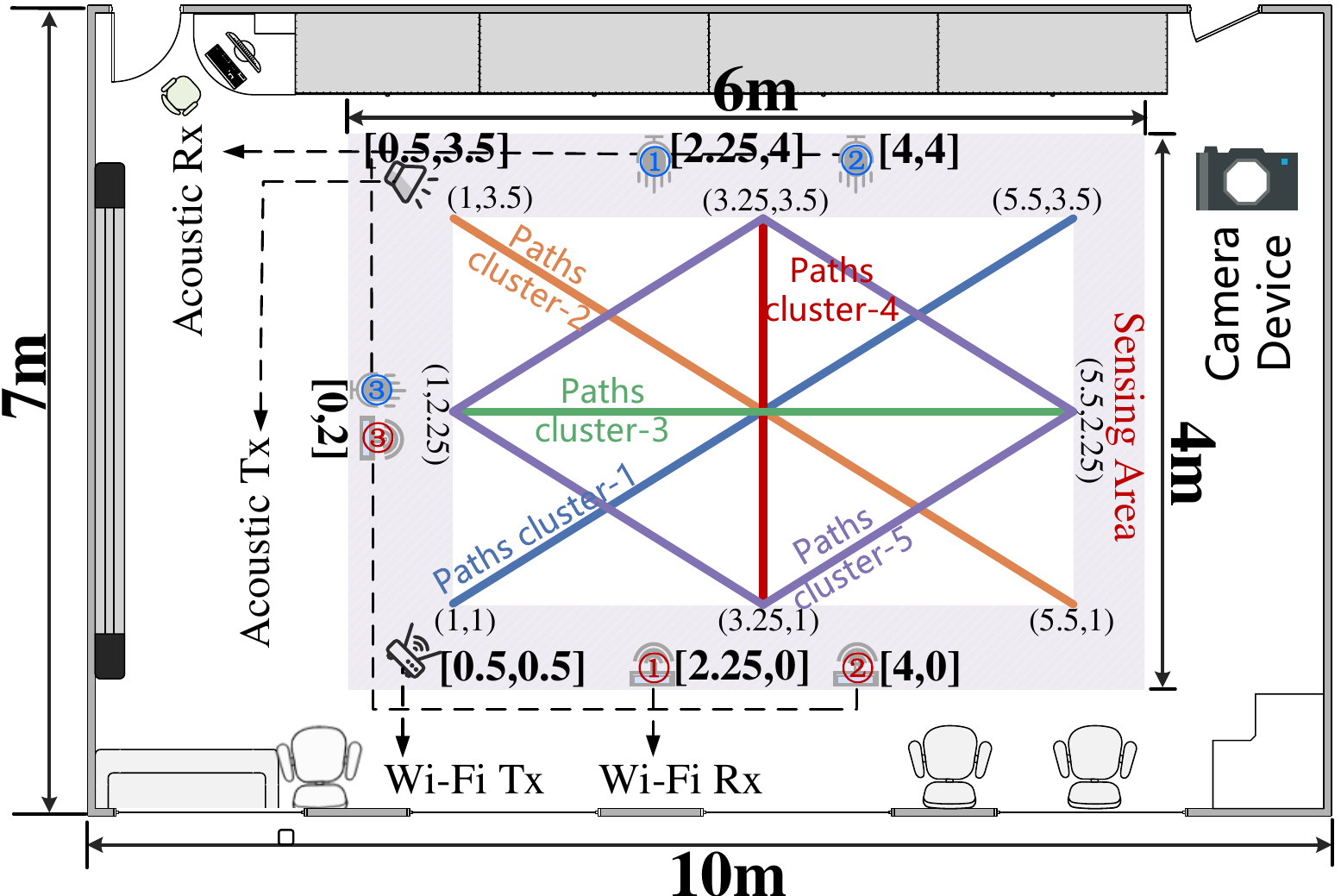}}	
	\subfigure[]{		
		\label{fig2d} 
		\includegraphics[height=1.3in]{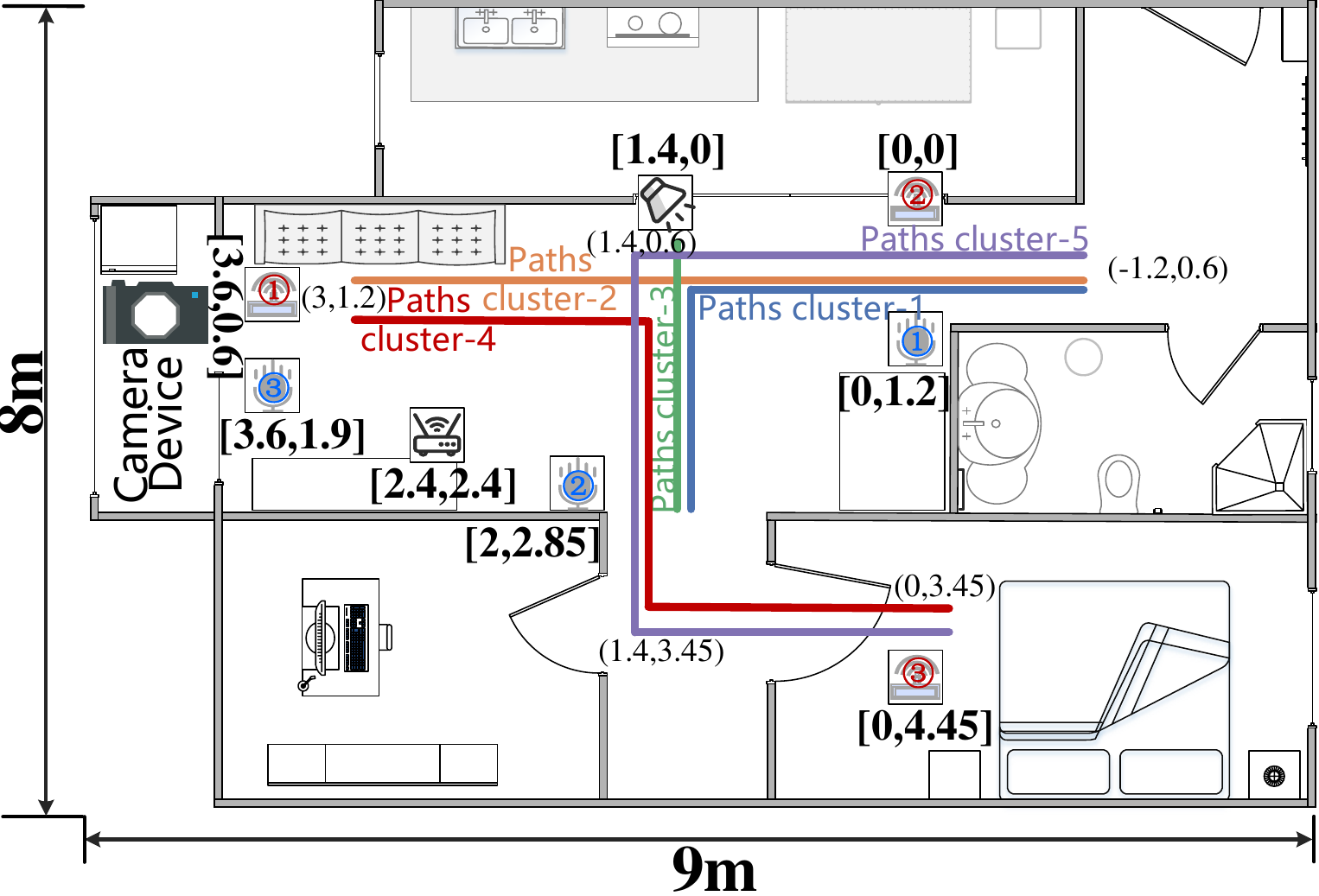}}	
    \subfigure[]{		
		\label{fig2f} 
		\includegraphics[height=1.3in]{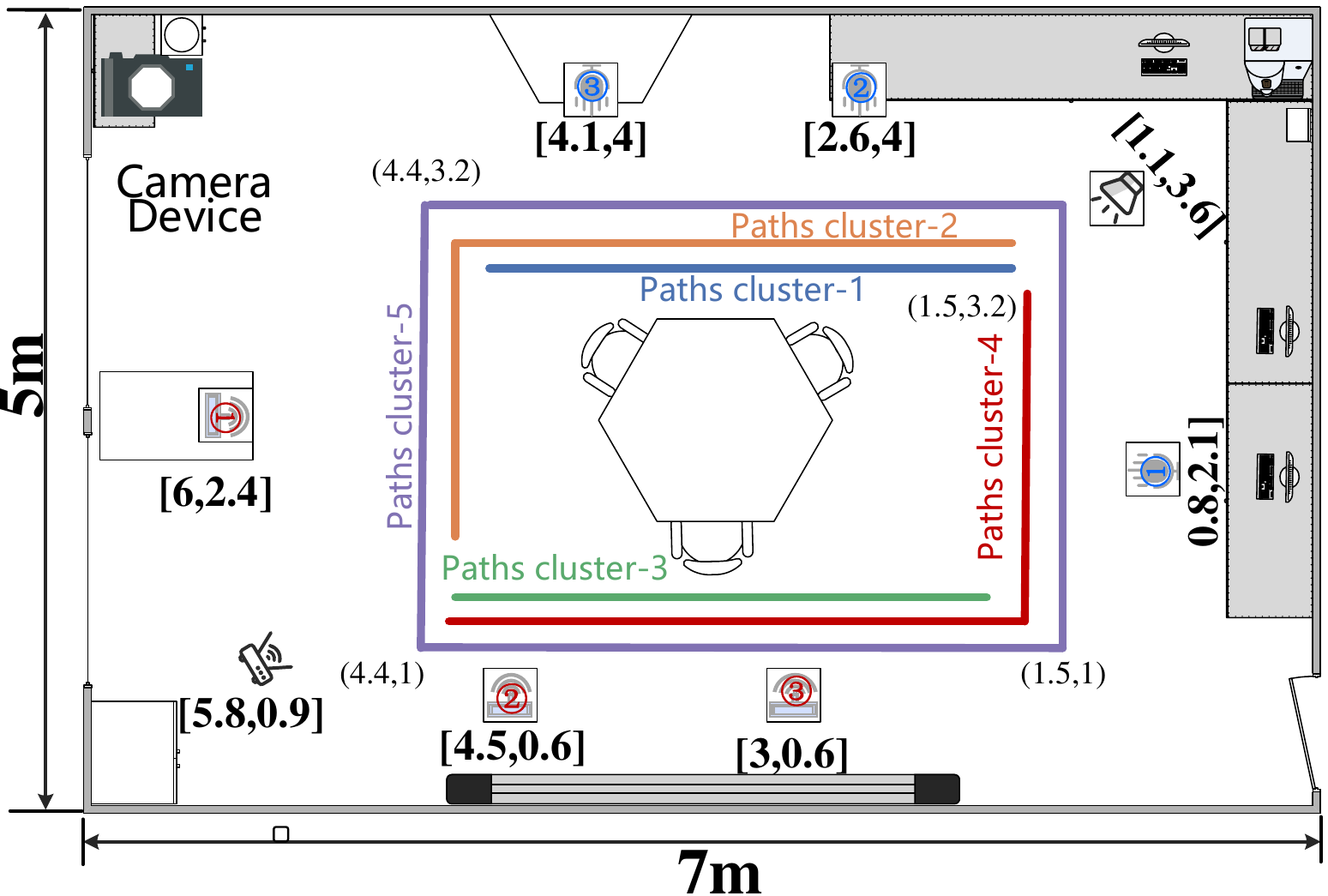}}
	
	\caption{Dataset collection settings. (a) Laboratory  configuration. (b) Home configuration. \textcolor{blue}{(c) Meeting-room configuration.} (d) Laboratory trajectories. (e) Home trajectories.  \textcolor{blue}{(f) Meeting-room trajectories.}
	}
	\label{fig2} 
\end{figure}

\textbf{(1) Wi-Fi devices.} We use four mini-PCs equipped with Intel 5300 wireless network interface cards (NICs), each connected to three external antennas with $2.5\,\mathrm{cm}$ spacing. One node is configured as a transmitter, and the remaining three act as receivers. The device placement strategy follows common practices in Wi-Fi sensing studies~\cite{qian2017widar,ma2025adapttrack}, while adapting to the constraints of each scene. \textcolor{blue}{As shown in Fig.~\ref{fig2}, regular layouts are adopted in the open laboratory and meeting-room scenario, while irregular layouts are used in the residential environment.}

To capture gait-related dynamics from both the torso and the limbs, all Wi-Fi nodes are mounted at a height of  $0.8\,\mathrm{m}$. The transmitter uses a single antenna to continuously broadcast packets (IEEE 802.11n) on channel $64$ ($5.31\,\mathrm{\,GHz}$) with a sampling rate of $1\,\mathrm{kHz}$. Each receiver listens on the same channel using all three antennas. CSI is acquired by the Linux CSI Tool~\cite{halperin2011tool}, where each packet is characterized by measurements from 30 subcarriers.

\textbf{(2) Acoustic devices.} The acoustic sensing system consists of four commodity audio devices connected to four laptops, one speaker (JBL PS3500) as a transmitter and three microphones (UGREEN, $16$-bit, $48\,\mathrm{\,kHz}$) as receivers, installed at the same height as the Wi-Fi nodes. Unlike prevailing monostatic acoustic setups~\cite{xu2019acousticid,li2025synergizing,wang2024agr}, we employ a bistatic multi-node configuration with separate transmitter and receivers. The speaker emits a single-tone continuous-wave (CW) signal at $19\,\mathrm{kHz}$, and the microphones record the reflected signals at $48\,\mathrm{kHz}$.

This design is motivated by sensing geometry and practical deployment constraints. Specifically, a monostatic setup represents a degenerate case of the Fresnel zone model~\cite{wang2016human} commonly used in Wi-Fi sensing, where the elliptical sensitivity regions collapse into concentric circles centered at a single node. Given the higher directionality and stronger path loss inherent in acoustic propagation, such co-location provides limited spatial coverage in cluttered indoor environments. To better emulate realistic home scenarios and expand the sensing field, we decouple the transmitter and the receivers, adopting a deployment strategy conceptually aligned with multi-node Wi-Fi networks. 

\textbf{(3) Camera device.} We use a DSLR camera (Canon EOS 80D) as the reference modality to obtain ground-truth trajectories. It captures 1080p video at $30\,\mathrm{fps}$ from a height of $1\,\mathrm{m}$. AprilTag~\cite{AprilTag2011} fiducial markers are deployed for spatial calibration, and the camera's extrinsic parameters are estimated via tag-corner correspondences. Ground-truth trajectories are reconstructed by back-projecting per-frame foot-ground contact points, identified through human pose estimation, from the image plane to the 3-D world coordinate system.

\subsection{Scenarios and trajectories}
\textcolor{blue}{Data are collected across three representative scenarios: a controlled laboratory setting, a residential environment, and a meeting-room, covering different walking directions, path geometries, and visibility conditions (LoS/NLoS).}

\textbf{(1) Laboratory scenario.} In the controlled LoS laboratory environment, we establish a rectangular sensing region as illustrated in Figs.~\ref{fig2a} and~\ref{fig2c}. Position markers are placed at the four corners and the midpoints of each edge, forming eight reference anchors. Using different combinations of these locations, we construct five trajectory groups (``path clusters''), each defined by a subset of the start--end marker pairs and representing a family of related trajectories rather than a single fixed path. Although the straight segments in the schematic serve as guides, our data collection includes linear and curved walking patterns. Specifically, path clusters 1--4 each incorporate both linear and curved trajectories. Within these four clusters, the linear segments are assigned labels \#1--\#4, while their corresponding curved counterparts are labeled \#6--\#9. Path cluster 5 is dedicated to a more complex turning pattern, designated as label \#5. This hierarchical structure ensures that each cluster provides diverse gait signatures, ranging from simple straight walks to intricate maneuvers, resulting in nine unique path labels for comprehensive analysis.

All trajectories are bidirectional. The direction is encoded by the parity of the repetition index. The data naming convention adheres to Widar-style~\cite{qian2017widar} formatting, facilitating seamless integration with existing analytic pipelines and community-developed tools. For example, the file \textit{user1-1-1-r1.dat} identifies the signal recorded at receiver r1 of user 1 on path 1 for the first trial. In this scenario, walking data are collected from 20 participants. Each participant completes all nine paths with 60 trials per path, yielding 540 trials per participant and 10.8K trials in total.

\textbf{(2) Home scenario.} Following the same design principle, we define five path clusters that cover both LoS and NLoS regions, as shown in Figs.~\ref{fig2b} and ~\ref{fig2d}. In contrast to the previous scene, the home environment is spatially constrained and geometrically irregular due to walls and furniture. Consequently, trajectories are not strictly categorized into linear or curved. Instead, most paths naturally include turns and detours induced by the physical layout.

In this scenario, each path is repeated approximately 50 times. \textcolor{blue}{Nineteen volunteers participate, 15 of whom also appear in the laboratory dataset.} Due to occlusion in the residential environment, vision-based ground-truth is not available for all trajectories. Therefore, accurate trajectory annotations are provided only for clusters 1 and 3.

\textcolor{blue}{\textbf{(3) Meeting-room.} We collect an additional dataset in a small meeting-room environment (Fig.~\ref{fig2e}), featuring a relatively open layout with partial obstructions from furniture. The Wi-Fi and acoustic nodes are placed diagonally across the sensing area, with major obstacles near the center. This configuration introduces diverse propagation paths while creating challenging conditions for single-modality sensing, as signals from one type of node may be significantly blocked when the subject moves across the obstacle. This leads to asymmetric visibility across modalities, providing a more rigorous testbed for evaluating multi-modal robustness and complementarity.}

\textcolor{blue}{Five path clusters are defined as shown in Fig.~\ref{fig2f}, clusters 1 and 2 are equipped with ground-truth trajectories. Data collection consists of two parts. First, following the protocol used in the previous scenarios, we collect data from 20 participants, of whom 13 also participated in the laboratory setting and 17 in the home setting. Each subject completes 50 trials per path cluster. Second, to explicitly investigate intra-person variability, 6 participants from this group are recorded on multiple days over one week with naturally varying clothing. The naming format \textit{user1-1-1-d1-r1.dat} encodes the recording date. This design introduces realistic temporal and appearance variations, enabling a systematic analysis of the stability of wireless gait signatures under daily changes.}

\begin{figure}[!t]	
	\centering	
	\subfigure[]{		
		\label{fig3a} 
		\includegraphics[width = 1.7in]{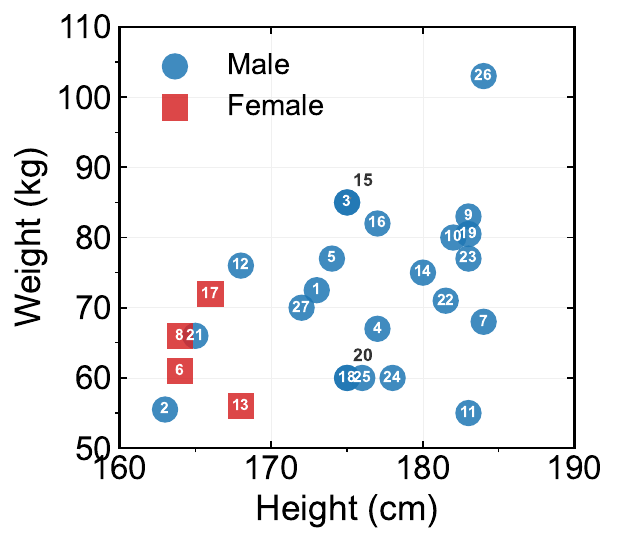}}	
	\subfigure[]{		
		\label{fig3b} 
		\includegraphics[width = 1.68in]{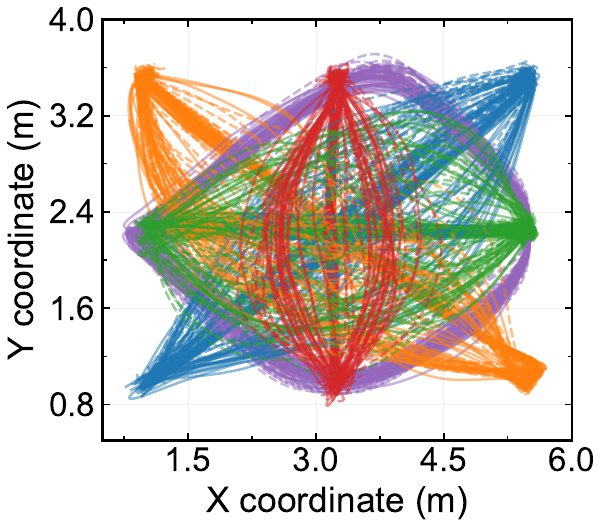}}	
    \subfigure[]{		
		\label{fig3c} 
		\includegraphics[width = 1.7in, height = 1.45in]{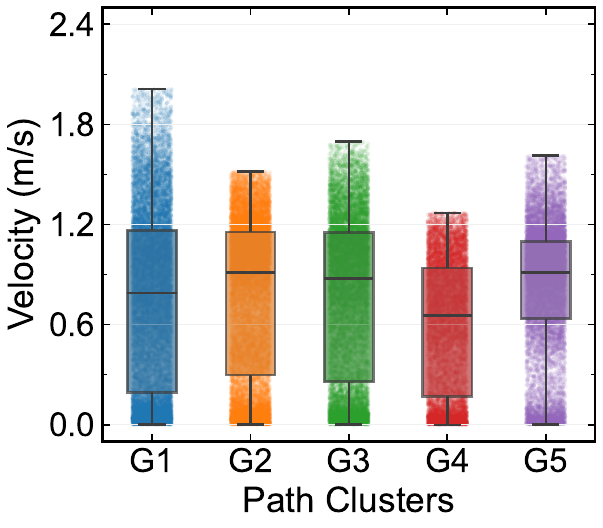}}	

    \begin{tikzpicture}[font=\footnotesize, x=1cm, y=1cm]
    \begin{scope}
        \node[black, font=\footnotesize\bfseries, anchor=west] at (0,0) {Legend for (b):};
        
        \begin{scope}[xshift=2.5cm]
            \draw[ultra thick, color={rgb,255:red,31; green,119; blue,180}] (0,0) -- (0.4,0) 
                node[right, black, xshift=-0.1cm] {G1 (\#1,\#6)};
        \end{scope}

        \begin{scope}[xshift=5.0cm]
            \draw[ultra thick, color={rgb,255:red,255; green,127; blue,14}] (0,0) -- (0.4,0) 
                node[right, black, xshift=-0.1cm] {G2 (\#2,\#7)};
        \end{scope}

        \begin{scope}[xshift=7.5cm]
            \draw[ultra thick, color={rgb,255:red,44; green,160; blue,44}] (0,0) -- (0.4,0) 
                node[right, black, xshift=-0.1cm] {G3 (\#3,\#8)};
        \end{scope}
    \end{scope}

    \begin{scope}[yshift=-0.45cm]
        \begin{scope}[xshift=0.0cm]
            \draw[ultra thick, color={rgb,255:red,214; green,39; blue,40}] (0,0) -- (0.4,0) 
                node[right, black, xshift=-0.1cm] {G4 (\#4,\#9)};
        \end{scope}

        \begin{scope}[xshift=2.5cm]
            \draw[ultra thick, color={rgb,255:red,148; green,103; blue,189}] (0,0) -- (0.4,0) 
                node[right, black, xshift=-0.1cm] {G5 (\#5)};
        \end{scope}

        \begin{scope}[xshift=5.0cm]
            \draw[thick, black] (0,0) -- (0.6,0) 
                node[right, black, xshift=-0.1cm] {Odd (Solid)};
        \end{scope}
        
        \begin{scope}[xshift=7.5cm]
            \draw[thick, black, dash pattern=on 3pt off 2pt] (0,0) -- (0.6,0) 
                node[right, black, xshift=-0.1cm] {Even (Dashed)};
        \end{scope}
    \end{scope}
    \end{tikzpicture}
    
	\caption{Volunteer attributes and motion trajectory diversity. (a) Volunteer Information. (b) Trajectories of $\text{User}_3$ (In Laboratory). (c) Walking speed distribution of different path clusters.
	}
	\label{fig3} 
\end{figure}
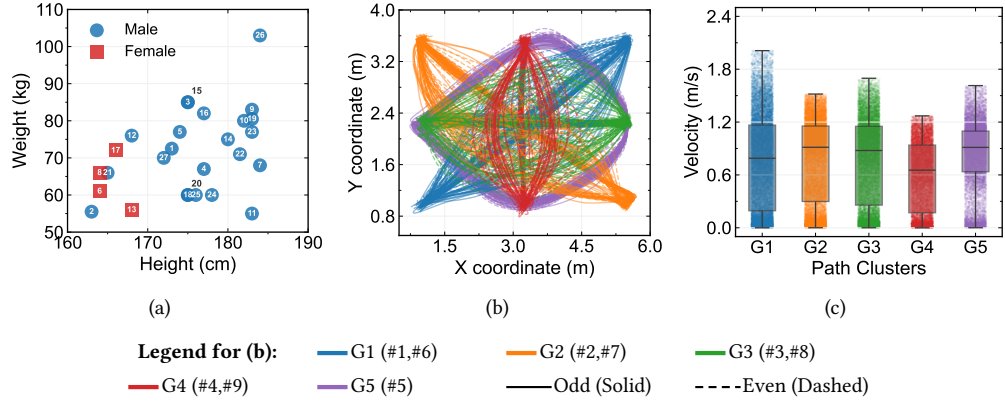

\subsection{Dataset Acquisition}
This study was conducted with institutional approval and oversight from our organization.

\textbf{(1) Volunteer recruitment.} \textcolor{blue}{We recruit 27 volunteers (4 female, 23 male) aged 20--35 years from our laboratory. Their heights range from $160\,\mathrm{cm}$ to $184\,\mathrm{cm}$ and weights from $55\,\mathrm{\,kg}$ to $105\,\mathrm{\,kg}$.} Gait is a physiological and habitual biometric trait, and body attributes such as height and weight can influence motion dynamics. Fig.~\ref{fig3a} illustrates the joint height-weight distribution, where each ID-labeled marker represents one participant, with the shape indicating gender. 

Before formal data collection, each participant completes a 10--15 minute familiarization session to acclimate to the environment and trajectory layouts. This warm-up period is designed to mitigate hesitation or unnatural movements, ensuring that the recorded data faithfully reflects habitual gait patterns.

\textbf{(2) Data collection procedure.} All sensing devices are connected to the same LAN through a router to support reliable data transmission and centralized control. A laptop serves as the coordinator, while Wi-Fi and acoustic devices run as clients.

For each trial, the Wi-Fi and acoustic transmitters emit continuously. The participants respond to two verbal cues: ``start'' and ``stop''. Upon the ``start'' instruction, all receiving devices begin recording and the camera simultaneously starts video capture. The volunteer then walks from the designated starting point to the endpoint of the selected trajectory. After reaching the destination, the volunteer remains stationary until the ``stop'' instruction, at which point all receivers terminate recording, and the trial is saved as one sample. Since these heterogeneous devices are not synchronized at the clock level, the cues are issued with slight buffers before and after the walking interval to prevent data truncation, and temporal alignment is performed during post-processing.

Although participants are required to start and end at specified markers, they are encouraged to maintain their preferred walking speed and path curvature. To highlight the spatial and kinematic richness of our dataset, we analyze the ground-truth trajectories and corresponding velocity profiles of a representative participant ($\text{User}_3$). As shown in Fig.~\ref{fig3b}, the five clusters cover different path geometries and motion directions, including long-range traversals, intersecting passes, and curved or turning trajectories, thus obtaining diverse kinematic patterns within the same sensing area. Fig.~\ref{fig3c} further quantifies this diversity: the velocity distributions vary substantially across clusters in both median speed and variability, with some clusters exhibiting broader spreads and higher-speed tails than others. 

\textbf{(3) Dataset statistics.} 
\textcolor{blue}{The XGait dataset comprises multi-modal recordings from 27 volunteers across three indoor scenarios.} For data release, Wi-Fi measurements are provided as raw CSI logs in \texttt{.dat} format. Each file stores packet-level channel measurements and can be parsed into CSI tensors of size $1000 \times 3 \times 30 \times t$ (packets $\times$ antennas $\times$ subcarriers $\times$ time). Acoustic measurements are released as raw audio waveforms in \texttt{.wav} format sampled at $48\,\mathrm{\,kHz}$. Vision data is provided as \texttt{.mp4} videos at 1080p resolution. To facilitate immediate research use, we also provide the corresponding ground-truth trajectories extracted from these videos, which are stored as 2-D spatial coordinate sequences in \texttt{.npy} format. 

\textcolor{blue}{Overall, XGait provides 22{,}288 valid Wi-Fi CSI recordings and 22{,}118 acoustic recordings. In contrast, the vision modality contains 15{,}257 usable samples, since videos are occasionally unavailable or unreliable (e.g., due to limited field of view and occlusions in NLoS regions) and are excluded during post-processing.} A complete summary of the counts of the samples and the storage footprints is provided in Table~\ref{tab:dataset_statistics}. 

\begin{table}[!t]
  \caption{\textcolor{blue}{Statistics of the XGait dataset.}}
  \label{tab:dataset_statistics}
  \centering
  \footnotesize
  \setlength{\tabcolsep}{3pt}

  \begin{tabular}{
      >{\centering\arraybackslash}m{2.5cm}  
      >{\centering\arraybackslash}m{2cm}
      >{\centering\arraybackslash}m{2cm}
      >{\centering\arraybackslash}m{2.1cm}
      >{\centering\arraybackslash}m{2cm}
  }
    \toprule
    \textbf{Modality} 
      & \textbf{Wi-Fi} 
      & \textbf{Acoustic} 
      & \textbf{Vision} 
      & \textbf{Ground-truth} \\
    \midrule
    File format 
      & \makecell[c]{user1-1-1-r1.dat}
      & \makecell[c]{user1-1-1-r1.wav}
      & user1-1-1.mp4
      & user1-1-1.npy \\
    Data representation
      & \makecell[c]{Raw CSI logs \\ ($1000\times3\times30\times t$)}
      & \makecell[c]{Audio waveform \\ ($48{,}000\times t$)}
      & video
      & \makecell[c]{trajectory points} \\
    \midrule
    \textbf{Valid Samples} \\
    \midrule
    Laboratory (20)
      & 10,816 (43.7GB)
      & 10,823 (18.8GB)
      & 10,730 (240GB)
      & 10,730 \\
    Home (19)
      & 4,753 (20.5GB)
      & 4,750 (8.6GB)
      & 1,848 (33.7GB)
      & 1,848 \\
    Meeting-room (20)
      & 5,005 (21.7GB)
      & 4,835 (7.8GB)
      & 2,000 (38.4GB)
      & 2,000 \\
    \makecell[c]{Meeting-room \\ (6, Different days)}
      & 1,710 (6.3GB)
      & 1,710 (3.6GB)
      & 679 (12.1GB)
      & 679 \\
    \textbf{All}
      & \textbf{22,284 (92.2GB)}
      & \textbf{22,118 (38.8GB)}
      & \textbf{15,257 (324.2GB)}
      & \textbf{15,257} \\
    \bottomrule
  \end{tabular}
\end{table}

\section{BENCHMARK PROCESSING PIPELINE}
\label{sec:benchmark_pipeline}
\begin{figure}[!t]	
	\centering	
	\includegraphics[width=15cm]{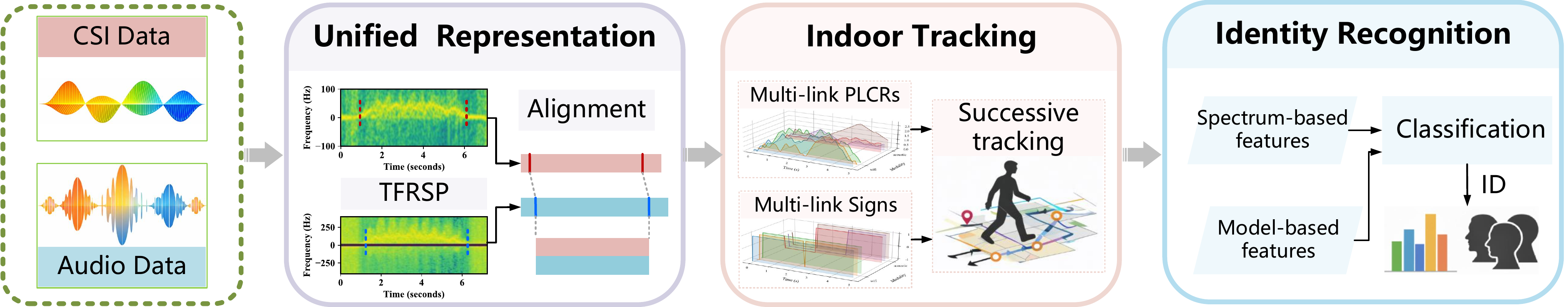}
	\caption{Benchmark processing pipeline.}
    \label{Fig4}
\end{figure}

\textcolor{blue}{To unify heterogeneous Wi-Fi CSI and acoustic signals, we develop a benchmark processing pipeline (Fig.~\ref{Fig4}) that maps both modalities into a shared Doppler spectrogram for velocity-based temporal alignment and motion representation, and further standardizes pre-processing and feature construction for indoor tracking and identity recognition.}

\subsection{Unified Representation}
\textbf{(1) Time--frequency analysis.} Due to substantial differences in carrier frequency, bandwidth, sampling rate, and propagation characteristics, directly fusing raw Wi-Fi CSI and acoustic measurements is impractical. We thus transform both modalities into a unified intermediate representation that highlights Doppler dynamics resulting from human walking. 

Human gait produces detectable Doppler shifts in both modalities. Inspired by this commonality, a Doppler-centric abstraction is adopted to generate spectrograms for each sensing link, serving as a consistent basis for cross-modality analysis. However, translating this abstraction into comparable spectra is non-trivial, as it requires harmonizing sampling rates and time--frequency resolutions across fundamentally different acquisition pipelines. For Wi-Fi, we apply CSI ratio~\cite{zeng2019farsense} pre-processing, followed by a $2\text{--}100\mathrm{\,Hz}$ band-pass filter. For acoustics, the received echo signal is first down-converted to baseband by quadrature demodulation. This is followed by low-pass filtering to isolate baseband components, which are then resampled from $48\,\mathrm{\,kHz}$ to $1\,\mathrm{kHz}$ to match the CSI packet rate. After resampling, we further suppress the strong near-DC component (within $\pm15\,\mathrm{Hz}$) caused by static reflections and residual carrier leakage around $0\,\mathrm{Hz}$. Otherwise, this dominant component could overwhelm weaker walking variations.

Following the modality-specific pre-processing, both Wi-Fi and acoustic streams are transformed into time--frequency spectra. We generate Doppler spectrograms using the time--frequency reassignment spectrum (TFRSP), which provides superior energy concentration in time and frequency compared to the standard short-time Fourier transform (STFT) spectrograms. With a uniform $1\,\mathrm{kHz}$ sampling rate, the resulting spectra achieve $1\,\mathrm{ms}$ temporal resolution and $1\,\mathrm{Hz}$ frequency resolution, allowing fine-grained characterization of rapid velocity variations during walking. Although the physical carrier wavelengths differ across modalities and thus yield different absolute Doppler ranges, the spectra share a consistent temporal grid and support a unified, velocity-oriented interpretation.

\textbf{(2) Spectrum alignment.} Without hardware synchronization, multi-link recordings from heterogeneous devices exhibit relative temporal shifts. Rigid synchronization, which is common in specialized laboratory equipment, is often impractical for commercial-off-the-shelf devices because of the absence of common clock interfaces and the unpredictable scheduling jitter in general operating systems. Rather than enforcing hardware synchronization, we adopt a PLCR-driven soft alignment strategy leveraging torso motion as a modality-invariant temporal anchor. 

In the Doppler spectrum, torso motion generally corresponds to the most stable and energetic component. For each link, a torso PLCR sequence is extracted to serve as alignment reference. Specifically, the dominant Doppler ridge is tracked over time using a dynamic programming algorithm that enforces local continuity across adjacent frames. The tracked frequency $f_D(t)$ is converted to PLCR as:
\begin{equation}
    \mathrm{PLCR}=-f_D(t)\lambda, 
\end{equation}
where $\lambda$ denotes the signal wavelength. To improve robustness and reduce computational cost, the resulting sequence is temporally down-sampled via block averaging and smoothed to obtain a stable envelope for correlation.

In the alignment phase, one link (typically Wi-Fi) is selected as the reference timeline. For all other links, the temporal offset is estimated by maximizing the normalized cross-correlation between their respective PLCR envelopes and the reference. The lag corresponding to the maximum correlation defines the global shift applied to all spectrograms associated with that link. Truncation or zero-padding is then employed to align these segments with the reference duration.

After coarse alignment, the effective motion window is detected by thresholding the smoothed reference envelope. This window is used to uniformly crop all modalities, removing static segments at the beginnings and ends. Finally, to avoid minor frame inconsistencies introduced by resampling and rounding, spectrograms are force aligned by truncating them to the minimum common length.

\subsection{Indoor Tracking}
\label{sec:Indoor_tracking}
This section outlines the methodology for recovering trajectories from multi-link PLCR observations across Wi-Fi and acoustics.  The concept of successive tracking of  PLCRs has been explored in previous work~\cite{qian2017widar,Wu2021GaitWay,tong2025stagr}, and a lightweight formulation is presented here to maintain compatibility with the proposed unified representation. 

\textbf{(1) Multi-link motion parameterization.} Following temporal alignment, each link $l$ provides a torso-level PLCR magnitude sequence $r_l(t)$. However, for tracking, magnitude alone is insufficient because it does not distinguish whether the subject is moving \emph{towards} or \emph{away from} the transceiver pair. Consequently,  a link-wise motion polarity is inferred from the \emph{sign} of the Doppler components, defined with respect to the propagation path of each specific link rather than a global coordinate system. The signed PLCR measurement, $\tilde{r}_l(t)$, is encoded as:
\begin{equation}
\tilde{r}_l(t) = s_l(t)\, r_l(t), \quad s_l(t)\in\{-1,+1\},
\end{equation}
where $s_l(t)$ is determined by comparing the energy asymmetry of positive and negative Doppler bins within the torso-dominant frequency band. All links from both modalities are subsequently integrated into a unified observation vector $\mathbf{z}(t)=[\tilde{r}_1(t),\ldots,\tilde{r}_L(t)]^\top$, which is associated with known transceiver coordinates $(\mathbf{p}^{\mathrm{tx}}_{l}, \mathbf{p}^{\mathrm{rx}}_{l})$.

\textbf{(2) Successive tracking.} Let $\mathbf{x}(t)\in\mathbb{R}^2$ and $\mathbf{v}(t)\in\mathbb{R}^2$ denote the subject's position and instantaneous velocity, respectively. For the link $l$, the total length of the bistatic path is defined as $d_l(t)=\|\mathbf{x}(t)-\mathbf{p}^{\mathrm{tx}}_{l}\|+\|\mathbf{x}(t)-\mathbf{p}^{\mathrm{rx}}_{l}\|$. The PLCR measurement corresponds to the time derivative $r_l(t)=\dot d_l(t)$. Through a first-order linearization, the signed PLCR is expressed as the projection of the velocity vector onto the  unit direction determined by the link geometry:
\begin{equation}
\tilde{r}_{l}(t) \approx 
\mathbf{a}_{l}^\top(\mathbf{x}(t))\, \mathbf{v}(t),
\quad
\mathbf{a}_{l}(\mathbf{x})=
\frac{\mathbf{x}-\mathbf{p}^{\mathrm{tx}}_{l}}{\|\mathbf{x}-\mathbf{p}^{\mathrm{tx}}_{l}\|}
+
\frac{\mathbf{x}-\mathbf{p}^{\mathrm{rx}}_{l}}{\|\mathbf{x}-\mathbf{p}^{\mathrm{rx}}_{l}\|}.
\label{eq:plcr_projection}
\end{equation}
Stacking the observations from all $L$ links gives the linear system:
\begin{equation}
\mathbf{z}(t) \approx \mathbf{A}(\mathbf{x}(t))\,\mathbf{v}(t),
\end{equation}
where $\mathbf{A}(\mathbf{x}(t))\in\mathbb{R}^{L\times 2}$ formed by stacking the row vectors $\mathbf{a}_l^\top(\mathbf{x}(t))$. Given an estimate of $\mathbf{x}(t)$, the velocity $\mathbf{v}(t)$ is solved by weighted least squares:
\begin{equation}
\hat{\mathbf{v}}(t)=\arg\min_{\mathbf{v}}\ \|\mathbf{W}^{1/2}\big(\mathbf{A}(\mathbf{x}(t))\mathbf{v}-\mathbf{z}(t)\big)\|_2^2,
\quad
\hat{\mathbf{v}}(t)=\big(\mathbf{A}^\top \mathbf{W}\mathbf{A}\big)^{-1}\mathbf{A}^\top \mathbf{W}\mathbf{z}(t),
\label{eq:ls_velocity}
\end{equation}
where $\mathbf{W}$ denotes a diagonal weight matrix based on link reliability.
The subject's position is then updated using a discrete-time motion model:
\begin{equation}
\mathbf{x}(t{+}1)=\mathbf{x}(t)+\hat{\mathbf{v}}(t)\Delta t,
\label{eq:state_update}
\end{equation}
with $\Delta t$ being the sampling interval of the aligned PLCRs.
Initialized from a known starting point, this procedure is iterated to reconstruct the trajectory $\hat{\mathbf{x}}(t)$ and the velocity profile $\hat{\mathbf{v}}(t)$. By aggregating heterogeneous links in a single projection system, the tracker naturally benefits from complementary viewpoints of Wi-Fi and acoustics, providing the necessary spatial context for subsequent gait analysis.

\subsection{Identity Recognition}
\label{sec:Gait Recognition}
To illustrate the usability of XGait for identity recognition, the benchmark supports two feature extraction paradigms that reflect common practices in wireless gait sensing.

\textbf{(1) Spectrum-based features.} The first paradigm performs gait recognition directly on spectrograms using deep learning models. Time--frequency representations from different modalities are used independently or jointly as input to neural networks, enabling an evaluation of modality complementarity at the representation level. This method aligns with the mainstream of recent wireless gait recognition studies~\cite{yang2025environment, zhang2021wi,wang2025muid}. Notably, it typically operates without incorporating the subject's position, largely because effectively fusing dynamic trajectory data with high-dimensional spectrograms within a deep learning framework remains an open issue. 

In the benchmark, the spectrograms are resampled along the time dimension using the same grid adopted for PLCR estimation, while the native frequency resolution of each modality is preserved.  For a  $1\,\mathrm{s}$ gait segment, the resulting Wi-Fi and acoustic spectrograms have dimensions of $201\times T_s$ and $801\times T_s$, respectively, where the first dimension corresponds to frequency bins and the second to time frames. When multiple links are available, each link's spectrogram is treated as one channel and stacked along the channel dimension, resulting in a multi-channel spectrogram tensor per modality. Since gait trials vary in duration across paths and walking speeds, each sample is converted to a fixed-length tensor by trimming overly long segments and zero-padding shorter ones along the time dimension, enabling efficient batching for training and evaluation.

\textbf{(2) Model-based features.} The second paradigm addresses the limitations of raw spectral analysis by representing gait through explicit motion descriptors. These descriptors are derived by integrating Doppler spectra with tracking outputs to recover physically interpretable gait patterns. Compared with purely spectrum-driven learning, this approach leverages the subject's trajectory to map link observations into a structured kinematic space, thereby transforming location-dependent Doppler energy into a normalized body representation.

Representative methods in this line include the AcousticID system~\cite{xu2019acousticid}, the body-coordinate velocity profile (GBVP)~\cite{zhang2021gaitsense}, and the polar-coordinate velocity profile (PPVP)~\cite{tong2025stagr}. AcousticID extracts cycle-level attributes (e.g., gait length and cycle duration) from path-dependent signals, which are highly sensitive to the walking trajectory and sensing geometry. GBVP is particularly relevant for benchmarking as it models body-centric velocity patterns by jointly explaining multi-link measurements through optimization. However, solving the high-dimensional optimization problem in GBVP becomes computationally prohibitive for acoustic sensing due to the wider frequency range. Therefore, we adopt PPVP as the primary model-based descriptor. PPVP projects Doppler measurements into a polar coordinate system that explicitly encodes motion direction and velocity. With $1^\circ$ angular resolution and 40 velocity bins, the resulting PPVP feature size is $360 \times 40 \times T_s$, where $T_s$ denotes the number of trajectory points.

\begin{table}[t]
  \caption{Learning network candidates.}
  \label{tab:model_zoo}
  \centering
  \small
  \begin{tabular}{
    >{\centering\arraybackslash}p{3.5cm}
    >{\centering\arraybackslash}p{1.5cm}
    >{\centering\arraybackslash}p{4cm}
    >{\centering\arraybackslash}p{3.cm}
  }
    \toprule
    \textbf{Model (ID)} & \textbf{Backbone} & \textbf{Temporal Modeling} & \textbf{Fusion Mechanism} \\
    \midrule
    CNN (M1) &
    CNN &
    -- &
    Late fusion (concat) \\
    CNN + LSTM (M2) &
    CNN &
    LSTM &
    Late fusion (concat) \\
    CNN + Transf. (M3) &
    CNN &
    Transf. enc. &
    Late fusion (concat) \\
    ResNet + Transf. (M4) &
    ResNet &
    Transf. enc. &
    Late fusion (concat) \\
    Cross-attn Transf. (M5) &
    CNN &
    Self-attn + Cross-attn &
    Cross-attention fusion \\
    \bottomrule
  \end{tabular}
\end{table}

\textbf{(3) Learning model candidates.} To support systematic evaluation across different features and fusion strategies, the benchmark includes a set of representative learning models as summarized in Table~\ref{tab:model_zoo}. Rather than advocating for a specific architecture, these models are selected to span common design choices in wireless sensing and time--frequency-based activity analysis. These cover: (i) convolutional backbones for spatial-spectral pattern extraction, (ii) recurrent and Transformer-style architectures for temporal modeling, and (iii) different cross-modality fusion mechanisms, such as late fusion and cross-attention. This design enables a controlled analysis of how model capacity and fusion strategy interact with modality complementarity across diverse learning frameworks.

\subsection{Implementation}
The complete benchmark processing pipeline will be released alongside the XGait dataset to support reproducible evaluation. The system is implemented in a hybrid Python--MATLAB architecture. For computationally intensive time--frequency analysis, the pipeline invokes optimized MATLAB routines (e.g., TFRSP) through a lightweight runtime interface. The implementation is cross-platform, supporting both Linux and Windows environments, and is compatible with GPU acceleration to facilitate large-scale training and evaluation. 


The benchmark is intended to serve as a standardized reference pipeline rather than a finalized solution. Its objective is to provide a consistent framework for studying the complementarity of Wi-Fi and acoustic sensing while enabling fair comparisons across diverse modeling and fusion strategies. This baseline is expected to facilitate future research in areas such as advanced synchronization, robust tracking formulations, and advanced representation learning on XGait dataset.

\section{VALIDATION}
\label{sec:validation}
This section validates the utility of XGait and the associated benchmark pipeline through two primary tasks: \textit{human tracking} and \textit{identity recognition}. The objective is not to establish a new state-of-the-art performance, but to show that: (i) the multi-modal dataset provides rich, complementary information that reliably supports both tasks in realistic indoor conditions, and (ii) the distinct behaviors of Wi-Fi and acoustic sensing can be systematically compared and fused within a unified analytical framework.

\subsection{Evaluation Setups}
Three sensing configurations are considered for evaluation: Wi-Fi only, acoustic only, and fusion. To ensure a fair comparison, all methods utilize the identical pre-processing and alignment procedure detailed in Section~\ref{sec:benchmark_pipeline}.

\begin{figure}[!t]	
	\centering	
	\includegraphics[width=15cm]{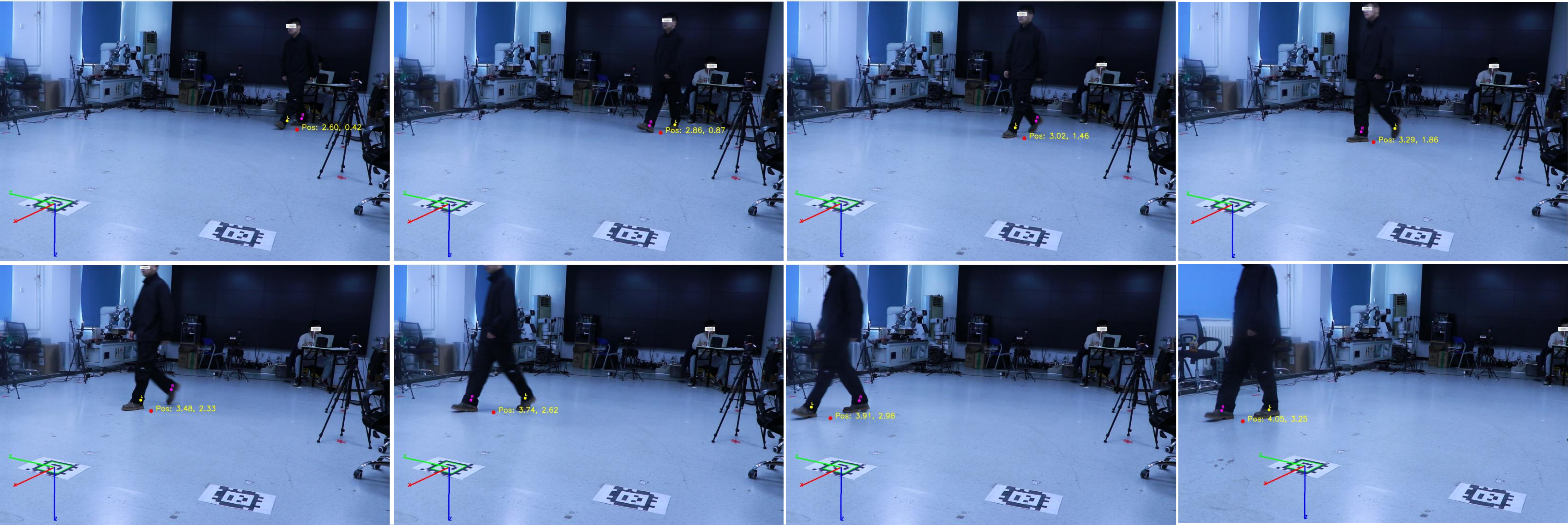}
	\caption{Ground-truth trajectories extraction.}
    \label{fig5}
\end{figure}

\textbf{(1) Human tracking setups.} Gait videos are used to provide ground-truth trajectories for tracking evaluation. We adopt a vision-based pipeline that combines AprilTag~\cite{AprilTag2011} fiducial markers with YOLOv8 pose estimation. Initially, tags placed within the environment are detected to solve a calibrated perspective-n-point (PnP) problem, yielding the camera pose $[R|t]$ in the world coordinate frame. Subsequently, each video frame is fed into the pose model to infer human key points, from which to extract the 2-D pixel coordinates $(u,v)$ of the subject's feet. The geometric center of these coordinates is defined as the projection point in the image plane, as shown in Fig.~\ref{fig5}. Finally, utilizing inverse perspective projection, this 2-D point is back-projected into the 3-D world coordinate system using the estimated camera pose and intrinsic parameters, thereby reconstructing the physical walking trajectory.

\textcolor{blue}{Given the scale of XGait, tracking performance is evaluated on a representative subset with reliable ground-truth annotations. We select 3 males ($\text{User}_{1,2,3}$) and 1 female ($\text{User}_{13}$) who appear across all three scenarios, chosen to reflect diversity in gender as well as physical characteristics (i.e., height and body build), including 2,157 trials from the laboratory dataset and 400 trials each from the home and meeting-room datasets.}

\textbf{(2) Identity recognition setups.} For gait recognition, wireless recordings are treated as labeled samples. Performance is assessed under both Doppler spectrum and model-based feature paradigms described in Section~\ref{sec:Gait Recognition}. The evaluation encompasses all available wireless samples with identity labels in each scenario. The dataset is partitioned into training, validation, and test sets according to specific protocols in which the test samples are strictly disjoint from the training samples. A subject-closed setting is adopted, where identities in the test set are present during training, reflecting common deployment scenarios such as monitoring household occupants.

\subsection{Human Tracking Validation}
As introduced in Section~\ref{sec:Indoor_tracking}, the tracker reconstructs 2-D trajectories from multi-link torso PLCR curves and their associated motion polarities. The following analysis evaluates performance in different environments to highlight the distinct strengths of each modality and the synergistic effects of their fusion.

\begin{figure}[!t]	
	\centering	
	\subfigure[]{		
		\label{fig6a} 
		\includegraphics[width=1.8in]{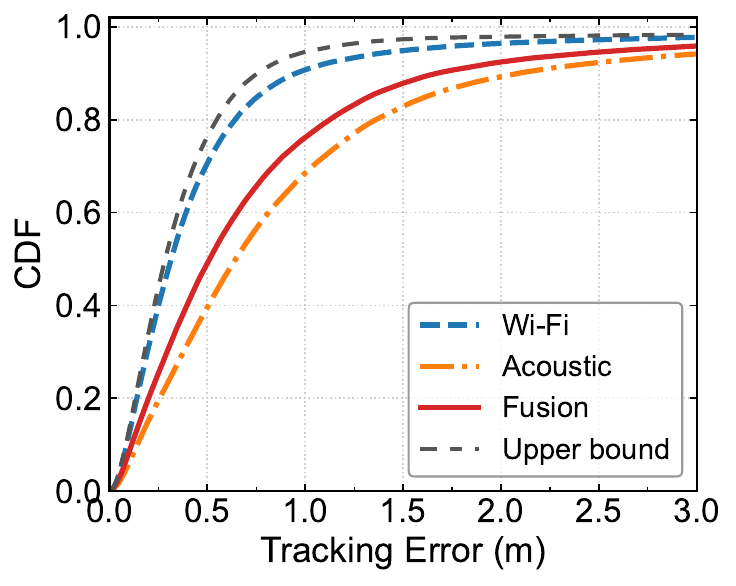}}	
	\subfigure[]{		
		\label{fig6b} 
		\includegraphics[width=1.8in]{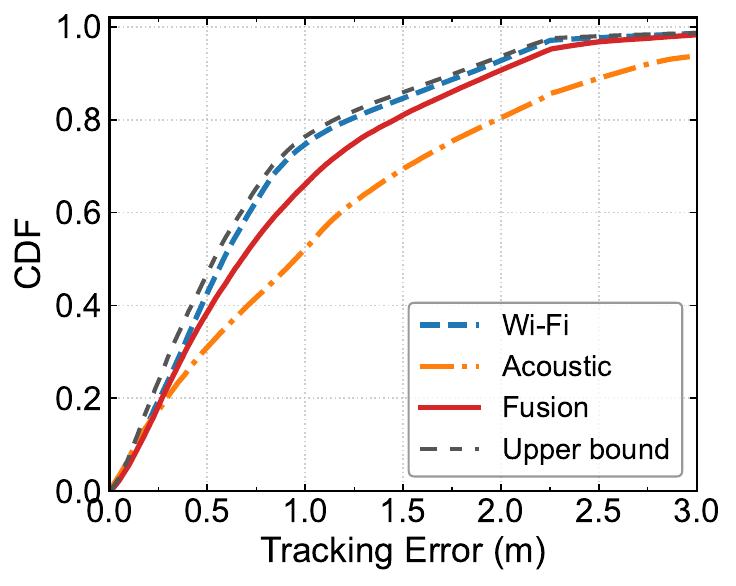}}	
    \subfigure[]{		
		\label{fig6c} 
		\includegraphics[width=1.8in]{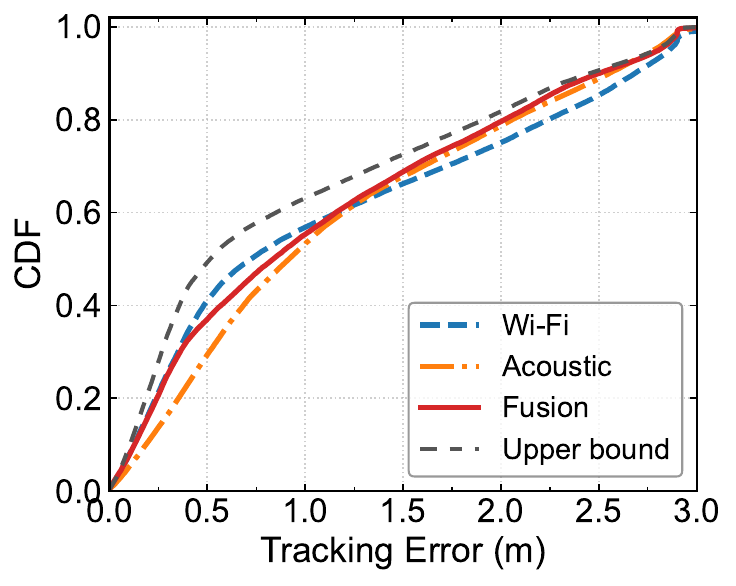}}	
	\caption{\textcolor{blue}{Overall tracking results. (a) CDF of position errors in Laboratory. (b) CDF of position errors in Home. (c) CDF of position errors in Meeting-room.}
	}
	\label{fig6} 
\end{figure}

\textbf{(1) Validation in the laboratory scene.} Results in this scenario reveal a clear advantage for Wi-Fi-based tracking. \textcolor{blue}{As shown in Fig.~\ref{fig6a}, Wi-Fi achieves higher tracking accuracy, with approximately 90\% of localization points exhibiting an error within $1\,\mathrm{m}$. In contrast, acoustic tracking is less reliable, with only 70\% of the samples meeting the same threshold. As a result, the fused tracking is often constrained by the weaker acoustic modality.} To establish an empirical performance ceiling, an oracle-based reference is computed by selecting the minimum error achieved among the three configurations for each individual trial. This reference represents the best possible performance attainable through an ideal modality selection. The proximity of the Wi-Fi curve to this bound is consistent with the physical limitations of acoustic sensing, where strong directionality and rapid signal attenuation increase sensitivity to deployment geometry and environmental noise.

Despite the average dominance of Wi-Fi, fusion delivers tangible gains on a significant subset of trajectories. \textcolor{blue}{At the trajectory level, fusion achieves the lowest mean error on 440 trajectories (20.4\%). More broadly, it outperforms acoustic-only tracking on 1375 (63.75\%) trajectories and Wi-Fi-only tracking on 543 (25.17\%) trajectories.} These results indicate that even when one modality is globally stronger, complementary observations can rectify specific failures caused by localized motion patterns. 

\begin{figure}[!t]	
	\centering	
	\subfigure[]{		
		\label{fig7a} 
		\includegraphics[width=2.8in]{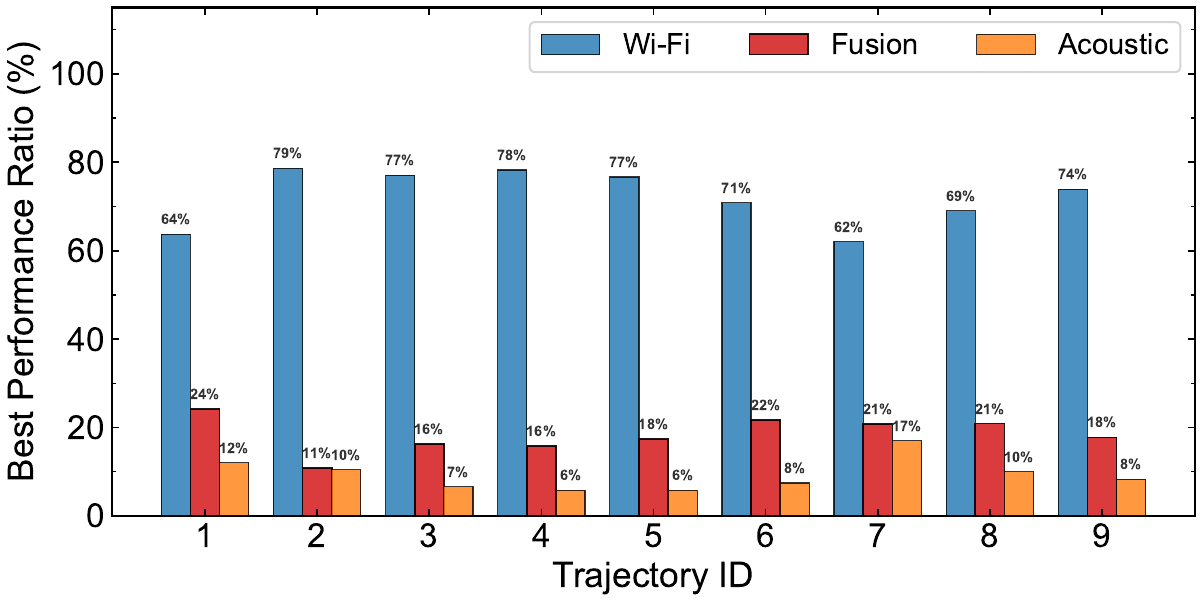}}	
	\subfigure[]{		
		\label{fig7b} 
		\includegraphics[width=2.7in]{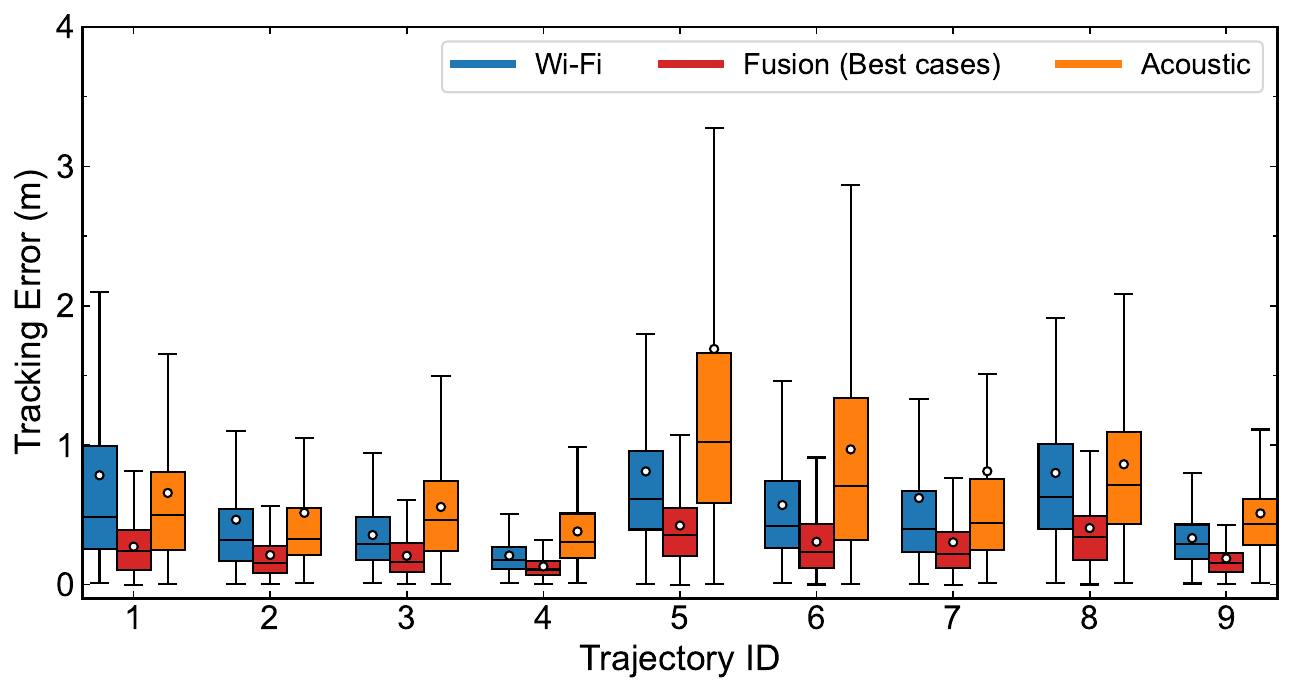}}	
	\caption{\textcolor{blue}{Different trajectories performance in Laboratory. (a) Best performance ratio. (b) Tracking error box chart. }
	}
	\label{fig7} 
\end{figure}

\textcolor{blue}{The breakdown by trajectory complexity in Fig.~\ref{fig7a} confirms these performance trends across different motion patterns. For simple linear trajectories (IDs \#1--\#4), Wi-Fi maintains a clear advantage, though its dominance fluctuates. Notably, in trajectory \#1, this advantage is slightly reduced, as portions of the path lie at the maximum distance from the transceivers, where the lower SNR can degrade performance. Excluding this case, the fusion-win ratio for linear paths remains around 10\%. In contrast, for more complex trajectories involving curves and turns (IDs \#5--\#9), the fusion-win ratio increases to 18\%--21\%. This trend suggests that multi-modal integration is especially effective in resolving local ambiguities when a single modality becomes vulnerable to heading changes.}

\begin{figure}[!t]	
	\centering	
	\subfigure[]{		
		\label{fig8a} 
		\includegraphics[width=1.5in]{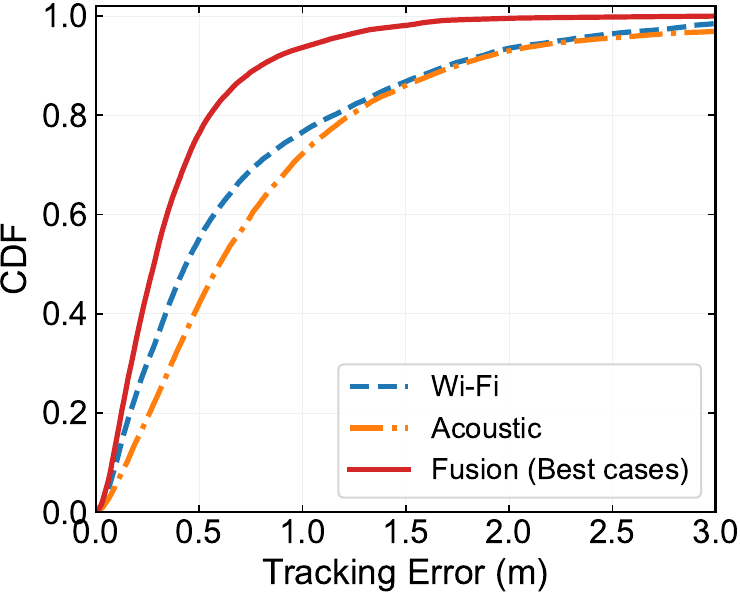}}	
	\subfigure[]{		
		\label{fig8b} 
		\includegraphics[width=1.5in]{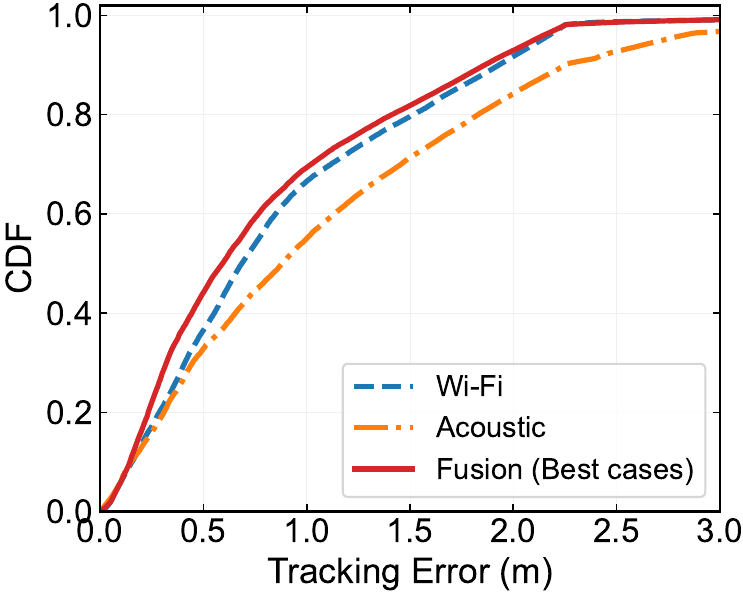}}	
    \subfigure[]{		
		\label{fig8c} 
		\includegraphics[width=1.5in]{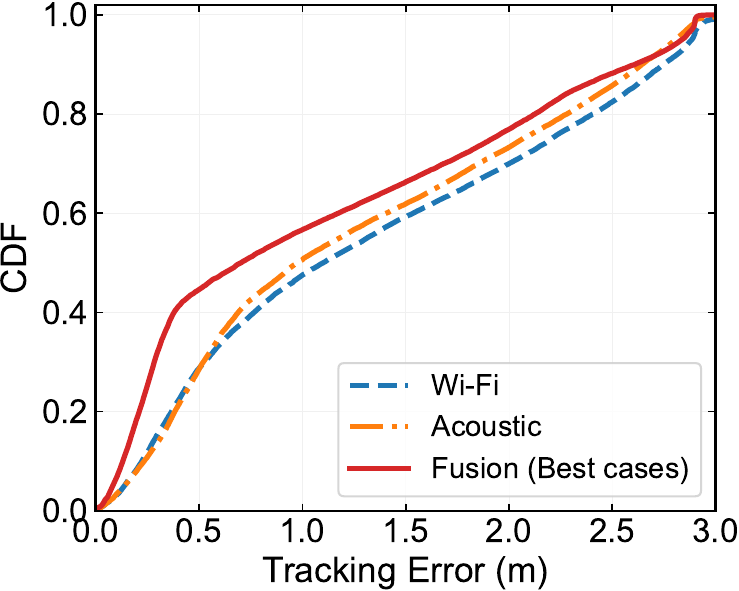}}
    \subfigure[]{		
		\label{fig8d} 
		\includegraphics[width=1.5in, height=1.22in]{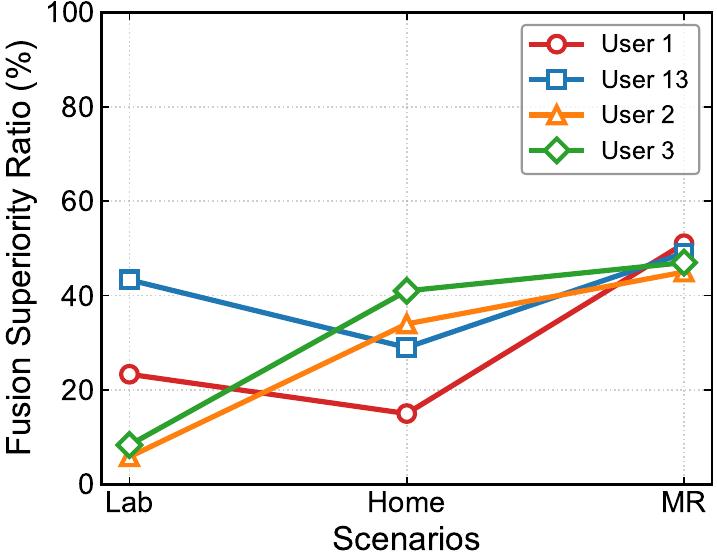}}
	\caption{\textcolor{blue}{Best fusion cases. (a) CDF of position errors in Laboratory. (b) CDF of position errors in Home. (c) CDF of position errors in Meeting-room. (d) Fusion superiority ratio across different users.}
	}
	\label{fig8} 
\end{figure}

\begin{figure}[!t]	
	\centering	
	\subfigure[]{		
		\label{fig9a} 
		\includegraphics[width=1.5in]{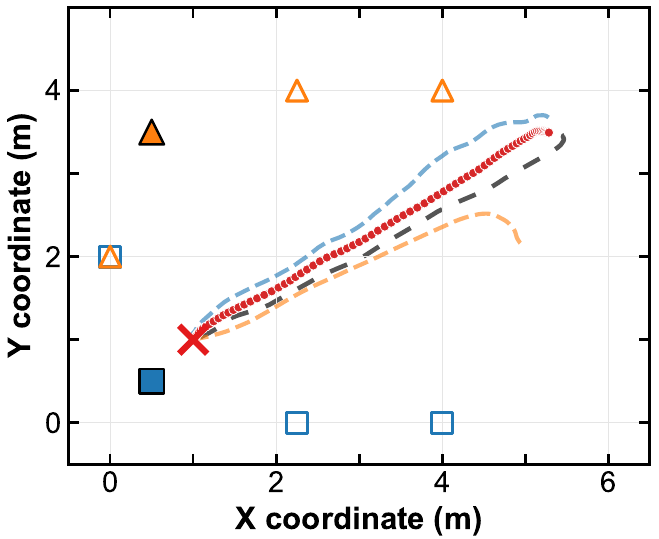}}	
	\subfigure[]{		
		\label{fig9b} 
		\includegraphics[width=1.5in]{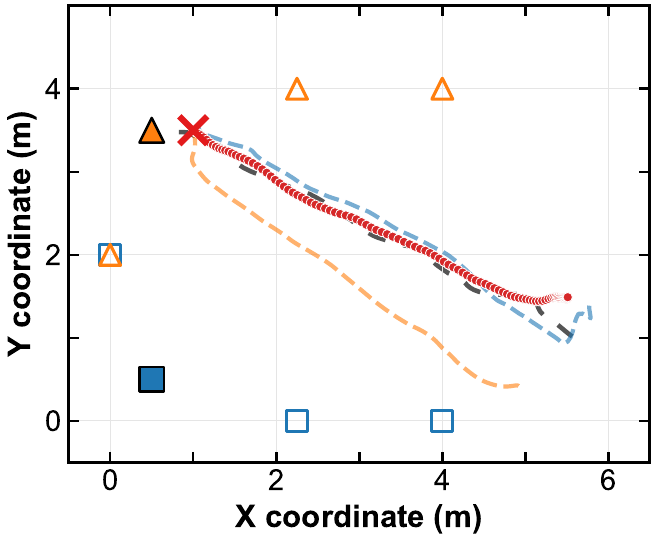}}	
    \subfigure[]{		
		\label{fig9c} 
		\includegraphics[width=1.5in]{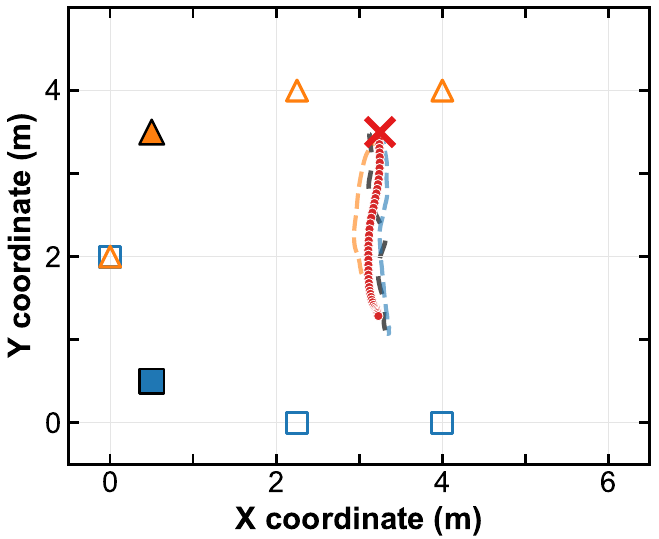}}
    \subfigure[]{		
		\label{fig9d} 
		\includegraphics[width=1.5in]{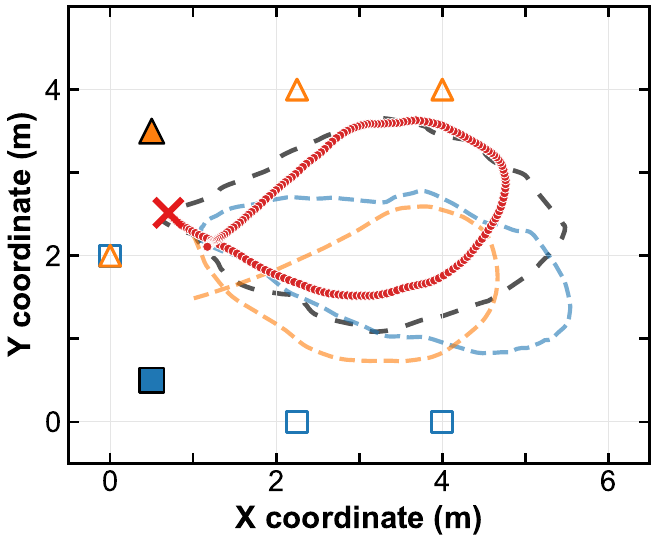}}
    \subfigure[]{		
		\label{fig9e} 
		\includegraphics[width=1.5in]{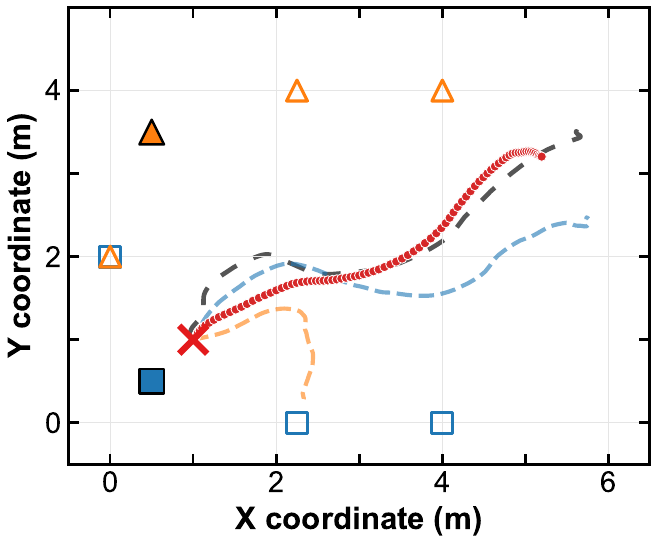}}
    \subfigure[]{		
		\label{fig9f} 
		\includegraphics[width=1.5in]{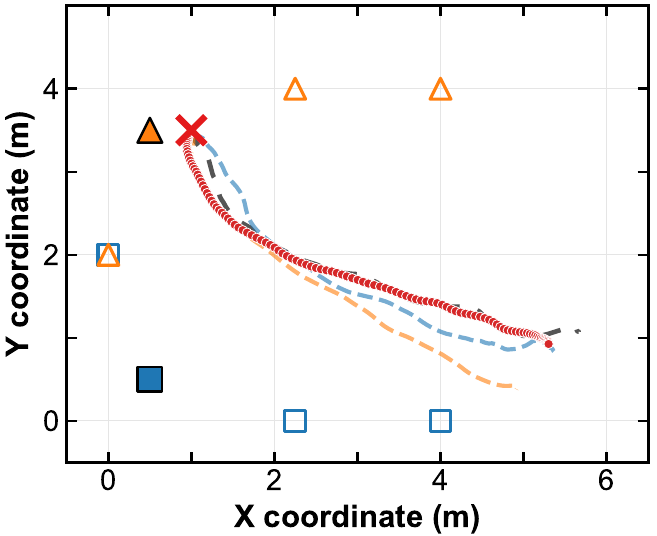}}
    \subfigure[]{		
		\label{fig9g} 
		\includegraphics[width=1.5in]{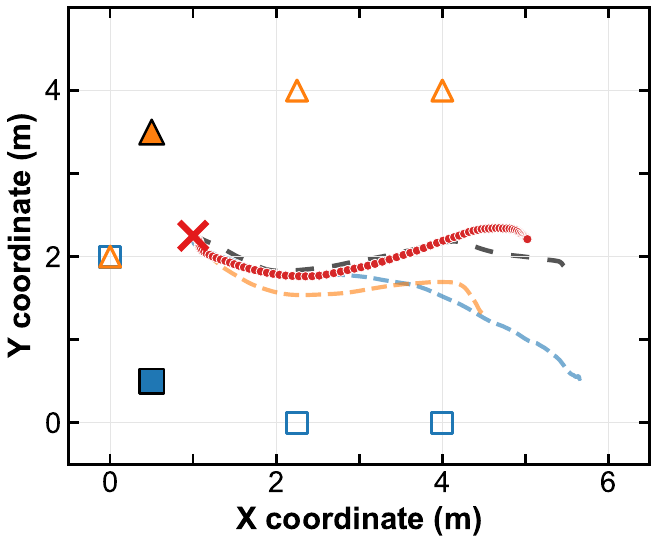}}
    \subfigure[]{		
		\label{fig9h} 
		\includegraphics[width=1.5in]{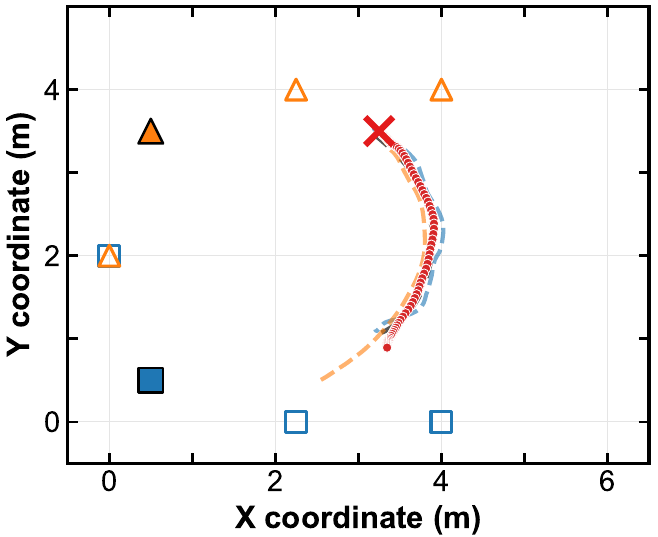}}
    \subfigure[]{		
		\label{fig9i} 
		\includegraphics[width=1.5in,height = 1.15in]{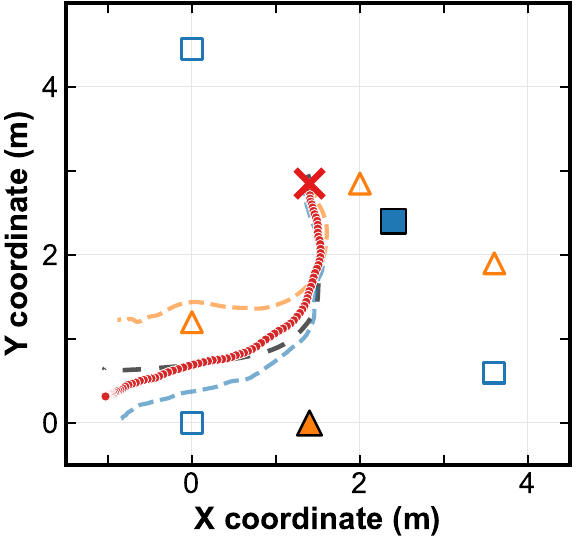}}
    \subfigure[]{		
		\label{fig9j} 
		\includegraphics[width=1.5in,height = 1.15in]{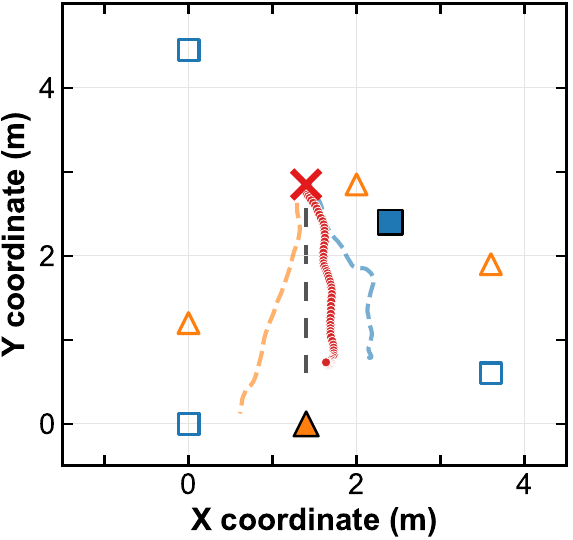}}
    \subfigure[]{		
		\label{fig9k} 
		\includegraphics[width=1.5in,height = 1.15in]{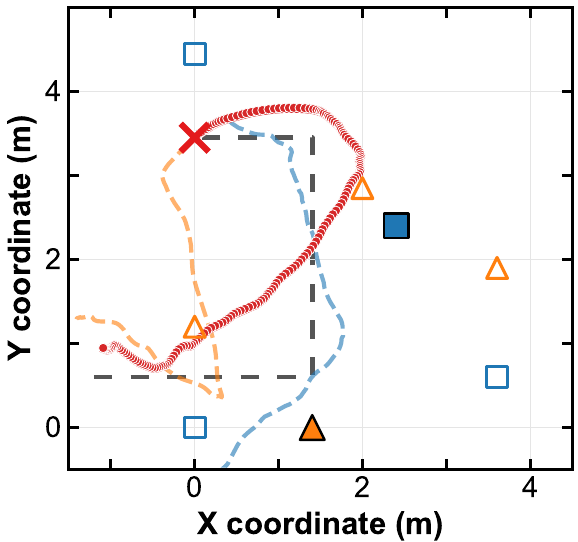}}
    \subfigure[]{		
		\label{fig9l} 
		\includegraphics[width=1.5in,height = 1.15in]{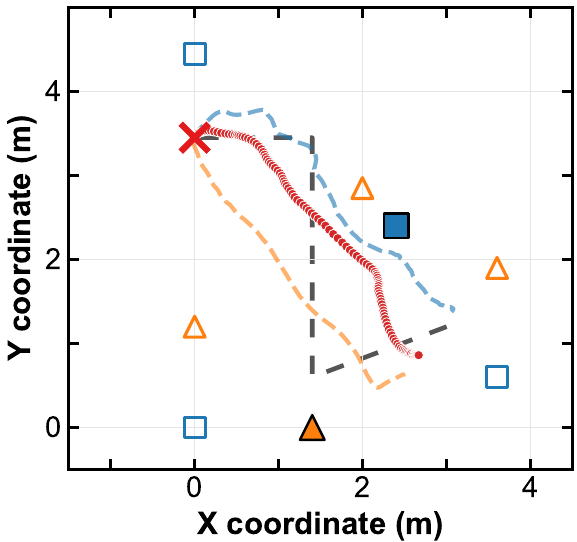}}
    \subfigure[]{		
		\label{fig9m} 
		\includegraphics[width=1.5in]{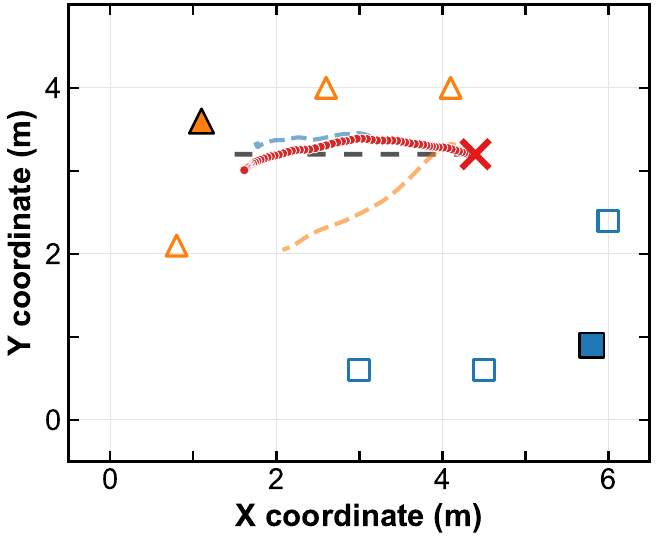}}
    \subfigure[]{		
		\label{fig9n} 
		\includegraphics[width=1.5in]{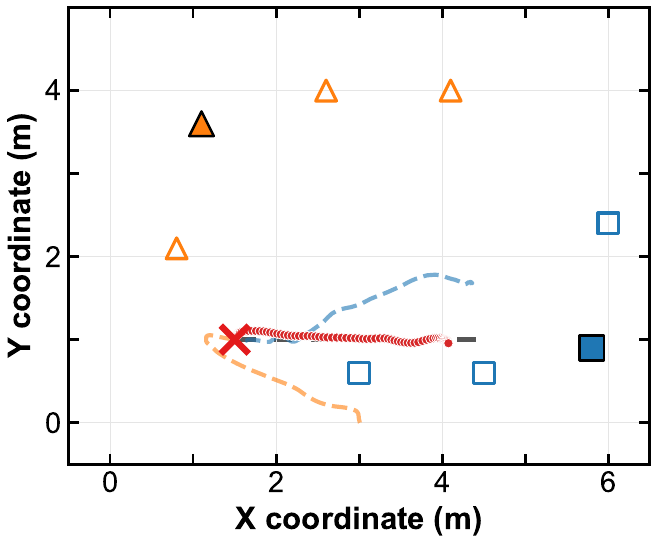}}
    \subfigure[]{		
		\label{fig9o} 
		\includegraphics[width=1.5in]{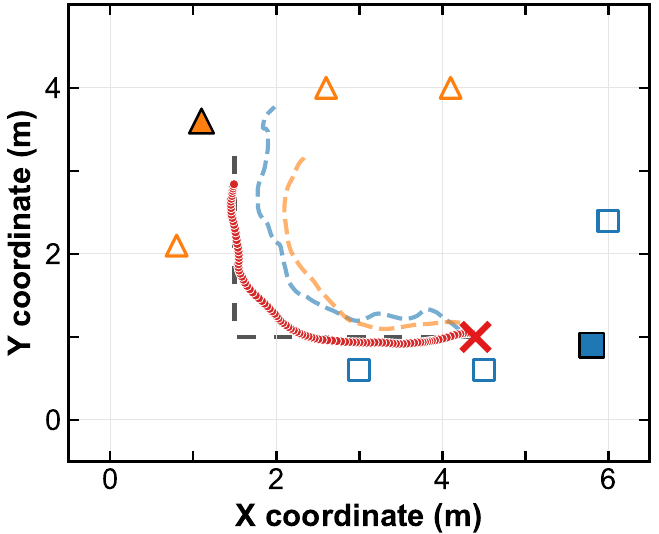}}
    \subfigure[]{		
		\label{fig9p} 
		\includegraphics[width=1.5in]{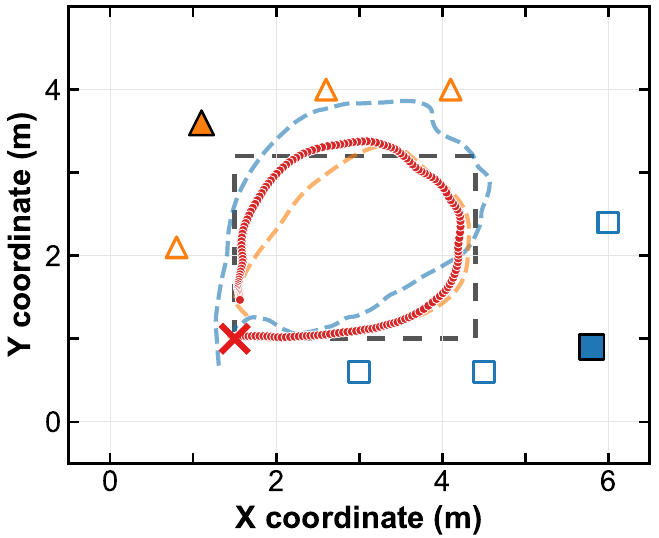}}

    \vspace{5pt} 

    \begin{tikzpicture}[font=\small, x=1cm, y=1cm]
        \begin{scope}
            \draw[thick, gray!100, dash pattern=on 3pt off 2pt] (0,0) -- (0.5,0) 
                node[right, black, xshift=-0.1cm] {GT};
                
            \begin{scope}[xshift=1.6cm]
                \draw[ultra thick, red!80] (0,0) -- (0.5,0) node[right, black, xshift=-0.1cm] {Fusion};
                \filldraw[red!80] (0.25,0) circle (1.8pt);
                \fill[white] (0.25,0) circle (0.8pt); 
            \end{scope}

            \begin{scope}[xshift=3.7cm]
                \draw[thick, wifiblue, dashed] (0,0) -- (0.5,0) node[right, black, xshift=-0.1cm] {WiFi};
            \end{scope}

            \begin{scope}[xshift=5.4cm]
                \draw[thick, orange, dash pattern=on 4pt off 1.5pt on 1pt off 1.5pt] (0,0) -- (0.5,0) 
                    node[right, black, xshift=-0.1cm] {Acoustic};
            \end{scope}
        \end{scope}

        \begin{scope}[yshift=-0.4cm]
            \node[draw=black, fill=wifiblue, inner sep=1.8pt] at (0,0) {};
            \node[draw=wifiblue, fill=white, thick, inner sep=1.8pt] at (0.4,0) {};
            \node[right, black] at (0.6,0) {WiFi TX/RX};

            \begin{scope}[xshift=3.1cm]
                \node[draw=black, fill=orange, regular polygon, regular polygon sides=3, inner sep=1pt] at (0,0) {};
                \node[draw=orange, fill=white, thick, regular polygon, regular polygon sides=3, inner sep=1pt] at (0.4,0) {};
                \node[right, black] at (0.6,0) {Acoustic TX/RX};
            \end{scope}

            \begin{scope}[xshift=6.7cm]
                \draw[ultra thick, red!80] (-0.1,0.1) -- (0.1,-0.1);
                \draw[ultra thick, red!80] (-0.1,-0.1) -- (0.1,0.1);
                \node[right, black] at (0.1,0) {Start};
            \end{scope}
        \end{scope}
    \end{tikzpicture}

	\caption{Best fusion cases. (a)-(h) Results in Laboratory. (i)-(l) Results in Home. \textcolor{blue}{(m)-(p) Results in Meeting-room.}
	}
	\label{fig9} 
\end{figure}
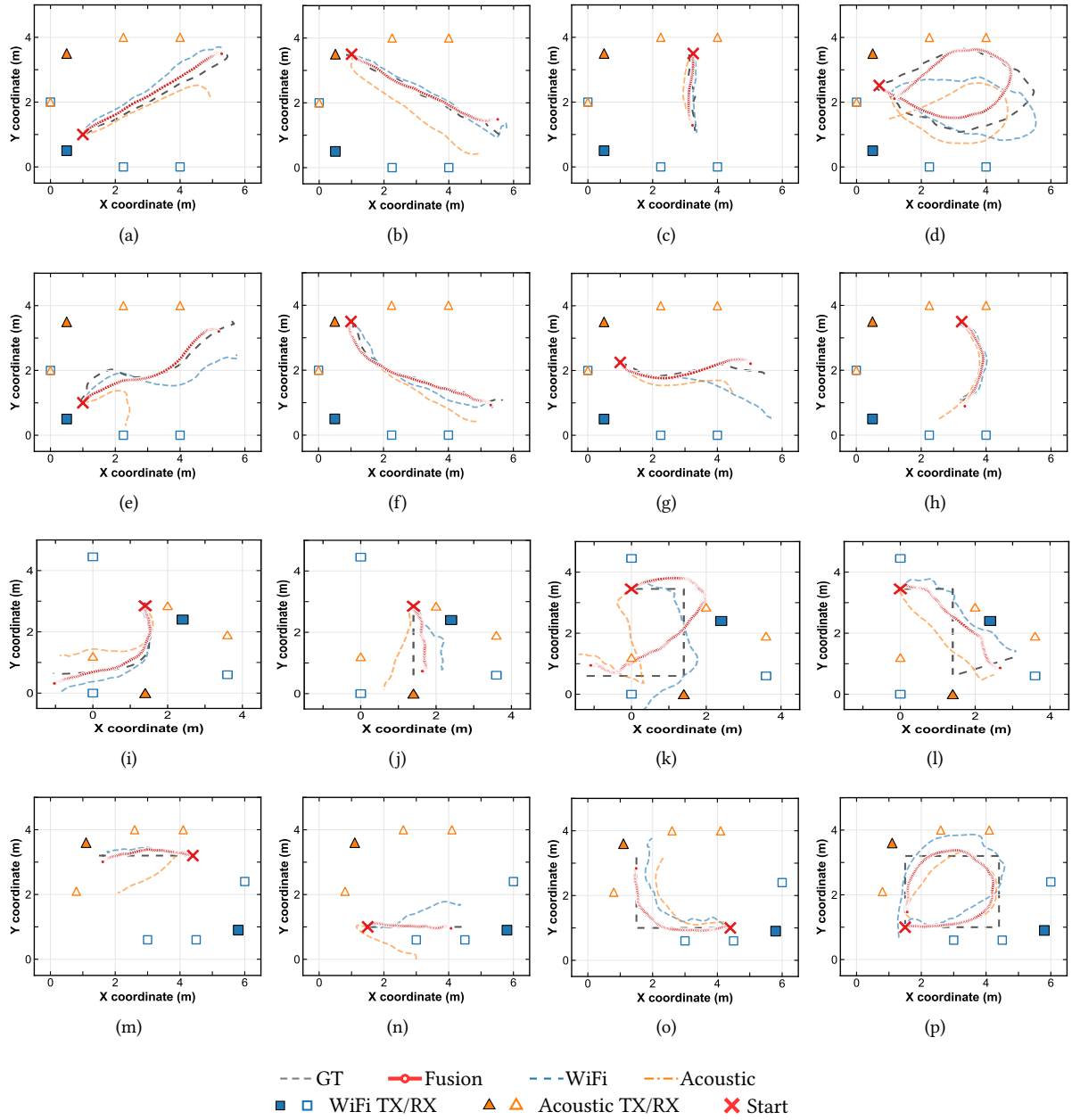

\textcolor{blue}{To better quantify these improvements, analysis focuses on the subset of 440 trajectories where fusion achieves the lowest error. As shown in the error CDF in Fig.~\ref{fig8a}, fusion increases the proportion of trajectories within a $1\,\mathrm{m}$ mean error to 91\%, while both Wi-Fi and acoustic sensing remain below 80\%. Beyond improving accuracy, fusion also reduces error fluctuation by compensating for modality-specific failures. This stabilizing effect is illustrated in Fig.~\ref{fig7b}. When one modality suffers from sporadic degradation, the complementary modality anchors the estimate, maintaining trajectory smoothness and consistency. Qualitative examples in Figs.~\ref{fig9a}--\ref{fig9h} further support these observations, showing that fusion better preserves global trajectory geometry, particularly in segments affected by local ambiguities or transient signal dropouts. By effectively bridging these gaps, fusion enables more robust and physically consistent trajectory reconstruction.}

\textbf{(2) Validation in the home scene.} The residential setting introduces increased multipath interference and layout irregularities, leading to greater performance fluctuations even for Wi-Fi sensing. \textcolor{blue}{As shown in Fig.~\ref{fig6b}, the fraction of Wi-Fi localization points within $1\,\mathrm{m}$ drops below 75\%, representing a decrease of 15 percentage points relative to the laboratory scenario. A similar degradation is observed for acoustic sensing, indicating that both modalities are susceptible to the complexities of home environments. Despite this, Wi-Fi maintains its relative advantage over acoustics. Notably, fusion performance shows improved alignment with the leading modality, as the gap between the fusion and Wi-Fi curves narrows compared to the laboratory setting.}

\textcolor{blue}{This increased environmental complexity further highlights the importance of multi-modal integration. In the residential environment, fusion achieves the lowest mean error in 29.75\% of the trajectories, outperforming Wi-Fi and acoustic tracking in 37.7\% and 83\% of the cases, respectively. These results represent a substantial improvement over the approximately 20\% win ratio observed in the laboratory. However, although the frequency of optimal cases increases, the marginal gain in terms of absolute error reduction appears to diminish. This is further illustrated in Fig.~\ref{fig8b}, where the reduced gap between the fusion and Wi-Fi curves indicates that the severe degradation of both individual modalities limits the scope for further precision gains. In this context, fusion acts as a stabilizer that effectively maintains the performance upper bound across challenging home layouts. Qualitative inspections in Figs.~\ref{fig9i}--\ref{fig9l}, including LoS and NLoS transitions, further confirm that the benefits of fusion become more pronounced as environmental complexity increases.}

\textcolor{blue}{\textbf{(3) Validation in the meeting-room scene.} This scenario presents the most challenging conditions in our evaluation. Although the residential layout in Fig.~\ref{fig2} is more spatially fragmented, the meeting room introduces a more severe sensing environment due to the large centralized conference table. Crucially, as the transceiver height is comparable to the table height, direct propagation paths are consistently blocked, resulting in persistent NLoS conditions and dynamic shadowing. As illustrated in Fig.~\ref{fig6c}, this leads to a sharp degradation in single-modality performance, with the fraction of localization points within a $1\,\mathrm{m}$ error threshold dropping to approximately 45\% for Wi-Fi and 40\% for acoustics, as their performance curves converge under severe multipath interference.}

\textcolor{blue}{Under these extreme conditions, the benefit of multi-modal integration becomes increasingly evident. Fusion achieves the lowest mean error in 48\% of the trajectories, outperforming Wi-Fi-only tracking in 71.25\% and acoustic-only tracking in 63.75\% of the cases. This represents a more pronounced improvement than in the laboratory and residential settings. As further shown in Fig.~\ref{fig8c}, while single modalities are prone to catastrophic failures under persistent occlusion, fusion maintains significantly higher resilience by leveraging complementary signatures to bridge sensing gaps. In addition, consistent performance across different individuals is observed in Fig.~\ref{fig8d}, reaching its peak in the meeting-room scenario. While minor fluctuations are present at the home stage for certain subjects, the overall trend remains consistent across both scenario-level and subject-level analyses: as environmental complexity increases from the laboratory to the home and meeting-room settings, the benefits of fusion become progressively more pronounced. The representative trajectories are shown in Figs.~\ref{fig9m}--\ref{fig9p}.}

\begin{figure}[!t]	
	\centering	
	\subfigure[]{		
		\label{fig10a} 
		\includegraphics[width=1.85in]{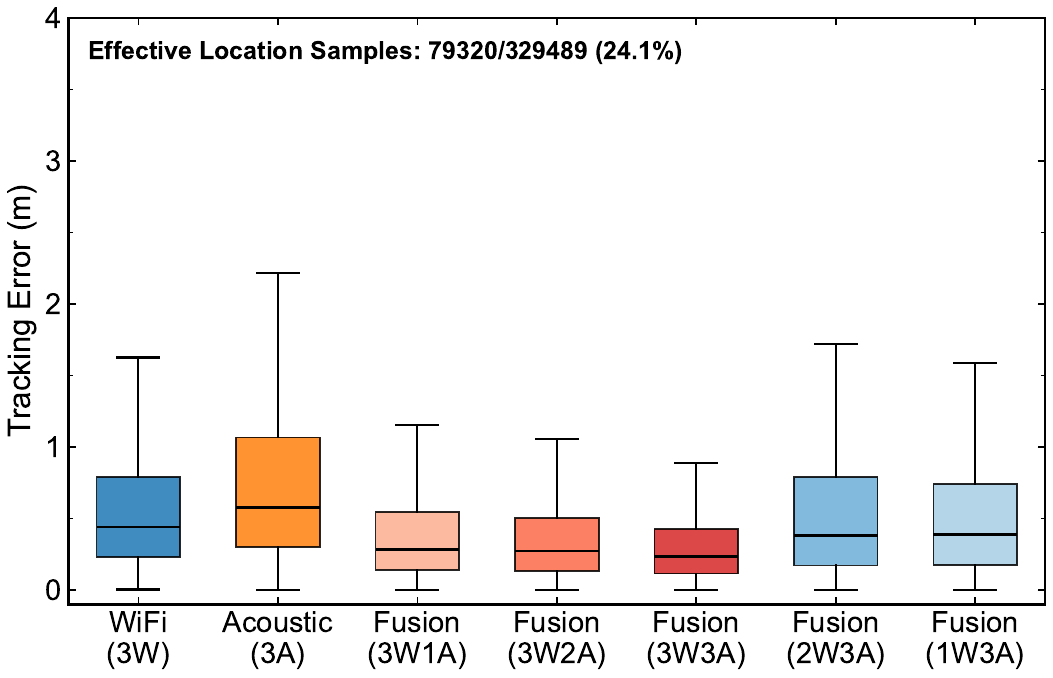}}	
    \hspace{0.15in}
	\subfigure[]{		
		\label{fig10b} 
		\includegraphics[width=1.85in]{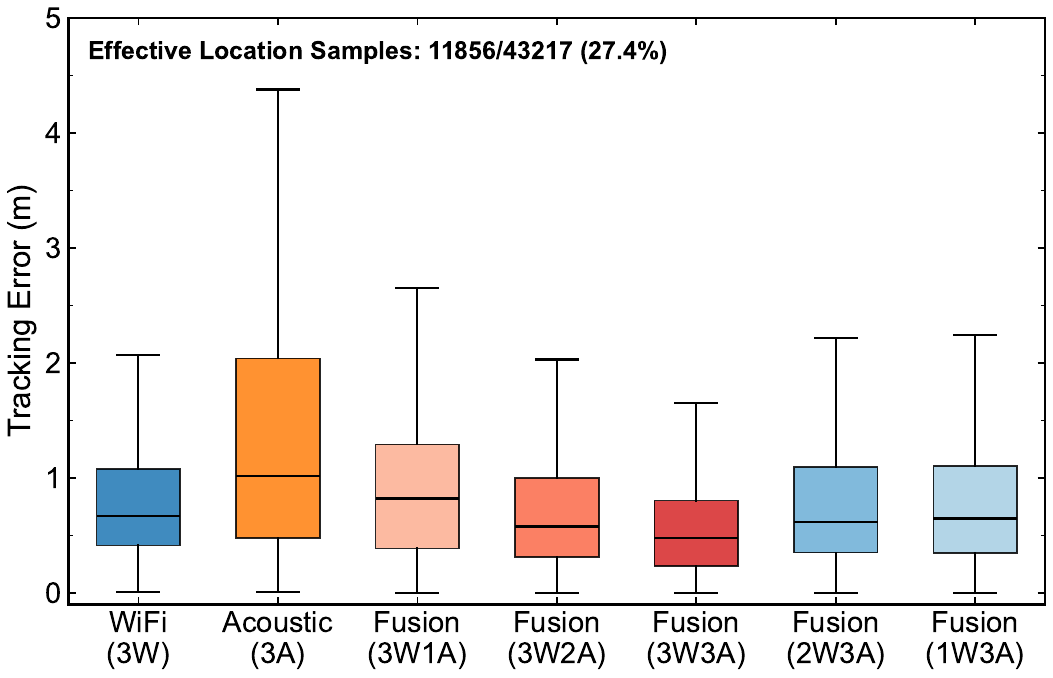}}	
    \hspace{0.15in}
    \subfigure[]{		
		\label{fig10c} 
		\includegraphics[width=1.85in]{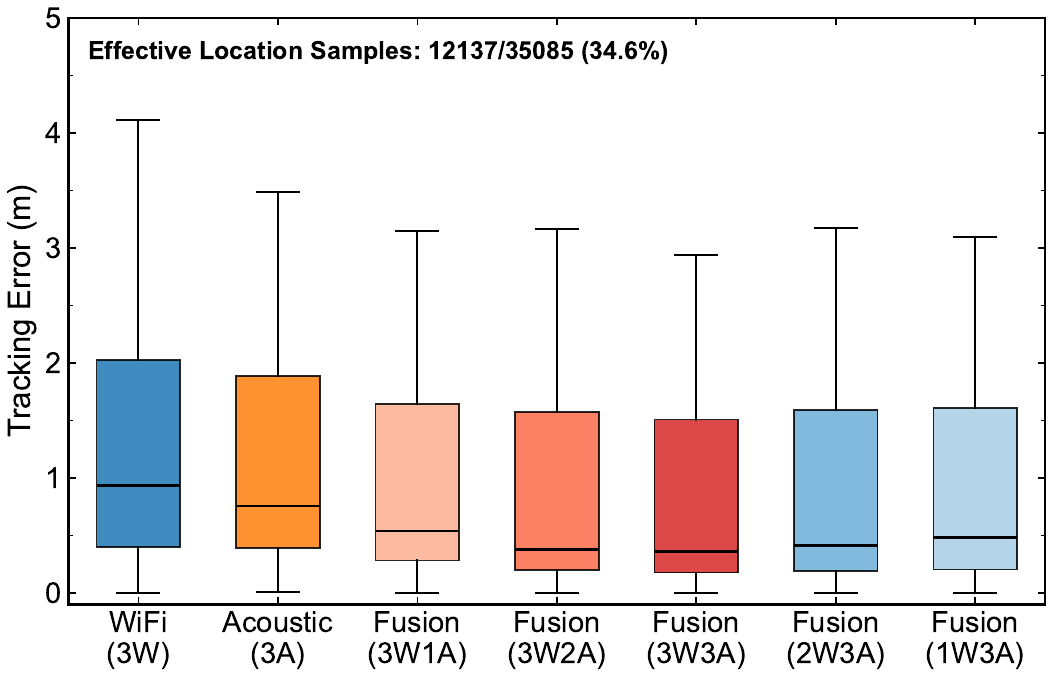}}	
	\caption{\textcolor{blue}{Impact of node density. (a) Laboratory scene. (b) Home scene. (c) Meeting-room scene.}
	}
	\label{fig10} 
\end{figure}

\textbf{(4) Impact of node density.} The influence of node density is examined by varying the number of links available for fusion in all scenarios. Starting from three links for each single modality, we progressively add one to three links from the complementary modality and evaluate the resulting trajectory errors. The experimental results in Fig.~\ref{fig10} show that fusion (3 Wi-Fi and 3 acoustic links, denoted 3W3A) consistently achieves the lowest median error and the most compact distribution across all configurations. The significant shrinkage of the interquartile range and whiskers compared to single-modality configurations proves the critical role of multi-modal redundancy in suppressing transient localization drifts. 

\textcolor{blue}{A cross-scenario comparison further reinforces the observation that the benefit of multi-modal integration increases with environmental complexity. This observation is consistent with the trend in fusion win ratio, as the system relies more heavily on complementary modalities to sustain reliable tracking under degraded signal conditions. Specifically, the proportion of effective localization samples increases from 24.1\% in the laboratory to 34.6\% in the meeting-room scenario. While Wi-Fi provides the primary global constraint due to its lower baseline error, acoustic sensing complements it by mitigating instantaneous multipath biases. This synergy creates geometric complementarity that ensures robust performance even in the meeting-room scenario characterized by severe occlusion, where single-modality approaches are prone to catastrophic failures.}

\textbf{(5) Key insights.} In summary, while Wi-Fi remains the more reliable baseline for indoor tracking, acoustic sensing offers essential stability in complex scenarios. Fusion exhibits a ``selective gain'' behavior, where its advantages are most evident in complex paths and challenging environments. These findings validate the complementary value of combining Wi-Fi and active acoustics for PLCR-based tracking and motivate further exploration into adaptive fusion strategies that can dynamically achieve optimal per-trajectory performance.

\subsection{Gait Recognition Validation}
Gait recognition is evaluated in different environments, considering both spectrum-driven and model-based feature paradigms, as introduced in Section~\ref{sec:Gait Recognition}.

\begin{figure}[!t]	
	\centering	
	\subfigure[]{		
		\label{fig11a} 
		\includegraphics[width=1.5in]{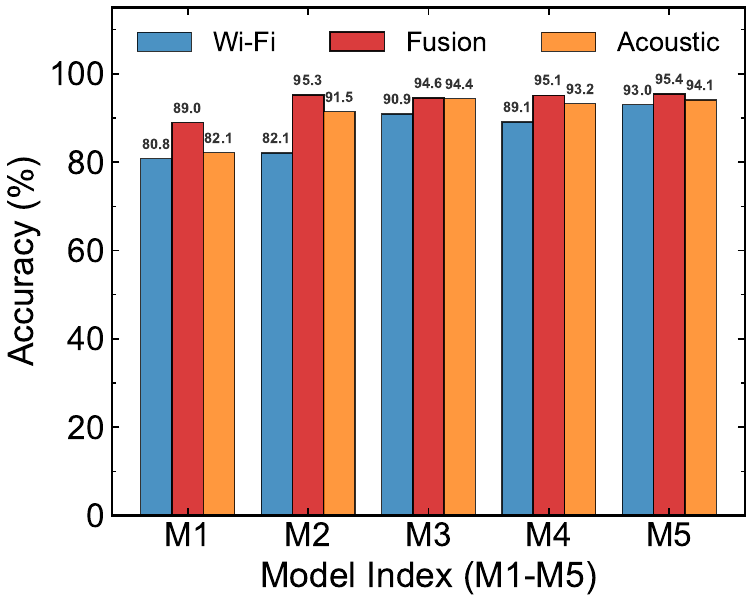}}	
	\subfigure[]{		
		\label{fig11b} 
		\includegraphics[width=1.5in]{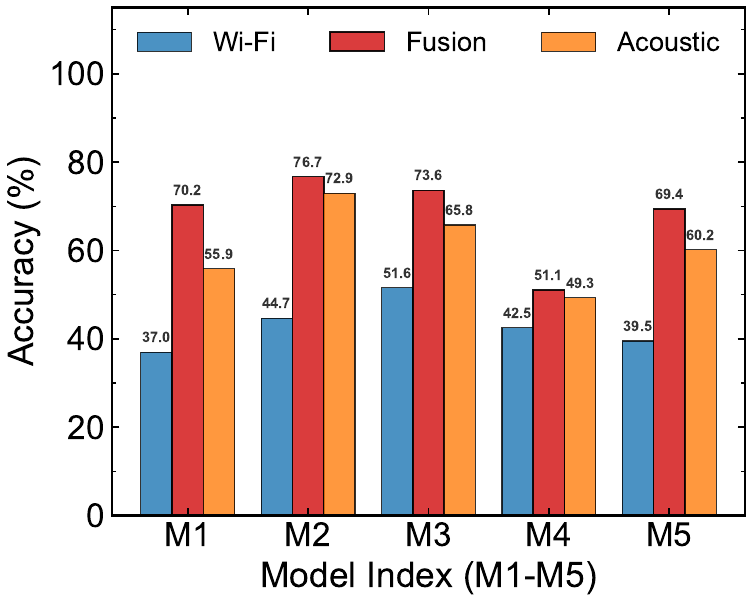}}	
    \subfigure[]{		
		\label{fig11c} 
		\includegraphics[width=1.5in]{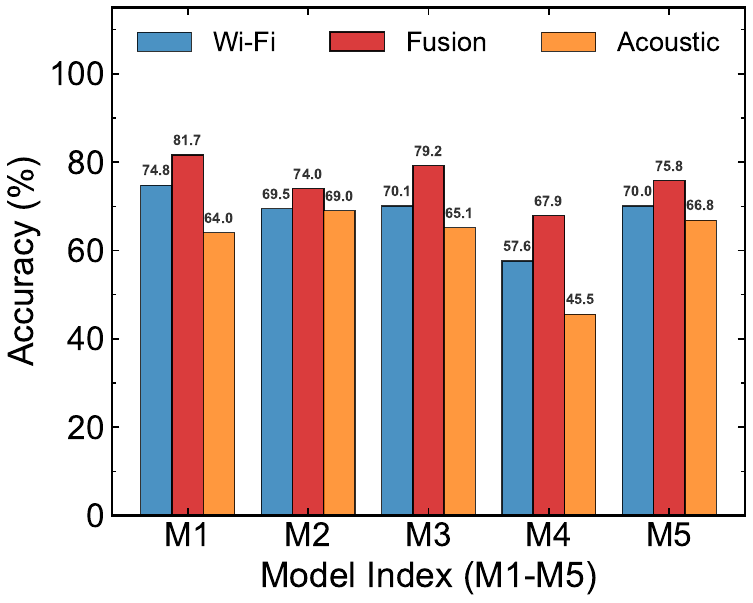}}	
    \subfigure[]{		
		\label{fig11d} 
		\includegraphics[width=1.5in]{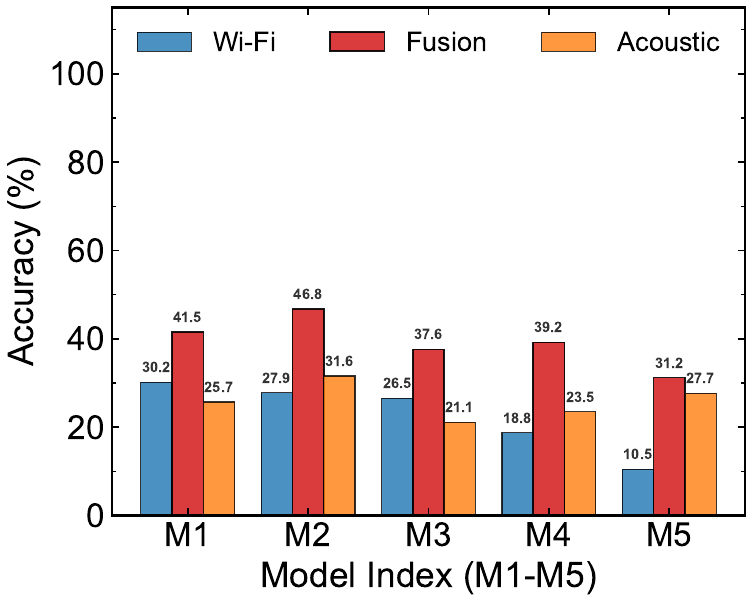}}
	\caption{Overall classification results in Laboratory. (a) Doppler-based learning. (b) Cross-trajectory Doppler-based learning. (c) PPVP-based learning. (d) Cross-trajectory PPVP-based learning.
	}
	\label{fig11} 
\end{figure}

\textbf{(1) Validation in the laboratory scene.} Recognition performance is first assessed using a random split ($6:2:2$). As shown in Figs.~\ref{fig11a} and~\ref{fig11c}, multi-modal fusion consistently achieves the highest accuracy across all benchmarks. This indicates that the two modalities capture complementary biometric cues that exceed the capacity of either modality in isolation. Although Wi-Fi-based models typically surpass acoustic-based results when using PPVP features, this outcome is expected because PPVP relies on tracking outputs where Wi-Fi currently maintains a precision advantage.

The cross-trajectory generalization is further investigated by partitioning the dataset according to spatial paths: the model is trained on the trajectories \#2--\#8, tuned on the trajectory \#9, and evaluated on the previously unseen trajectory \#1. As illustrated in Figs.~\ref{fig11b} and~\ref{fig11d}, this spatially independent evaluation presents a significantly higher performance bar. Specifically, Wi-Fi models exhibit a pronounced performance degradation in spectrum-driven settings, whereas acoustic models demonstrate superior robustness. This discrepancy suggests that acoustic signals provide a higher-dimensional feature space in the Doppler domain. Due to the much shorter wavelength of sound compared to Wi-Fi signals, acoustic Doppler signatures offer finer spectral granularity, capturing subtle gait dynamics that are less sensitive to specific trajectory geometries.

When comparing the two feature paradigms, spectrum-based features generally yield higher absolute accuracy and greater resilience to distribution shifts than PPVP. In contrast, PPVP is more sensitive to cross-trajectory variations because it inherently inherits errors from the tracking stage. Interestingly, in the random-split protocol, PPVP-based learning models favor Wi-Fi features. This preference arises because the fidelity of PPVP relies heavily on the tracking precision, and Wi-Fi provides more stable trajectory estimations in this environment. Despite these localized differences, multi-modal fusion consistently enhances performance across both paradigms, confirming that modal complementarity is a robust asset regardless of the specific level of feature abstraction.

In addition to recognition accuracy, Fig.~\ref{fig14a} presents the CDFs of the false acceptance rate (FAR) and false rejection rate (FRR). Fusion consistently shifts both curves toward the lower-error region with a markedly steeper gradient, indicating that a significantly larger proportion of test samples achieve minimal FAR and FRR simultaneously. This leftward shift suggests that the two modalities make different mistakes across subjects, and combining them effectively suppresses the long-tail cases where a single modality exhibits high false alarms or false rejects. In contrast, Wi-Fi shows a heavier tail on FRR, while the acoustic modality exhibits intermediate performance, further highlighting that fusion improves not only average performance but also reliability by reducing worst-case misidentification behaviors.

\begin{figure}[!t]	
	\centering	
	\subfigure[]{		
		\label{fig12a} 
		\includegraphics[width=1.5in]{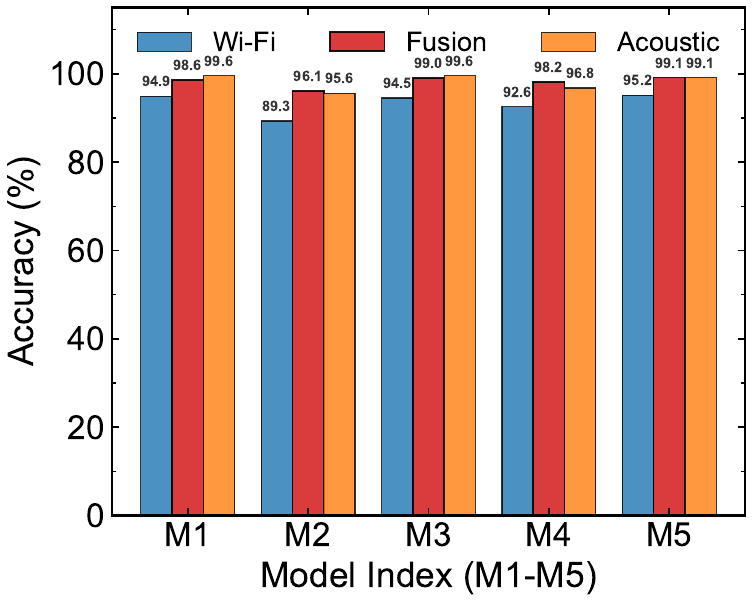}}	
	\subfigure[]{		
		\label{fig12b} 
		\includegraphics[width=1.5in]{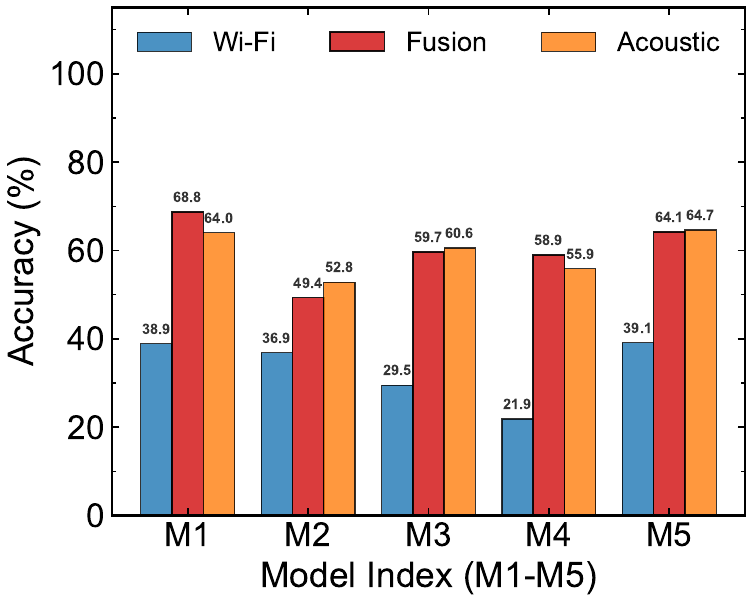}}	
    \subfigure[]{		
		\label{fig12c} 
		\includegraphics[width=1.5in]{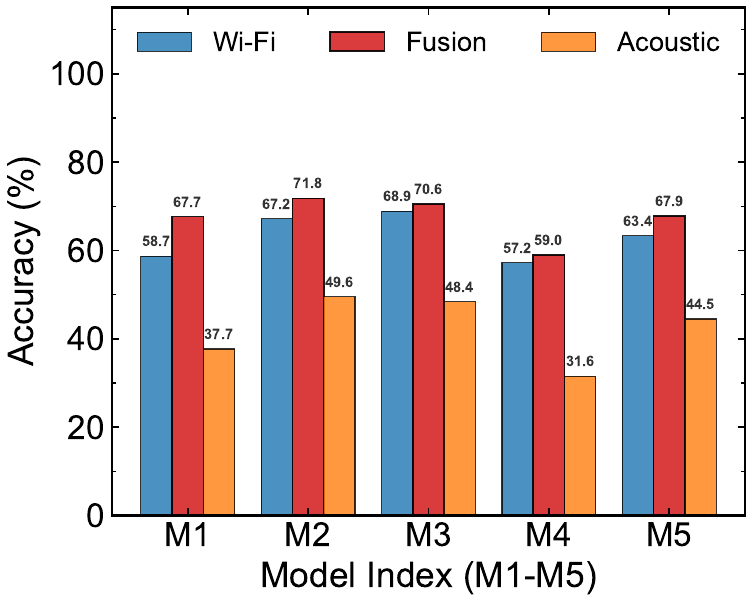}}	
    \subfigure[]{		
		\label{fig12d} 
		\includegraphics[width=1.5in]{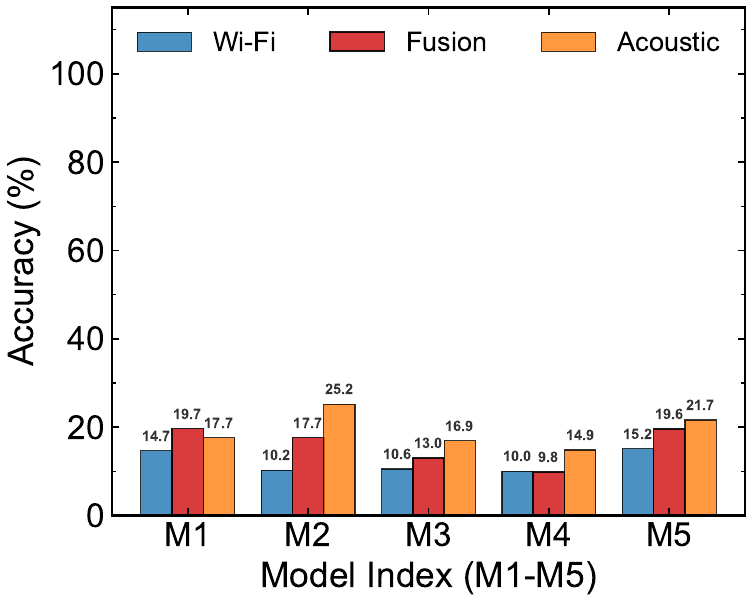}}
	\caption{\textcolor{blue}{Overall classification results in Home. (a) Doppler-based learning. (b) Cross-trajectory Doppler-based learning. (c) PPVP-based learning. (d) Cross-trajectory PPVP-based learning.}
	}
	\label{fig12} 
\end{figure}

\begin{figure}[!t]	
	\centering	
	\subfigure[]{		
		\label{fig13a} 
		\includegraphics[width=1.5in]{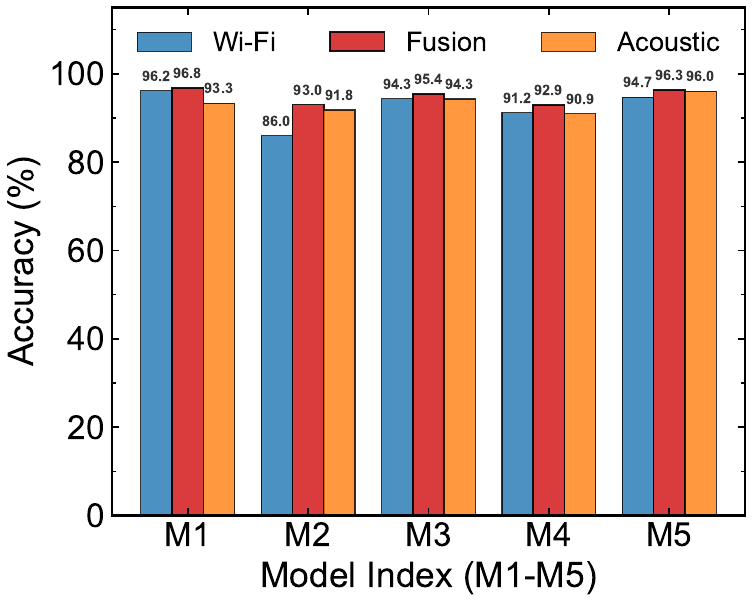}}	
	\subfigure[]{		
		\label{fig13b} 
		\includegraphics[width=1.5in]{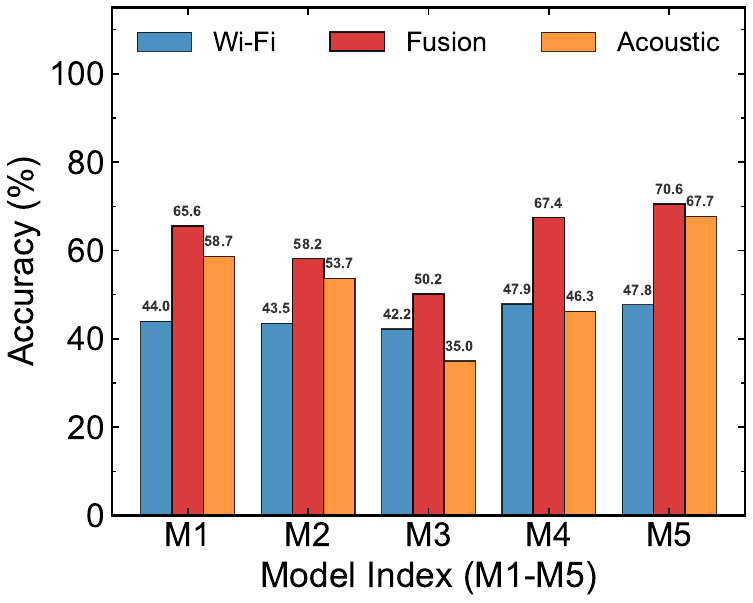}}	
    \subfigure[]{		
		\label{fig13c} 
		\includegraphics[width=1.5in]{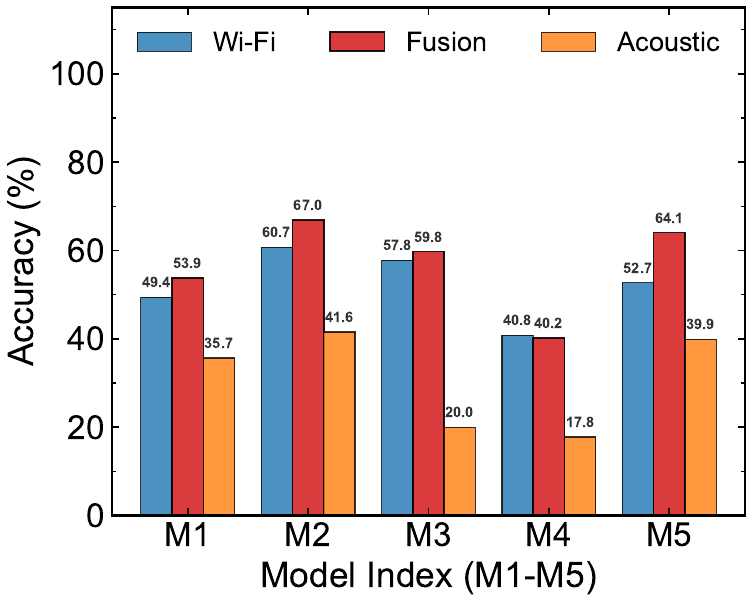}}	
    \subfigure[]{		
		\label{fig13d} 
		\includegraphics[width=1.5in]{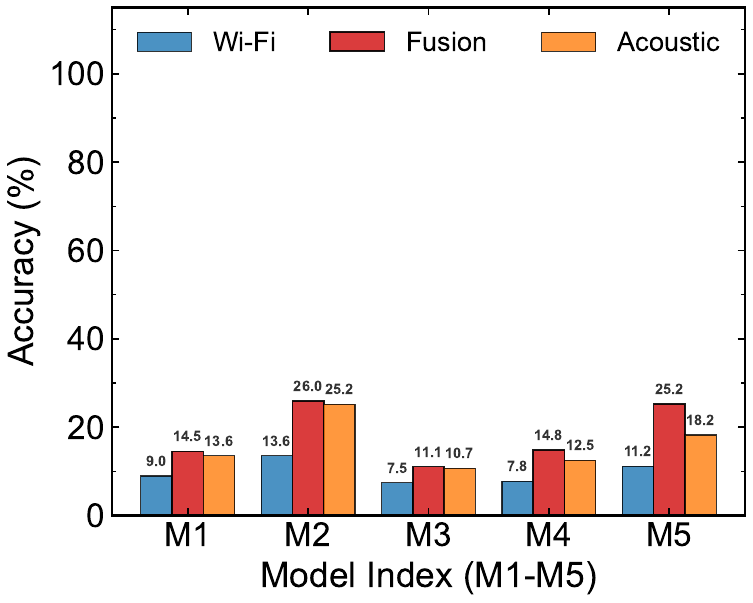}}
	\caption{\textcolor{blue}{Overall classification results in Meeting-room. (a) Doppler-based learning. (b) Cross-trajectory Doppler-based learning. (c) PPVP-based learning. (d) Cross-trajectory PPVP-based learning.}
	}
	\label{fig13} 
\end{figure}

\begin{figure}[!t]	
	\centering	
	\subfigure[]{		
		\label{fig14a} 
		\includegraphics[width=2.28in]{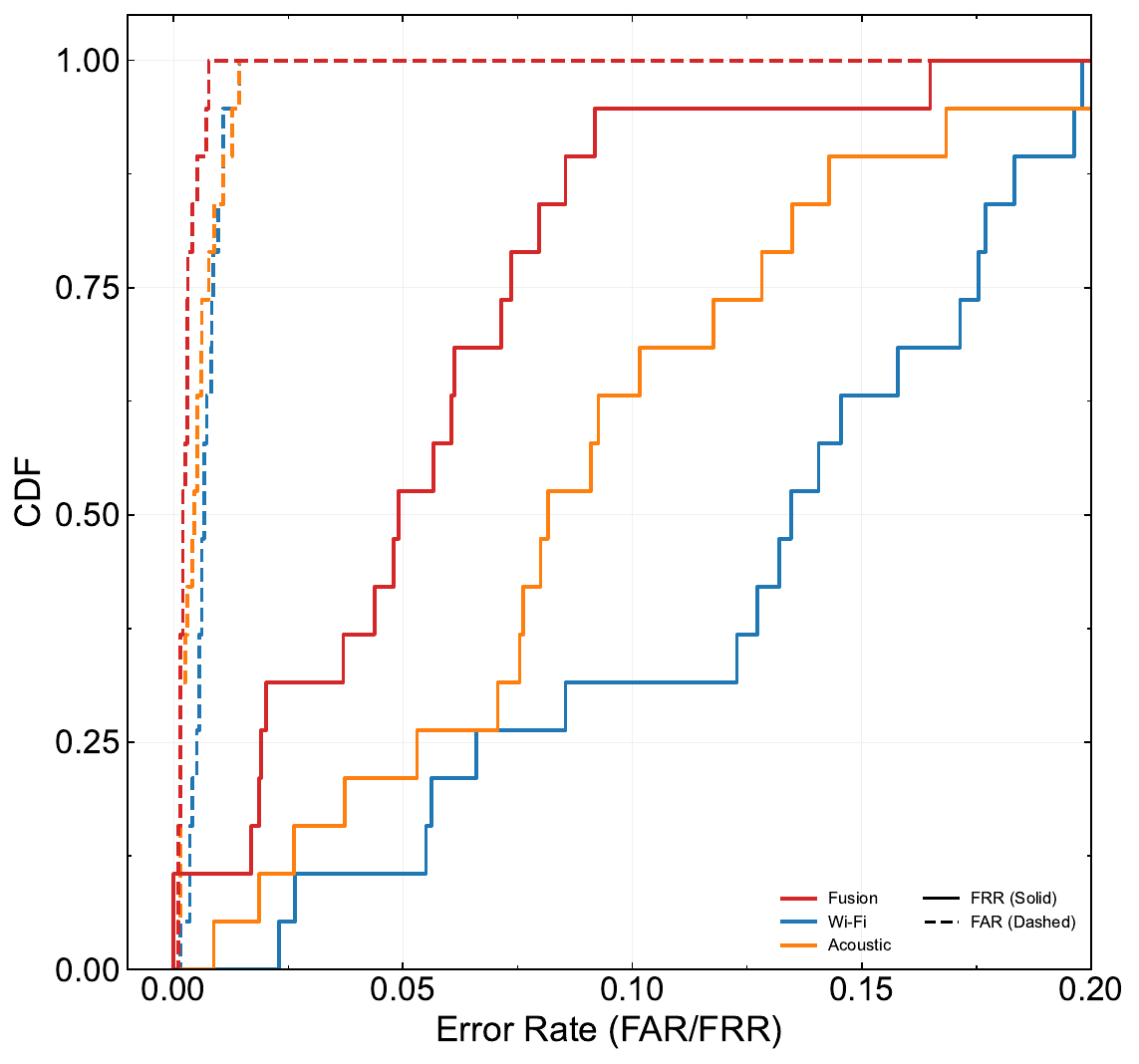}}
    \hspace{0.25in}
	\subfigure[]{		
		\label{fig14b} 
		\includegraphics[width=2.5in]{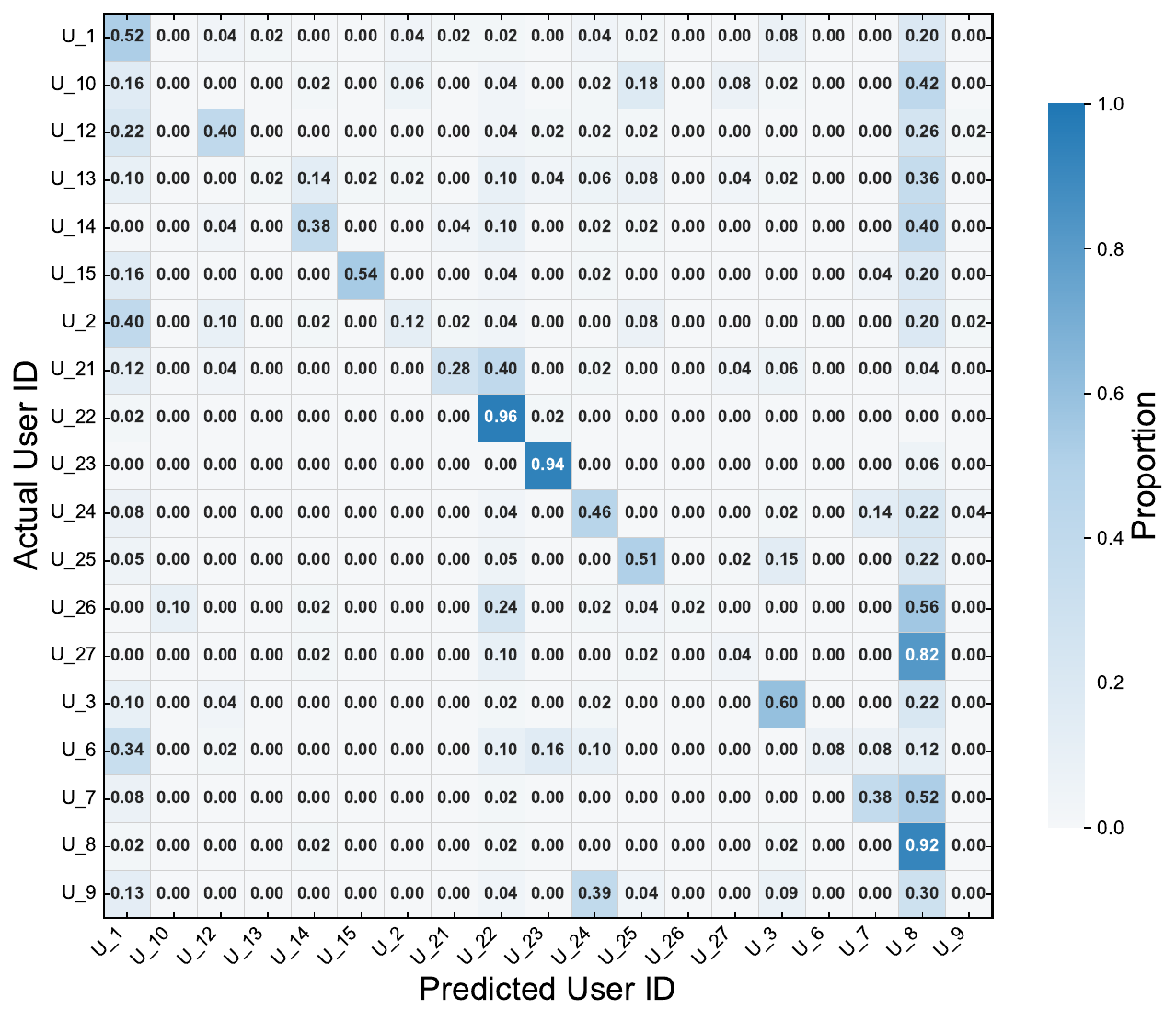}}	
    \subfigure[]{		
		\label{fig14c} 
		\includegraphics[width=2.5in]{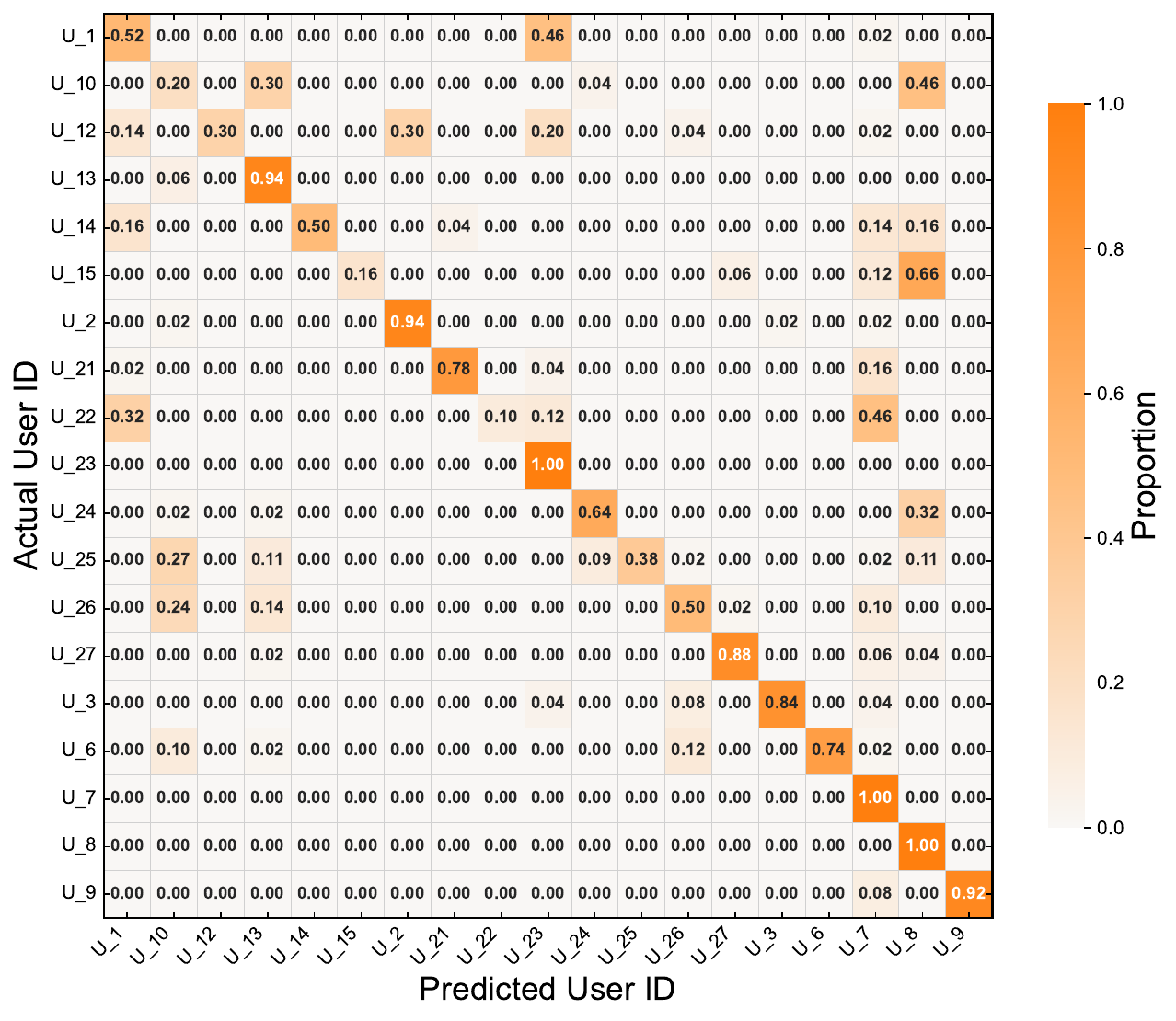}}	
    \hspace{0.03in}
    \subfigure[]{		
		\label{fig14d} 
		\includegraphics[width=2.5in]{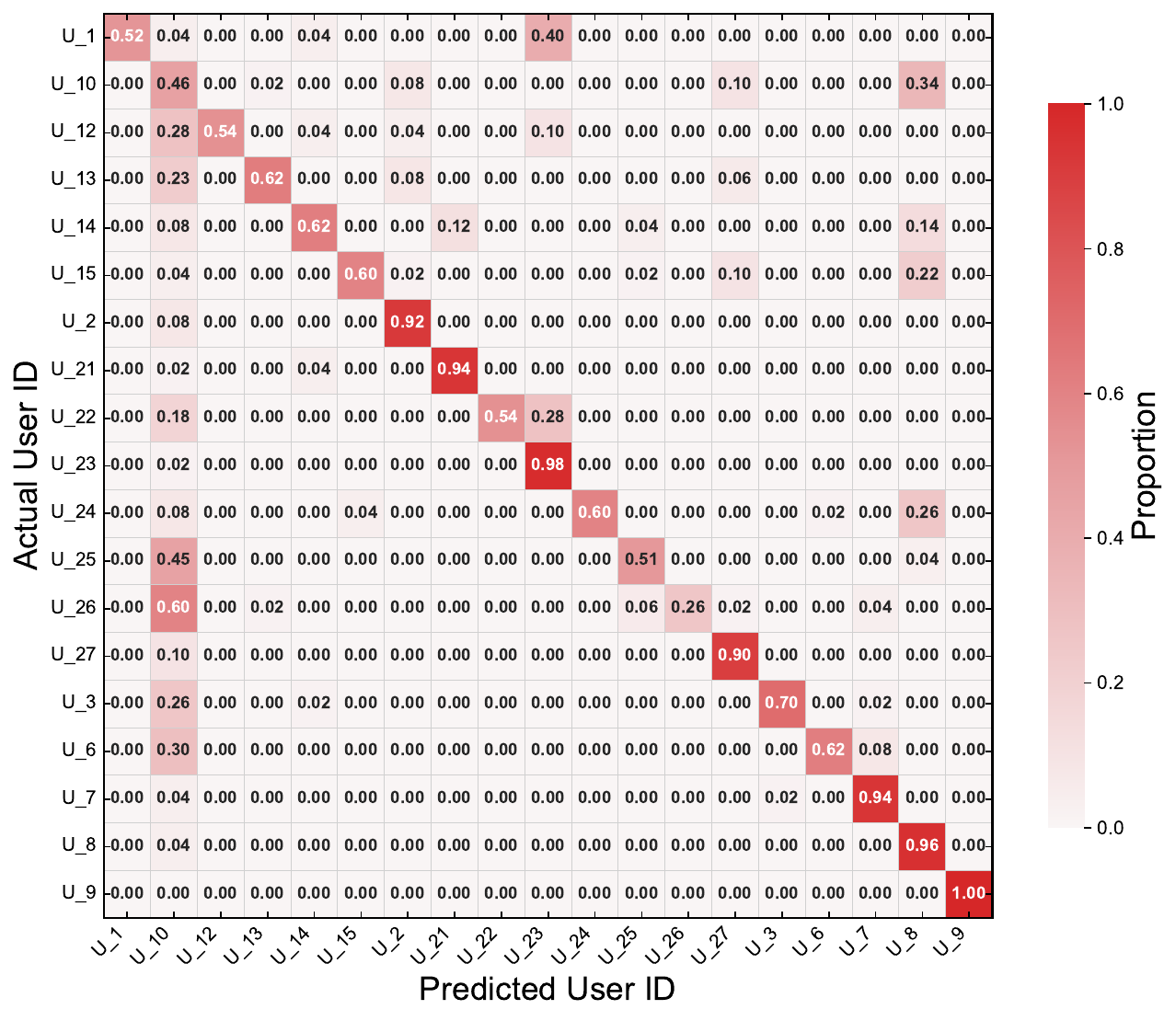}}
	\caption{\textcolor{blue}{Error estimation. (a) Error rate in laboratory. (b) Confusion matrix from Wi-Fi sensing in home. (c) Confusion matrix from acoustic sensing in home. (d) Confusion matrix from fusion sensing in home.}
	}
	\label{fig14} 
\end{figure}

\textbf{(2) Validation in the home scene.} Under the pooled random split protocol, spectrum-driven learning achieves near-saturated accuracy for both modalities, and fusion provides limited headroom because many backbones already operate close to a ceiling (e.g., Fig.~\ref{fig12a}). \textcolor{blue}{The overall performance in home settings shows a marginal improvement compared to laboratory environments. Setting aside the minor difference in participant count (19 vs.~20), we attribute this gain to the fewer movement trajectories in residential layouts. Although home scene is spatially more complex, it typically involves fewer distinct walking paths than the open lab setting. For the results shown in Fig.~\ref{fig12c} related to PPVP, Wi-Fi sensing predictably outperforms acoustics, while the fusion approach remains robust and consistently achieves the highest performance, maintaining its dominant position.}

When cross-trajectory generalization is enforced by training on trajectories \#1--\#3, validating on \#4, and testing on unseen \#5, the task becomes substantially harder and the modality gap widens, as illustrated in Figs.~\ref{fig12b} and~\ref{fig12d}. The Wi-Fi accuracy drops sharply in both feature paradigms, while the acoustic accuracy degrades less, indicating stronger robustness to path-induced distribution shift. \textcolor{blue}{Notably, under these challenging conditions, the fusion method does not consistently outperform the acoustic modality across all backbones, particularly in the PPVP-based features where overall accuracy remains relatively low. This suggests that severely degraded Wi-Fi features under extreme trajectory shifts may introduce noise into the joint representation, partially obscuring the potential gains of multi-modal integration. Nevertheless, fusion effectively prevents the recognition collapse observed in the Wi-Fi-only baseline by leveraging complementary cues from the acoustic modality, maintaining a stable performance floor even when one sensing channel fails.}

\textcolor{blue}{To evaluate the fine-grained recognition performance, the confusion matrices of one method (M1) from the Cross-trajectory Doppler-based learning framework are presented in Figs.~\ref{fig14b}--\ref{fig14d}, which provide detailed insights into classification behavior across 19 subjects. Both single modalities exhibit distinct performance blind spots, which are  effectively mitigated by fusion. Especially, although the average accuracy of the fusion method ($68.75\%$) yields a relatively modest improvement of less than $5\%$ over the best-performing single modality (Acoustic: $64.01\%$, Wi-Fi: $38.9\%$), its primary benefit lies in improved classification reliability across individuals. A comparative analysis reveals a clear hierarchy in modality robustness: Wi-Fi achieves over $50\%$ for only 7 subjects ($\text{User}_{1,3,8,15,22,23,25}$), while acoustic sensing doubles this coverage to 14 subjects. Fusion further extends this range, with only two subjects ($\text{User}_{10,16}$) falling below the $50\%$ threshold, demonstrating that it expands the system's operational envelope rather than merely averaging modality outputs.}

\textcolor{blue}{From the perspective of error distribution, both single modalities exhibit a tendency to misidentify multiple subjects as $\text{User}_1$ or $\text{User}_{8}$, a label absorption that fusion successfully alleviates. However, suppressing these dominant errors introduces a subtle trade-off, as it reshapes the feature space and amplifies other biases, such as increased confusion around $\text{User}_{10}$. By correlating these results with the anthropometric data in Fig.~\ref{fig3a}, we observe that fusion achieves exceptional accuracy (over $90\%$) for 7 subjects ($\text{User}_{2,7,8,9,21,23,27}$) with distinct physical combinations of height, weight, and gender. Conversely, among a group of four subjects with the same gender and similar body profiles ($\text{User}_{10,14,22,23}$), the recognition accuracies vary significantly (46\%, 62\%, 54\%, and 98\%), while mutual misclassification rates remain relatively low. This suggests that the fusion method can still achieve high accuracy even when anthropometric cues are less distinctive, without introducing excessive confusion among similar individuals. Additionally, for subjects with misclassifications rates exceeding $20\%$, such as cases where $\text{User}_{3,6,12,13,25,26}$ are misclassified as $\text{User}_{10}$, or $\text{User}_{10,15,24}$ as $\text{User}_{8}$, no clear correlation with physical attributes is observed, as substantial differences in their anthropometric characteristics do not prevent these misclassifications. Consequently, we conclude that while anthropometric characteristics are a critical factor in gait-based identity recognition, they are not the sole decisive factor, as the system successfully captures deeper, individual-specific motion dynamics beyond simple body morphology.}

\textcolor{blue}{\textbf{(3) Validation in the meeting-room scene.} In this scenario, the average recognition accuracies for both modalities remain largely consistent with the trends observed in the previous two settings. A divergence emerges in cross-trajectory recognition tasks, where the performance in the meeting-room exhibits a more pronounced decline compared to the home scene. Specifically, the performance degradation of acoustic-based models is slightly greater than that of Wi-Fi, consistent with our earlier observations in tracking tasks. This behavior is primarily attributed to the physical layout of the meeting-room, where centrally placed furniture such as tables and chairs introduces severe signal blockage. Such conditions are particularly detrimental to acoustic sensing, as identity-bearing reflections from leg movements become significantly attenuated when the subject moves to positions far from the transceivers. Despite these challenges, the fusion method remains effective under NLoS conditions, maintaining robust performance by compensating for the limitations of individual modalities.}

\begin{figure}[!t]	
	\centering	
	\subfigure[]{		
		\label{fig15a} 
		\includegraphics[width =1.5in]{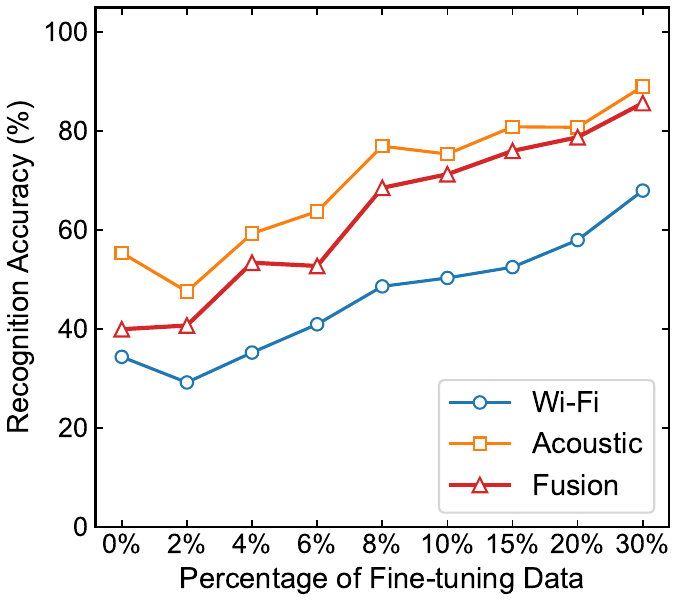}}	
    \subfigure[]{		
		\label{fig15b} 
		\includegraphics[width =1.5in]{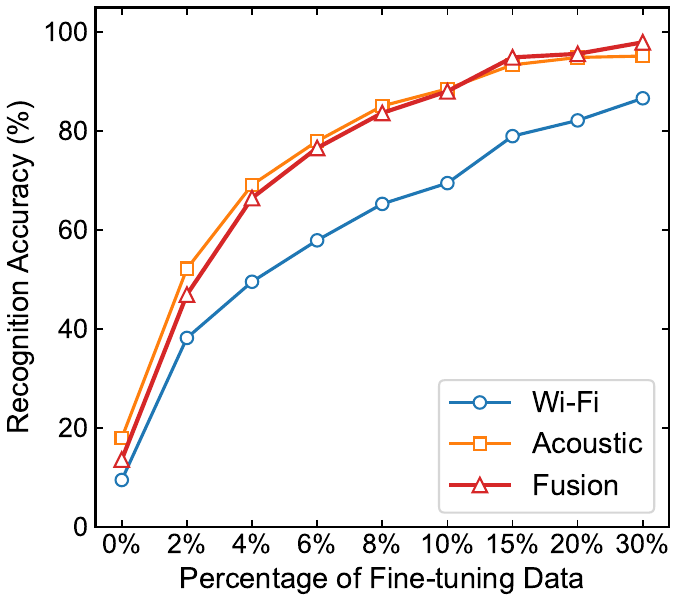}}	
    \subfigure[]{		
		\label{fig15c} 
		\includegraphics[width =1.5in]{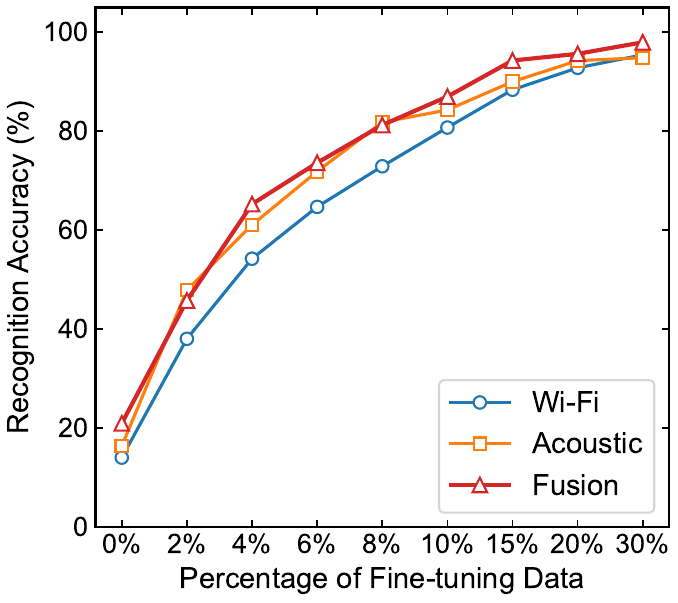}}	
	\subfigure[]{		
		\label{fig15d} 
		\includegraphics[width =1.5in, height = 1.36in]{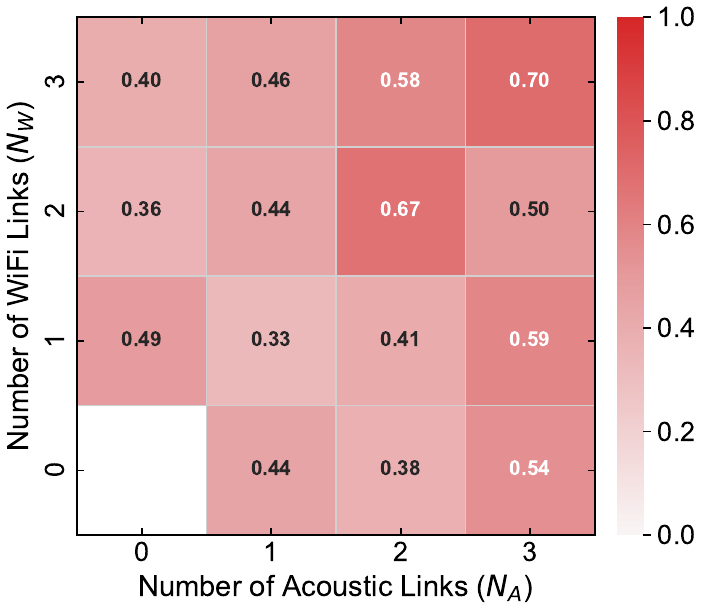}}

	\caption{\textcolor{blue}{Cross-scene validation and varying configurations. (a) Fine-tuning data ratio on clothing. (b) Fine-tuning data ratio in Home. (c) Fine-tuning data ratio in Meeting-room.} (d) Link density analysis.
	}
	\label{fig15} 
\end{figure}

\textcolor{blue}{To further examine longitudinal robustness and generalization under co-variate shifts, we conduct additional evaluations on a dedicated dataset. This subset comprises six subjects recorded over one week, during which participants naturally varied their attire. We first apply the model pre-trained on the source scenario in a zero-shot setting, and then track performance as the proportion of fine-tuning data increases from $2\%$ to $30\%$, as shown in Fig.~\ref{fig15a}. Notably, even without fine-tuning ($0\%$), it retains a reasonable level of recognition accuracy, indicating that the learned gait representations are inherently robust to clothing variations. However, a significant performance disparity is observed between modalities: the acoustic-based recognition substantially outperforms its Wi-Fi counterpart. This is primarily because acoustic Doppler shifts effectively capture the macro-dynamics of human motion envelopes, whereas Wi-Fi signals are highly susceptible to the varying electromagnetic absorption and scattering properties of different fabric materials.}

\textcolor{blue}{A notable phenomenon emerges during the early stage of fine-tuning, where recognition accuracy across all modalities exhibits a non-monotonic pattern, with a temporary decline at the 2\% data level before recovering. This behavior reflects the adaptation dynamics inherent in transfer learning. The introduction of a small amount of clothing-dependent data perturbs the previously learned feature distribution, leading to temporary misalignment before the model re-stabilizes. Furthermore, throughout the fine-tuning process, the fusion method consistently underperforms the acoustic modality. However, the performance gap steadily decreases as the amount of fine-tuning data increases. This indicates that, despite severe Wi-Fi degradation under clothing variation introducing modality-specific noise that constrains fusion, additional training data gradually improves cross-modal alignment and mitigates this imbalance.}

\textcolor{blue}{\textbf{(4) Cross-scene validation.} To evaluate cross-scene robustness, we focus on the ten subjects shared across all three scenes. Following the established evaluation protocol, model performance is tracked from a zero-shot baseline up to a fine-tuning ratio of 30\%, as illustrated in Figs.~\ref{fig15b} and~\ref{fig15c}. A key observation emerges when comparing these results with the prior clothing-variation experiments. Although the model maintained reasonable zero-shot accuracy under clothing variations within a fixed environment, its performance collapses to near-chance levels in cross-scene tests. This failure of zero-shot transfer can be attributed to a compounded domain shift. Unlike the disturbances caused by clothing, cross-scene deployment introduces changes in multipath geometry, room layout, and transceiver orientations. These factors substantially reshape the signal distribution, rendering spatial representations learned in the source domain largely incompatible with the target domain.}

\textcolor{blue}{Despite the initial collapse at the zero-shot stage, the recovery slope during fine-tuning is remarkably steep. With only 2\% of target data, the accuracy across all modalities increases sharply, indicating that the model re-calibrates its feature representations to the new spatial configuration. This adaptation highlights a key distinction between spatial and clothing co-variates. While the environmental shift poses a strong barrier to zero-shot generalization, its largely deterministic geometric structure enables the model to effectively characterize and suppress scene-specific multipath effects once a small number of local anchors are available. As a result, the performance ceiling in cross-scene adaptation eventually exceeds that of the clothing experiments. In contrast to clothing variations, which introduce stochastic and non-linear distortions directly into the gait signatures, environmental changes remain structurally consistent within a scene and can therefore be learned and effectively compensated for, leading to higher final accuracy after supervised calibration. Throughout this adaptation process, acoustic sensing demonstrates superior efficiency compared to Wi-Fi, likely due to the higher stability of Doppler shifts across different spatial layouts. More importantly, fusion consistently tracks or exceeds the performance of the best individual modality.}

\textbf{(5) Impact of node density.}  We evaluate the effect of node density by varying the number of Wi-Fi links (\(N_W\)) and acoustic links (\(N_A\)) for spectrum-driven recognition using CNN (M1) in the laboratory cross-trajectory scene (Fig.~\ref{fig15d}). Overall, increasing link diversity improves recognition, and the best performance is achieved when both modalities are sufficiently populated. In particular, the accuracy peaks at \(N_W{=}3, N_A{=}3\) with an accuracy of \(70.23\%\), indicating that dense multi-modal coverage provides redundant Doppler evidence beneficial for identity discrimination. However, the trend is not strictly monotonic, revealing interactions between modalities and link quality. For example, with \(N_W{=}2\), increasing \(N_A\) from 1 to 2 raises accuracy from \(43.75\%\) to \(67.19\%\), but a further increase to \(N_A{=}3\) drops it to \(49.57\%\), suggesting that additional links can introduce less reliable observations under certain geometries. Similarly, when \(N_A{=}3\), increasing \(N_W\) from 0 to 3 yields clear gain from \(54.39\%\) to \(70.23\%\), underscoring that balanced redundancy across modalities is critical. These results motivate future work on reliability-aware link selection or weighting when scaling to denser deployments.

\textbf{(6) Key insights.} The results suggest four takeaways. First, acoustic sensing provides superior and more stable identity cues compared to Wi-Fi, and their fusion consistently improves both recognition accuracy and error-tail reliability. Second, the benefit of fusion is small when the split is within-distribution (random split) but becomes pronounced under cross-trajectory generalization. Third, Doppler-based features are generally more robust to distribution shift, while PPVP is more sensitive because it propagates tracking errors and thus tends to favor the better-tracked modality. Fourth, cross-scene transfer is near-chance levels without adaptation, but modest target-scene fine-tuning quickly restores performance, with fusion remaining the most stable choice.

\section{DISCUSSION}
\label{sec:discussion}
\textcolor{blue}{This section discusses modality complementarity, reflects on the limitations of our design, and outlines potential research opportunities in the ubiquitous community.}

\subsection{\textcolor{blue}{Sensing Mechanisms and Synergy}}
\textcolor{blue}{\hspace*{1em}\textbf{(1) Task performance and physical grounding}. Our evaluations reveal a clear divergence, where Wi-Fi provides more stable tracking, while acoustic sensing achieves higher recognition precision. This difference stems from sensing granularity at the centimeter scale. The longer wavelength of Wi-Fi enables robust diffraction and reliable global motion tracking, whereas acoustic signals, being mechanical waves with intrinsically shorter wavelengths, provide higher sensitivity to micro-Doppler variations and thus capture fine-grained limb dynamics more effectively. Their integration leverages these complementary characteristics, delivering consistent performance and mitigating single-modality failures under varying input conditions.}

\textcolor{blue}{\textbf{(2) Spatial and geometric complementarity}. By projecting heterogeneous signals into a unified Doppler-frequency domain, multi-modal fusion enables richer spatial representations that go beyond single-modality observations. In wireless sensing, feature quality is inherently dependent on the relative orientation between the subject and the transceivers, causing individual modalities to capture only partial aspects of the motion dynamics. The combination of multiple modalities and nodes provides diverse geometric perspectives, enriching the representation of gait patterns. This effect is further influenced by the spatial arrangement of sensing nodes across different environments, where variations in node placement implicitly reshape the available viewpoints and thus the degree of geometric complementarity. Rather than serving solely as a fallback under occlusion, this spatial diversity enhances the completeness of motion characterization, allowing the system to capture both global movement trends and fine-grained dynamics across the sensing area.}

\textcolor{blue}{\textbf{(3) Dynamic reliability under NLOS conditions}. The advantage of fusion sensing also lies in its ability to adapt to time-varying reliability under challenging scenes. In practical indoor environments, signal quality fluctuates dynamically as the subject moves, leading to intermittent degradation in individual modalities. For example, Wi-Fi sensing may suffer from unstable phase and multipath distortion during certain motion segments, while acoustic sensing can experience severe attenuation when propagation paths are temporarily obstructed. Such effects become more pronounced in layouts (e.g., home and meeting-room) with centralized obstructions or asymmetric node placement, where LoS conditions vary significantly across space. These degradations are not constant but occur intermittently along a trajectory. Integration leverages this temporal diversity by continuously aggregating information from modalities with varying reliability, allowing the system to maintain stable sensing performance over time.}

\subsection{\textcolor{blue}{Limitations on Datasets and Methods}}
\textcolor{blue}{\hspace*{1em}\textbf{(1) Diversity of the data corpus}. While XGait establishes a multi-modal foundation, its data diversity remains limited. The current gender distribution is unbalanced, indicating the need for broader demographic coverage to reduce potential gait bias. Longitudinal variation is also underrepresented due to the high cost of repeated data collection. Nevertheless, the dataset includes weekly recordings for a subset of participants, and additional variability is introduced through natural attire differences across individuals and scenarios. In addition, current benchmarks focus on single-subject scenarios with limited behavioral interference. While common activities (e.g., smartphone usage) are included, more complex factors such as carrying heavy objects are not yet covered, leaving robustness under richer real-world conditions to be further evaluated.}

\textcolor{blue}{\textbf{(2) Idealization of node configuration}. The sensing nodes are deployed in a structured and controlled manner, with uniform height and inter-node distances of 1--2 m, facilitating stable initial validation. Although layouts vary across scenes, environments such as the laboratory and meeting-room remain more regular than real-world cluttered settings, and all nodes are stationary during data collection. While this design improves reproducibility, it also introduces a potential confounding factor when comparing scenarios. In particular, performance differences under increasing environmental complexity may be influenced not only by the environment itself, but also by variations in node deployment. As a result, the observed advantage of fusion in more complex settings may be partially entangled with changes in node layout. In practical deployments with irregular placement and dynamic occlusions, the robustness of the fusion model remains an open question.}

\textcolor{blue}{\textbf{(3) Methodological scope and generality}. The current approach adopts the Doppler domain as a unified representation, but its effectiveness is still influenced by system-level factors. Temporal synchronization relies on system-level alignment, which may introduce subtle misalignment and limit fine-grained fusion gains. Moreover, despite the scale of XGait, the evaluation covers only a subset of conditions. For instance, cross-scenario and cross-trajectory generalization in identity recognition are assessed on selected configurations rather than exhaustively. This work focuses on validating cross-modal complementarity rather than optimizing modality selection or weighting. As a result, no formal guarantee is provided that fusion consistently outperforms the best single modality. Given the coupling between sensing geometry and signal characteristics, developing a comprehensive analytical framework remains an important direction for future work.}

\subsection{\textcolor{blue}{Potential Research Opportunities}}
\textcolor{blue}{\hspace*{1em}\textbf{(1) Unified representations and learning}. Projecting CSI and acoustic signals into a shared time–frequency domain enables modality-invariant gait representations and normalized feature spaces. This unified abstraction captures fundamental motion dynamics, supporting not only identity recognition but also broader human activity understanding. It further motivates cross-modal learning methods that generalize across sensing modalities and tasks.}

\textcolor{blue}{\textbf{(2) Task-driven adaptive fusion.} The observed complementarity motivates dynamic, task-driven fusion strategies. Rather than treating tracking and recognition separately, future systems can integrate them within a joint multi-task framework. Geometric cues, such as trajectory and heading, can regularize and stabilize gait features, enabling adaptive weighting that prioritizes the most reliable modality under varying conditions. This unified approach supports consistent performance from coarse localization to fine-grained identification.}

\textcolor{blue}{\textbf{(3) Efficient edge-AIoT optimization}. Bridging laboratory results to real-world deployment requires advances in edge efficiency and domain generalization. The modular design enables lightweight implementations suitable for resource-constrained IoT devices while maintaining real-time performance. In addition, multi-scenario data supports research on few-shot adaptation and domain transfer. These directions improve robustness in diverse residential environments and facilitate practical, privacy-preserving AIoT applications such as elderly care and personalized home automation.}

\section{DATASET AND CODE AVAILABILITY}
\label{sec:dataset and code availability}
To support open science and ensure the reproducibility of our results, XGait dataset and its associated benchmark processing pipeline are hosted at \url{https://github.com/warrior-087/XGait}. 

\section{CONCLUSION}
\label{sec:conclusion}
This paper introduces XGait, a large-scale, multi-modal wireless gait dataset that synchronously captures multi-node Wi-Fi CSI and active acoustic signals alongside vision-based ground-truth trajectories. Designed to overcome the limitations of single-modality sensing, XGait supports both multi-link PLCR-based tracking and identity recognition across diverse, unconstrained walking paths.\textcolor{blue}{ To bridge the physical gap between heterogeneous sensing modalities, we design a unified benchmark pipeline that maps Wi-Fi and acoustic signals into a shared Doppler-centric representation, enabling principled cross-modal comparison and evaluation of fusion strategies.}

Extensive evaluations reveal a complementarity between the two modalities: Wi-Fi provides robust global constraints for tracking, whereas acoustic sensing offers finer granularity and greater stability for identity recognition. \textcolor{blue}{ Fusion provides notable improvements in accuracy and reliability in complex scenarios, particularly in challenging trajectories and domain shifts.} By establishing this reference foundation, XGait aims to catalyze future research into robust cross-scene adaptation, sparse-sensing alignment, and uncertainty-aware fusion architectures for next-generation wireless sensing systems. 

\section{ACKNOWLEDGMENTS}
This work is partially supported the National Key R\&D Program of China (No. 2025YFC3309002), the National Natural Science Foundation of China (No. 62472358, 62322601), and the Fundamental Research Funds for the Central Universities (No. G2026KY06261). The authors sincerely thank all volunteers for their participation in the data collection experiments. 
\bibliographystyle{ACM-Reference-Format}
\bibliography{refrences}

@String{Computing = "Computing" }

@String{Computer = "{IEEE} Computer" }

@String{Springer = "Springer-Verlag" }

@article{Wu2021GaitWay,
    author = {Wu, Chenshu and Zhang, Feng and Hu, Yuqian and Liu, K. J. Ray},
    title = {GaitWay: Monitoring and Recognizing Gait Speed Through the Walls},
    year = {2021},
    issue_date = {June 2021},
    publisher = {IEEE Educational Activities Department},
    address = {USA},
    volume = {20},
    number = {6},
    issn = {1536-1233},
    url = {https://doi.org/10.1109/TMC.2020.2975158},
    doi = {10.1109/TMC.2020.2975158},
    journal = {IEEE Transactions on Mobile Computing},
    month = jun,
    pages = {2186–2199},
    numpages = {14}
}

@article{lian2021echospot,
    author = {Lian, Jie and Lou, Jiadong and Chen, Li and Yuan, Xu},
    title = {EchoSpot: Spotting Your Locations via Acoustic Sensing},
    year = {2021},
    issue_date = {Sept 2021},
    publisher = {Association for Computing Machinery},
    address = {New York, NY, USA},
    volume = {5},
    number = {3},
    url = {https://doi.org/10.1145/3478095},
    doi = {10.1145/3478095},

    journal = {Proc. ACM Interact. Mob. Wearable Ubiquitous Technol.},
    month = sep,
    articleno = {113},
    numpages = {21}
}

@article{cai2021we,
    author = {Cai, Chao and Pu, Henglin and Wang, Peng and Chen, Zhe and Luo, Jun},
    title = {We Hear Your PACE: Passive Acoustic Localization of Multiple Walking Persons},
    year = {2021},
    issue_date = {June 2021},
    publisher = {Association for Computing Machinery},
    address = {New York, NY, USA},
    volume = {5},
    number = {2},
    url = {https://doi.org/10.1145/3463510},
    doi = {10.1145/3463510},
    journal = {Proc. ACM Interact. Mob. Wearable Ubiquitous Technol.},
    month = jun,
    articleno = {55},
    numpages = {24}
}

@inproceedings{wang2016human,
    author = {Wang, Hao and Zhang, Daqing and Ma, Junyi and Wang, Yasha and Wang, Yuxiang and Wu, Dan and Gu, Tao and Xie, Bing},
    title = {Human respiration detection with commodity wifi devices: do user location and body orientation matter?},
    year = {2016},
    isbn = {9781450344616},
    publisher = {Association for Computing Machinery},
    address = {New York, NY, USA},
    url = {https://doi.org/10.1145/2971648.2971744},
    doi = {10.1145/2971648.2971744},
    booktitle = {Proceedings of the 2016 ACM International Joint Conference on Pervasive and Ubiquitous Computing},
    pages = {25–36},
    numpages = {12},
    location = {Heidelberg, Germany},
    series = {UbiComp '16}
}

@article{wang2018wi,
  title={Wi-Fi CSI-based behavior recognition: From signals and actions to activities},
  author={Wang, Zhu and Guo, Bin and Yu, Zhiwen and Zhou, Xingshe},
  journal={IEEE Communications Magazine},
  volume={56},
  number={5},
  pages={109--115},
  year={2018},
  publisher={IEEE}
}

@article{zhang2021wi,
  title={Wi-PIGR: Path independent gait recognition with commodity Wi-Fi},
  author={Zhang, Lei and Wang, Cong and Zhang, Daqing},
  journal={IEEE Transactions on Mobile Computing},
  volume={21},
  number={9},
  pages={3414--3427},
  year={2021},
  publisher={IEEE},
  doi={10.1109/TMC.2021.3052314}
}

@inproceedings{zhai2021rise,
    author = {Zhai, Shuangjiao and Tang, Zhanyong and Nurmi, Petteri and Fang, Dingyi and Chen, Xiaojiang and Wang, Zheng},
    title = {RISE: robust wireless sensing using probabilistic and statistical assessments},
    year = {2021},
    isbn = {9781450383424},
    publisher = {Association for Computing Machinery},
    address = {New York, NY, USA},
    url = {https://doi.org/10.1145/3447993.3483253},
    doi = {10.1145/3447993.3483253},
    booktitle = {Proceedings of the 27th Annual International Conference on Mobile Computing and Networking},
    pages = {309–322},
    numpages = {14},
    location = {New Orleans, Louisiana},
    series = {MobiCom '21}
}

@article{ma2025adapttrack,
    author = {Ma, Dongliang and Sun, Zhuo and Wei, Zhiqiang and Guo, Yifan and Lei, Yangqian and Wang, Zhu and Yu, Zhiwen and Guo, Bin},
    title = {AdaptTrack: A Robust Tracking System for Complex Environments Based on WiFi Device Selection Strategy},
    year = {2025},
    issue_date = {September 2025},
    publisher = {Association for Computing Machinery},
    address = {New York, NY, USA},
    volume = {9},
    number = {3},
    url = {https://doi.org/10.1145/3749506},
    doi = {10.1145/3749506},
    journal = {Proc. ACM Interact. Mob. Wearable Ubiquitous Technol.},
    month = sep,
    articleno = {116},
    numpages = {29}
}

@article{jiang2026edge,
  title={Edge large language models: a comprehensive survey},
  author={Jiang, Shan and Zhou, Xuecheng and Zhang, Mingjin and Xu, Changfu and Liao, Guocheng and Chen, Jianguo and Cao, Jiannong},
  journal={CCF Transactions on Pervasive Computing and Interaction},
  pages={1--30},
  year={2026},
  publisher={Springer}
}

@article{al2026multimodal,
  title={A Multimodal benchmark of EMG and 3D hand motion for gesture recognition and biometric authentication},
  author={Al-Mari, Wejdan and Al-Marri, Hamda and Al-Jaber, Sheyma and Al-Tamimi, Alreem and Kunhoth, Jayakanth and Al-Maadeed, Somaya and Saleh, Moutaz and Akbari, Younes},
  journal={CCF Transactions on Pervasive Computing and Interaction},
  pages={1--18},
  year={2026},
  publisher={Springer}
}

@article{wang2026meco,
  title={MeCo: Enhancing LLM-Empowered Multi-Robot Collaboration via Similar Task Memoization},
  author={Wang, Baiqing and Cui, Helei and Zhang, Bo and Zheng, Xiaolong and Guo, Bin and Yu, Zhiwen},
  journal={arXiv preprint arXiv:2601.20577},
  year={2026}
}

@inproceedings{wang2024agr,
  title={AGR: acoustic gait recognition using interpretable micro-range profile},
  author={Wang, Penghao and Jiang, Ruobing and Liu, Chao and Luo, Jun},
  address={Vancouver,BC,Canada},
  booktitle={IEEE INFOCOM 2024-IEEE Conference on Computer Communications},
  doi={10.1109/INFOCOM52122.2024.10621283},
  pages={1201--1210},
  year={2024},
  publisher={IEEE}
}

@article{xu2025acoustic,
  title={Acoustic sensing mechanisms, technologies, and applications: a survey: W. Xu et al.},
  author={Xu, Wei and Wang, Zhu and Guo, Yifan and Ren, Zhihui and Xu, Yandi and Guo, Bin and Yu, Zhiwen and Zhou, Xingshe},
  journal={CCF Transactions on Pervasive Computing and Interaction},
  pages={1--28},
  year={2025},
  publisher={Springer}
}

@ARTICLE{li2025synergizing,
  author={Li, Mengning and Wang, Wenye},
  journal={IEEE Transactions on Mobile Computing}, 
  title={Synergizing Acoustic and Wi-Fi Signals for Device-Free Gesture Recognition}, 
  year={2025},
  volume={24},
  number={9},
  pages={8167-8179},
  doi={10.1109/TMC.2025.3558139},
  publisher={IEEE}
}

@article{halperin2011tool,
  title={Tool release: Gathering 802.11 n traces with channel state information},
  author={Halperin, Daniel and Hu, Wenjun and Sheth, Anmol and Wetherall, David},
  journal={ACM SIGCOMM computer communication review},
  volume={41},
  number={1},
  pages={53--53},
  year={2011},
  publisher={ACM New York, NY, USA}
}

@article{wang2022caution,
  title={CAUTION: A Robust WiFi-based human authentication system via few-shot open-set recognition},
  author={Wang, Dazhuo and Yang, Jianfei and Cui, Wei and Xie, Lihua and Sun, Sumei},
  journal={IEEE Internet of Things Journal},
  volume={9},
  number={18},
  pages={17323--17333},
  year={2022},
  publisher={IEEE}
}

@inproceedings{qian2018widar2,
    author = {Qian, Kun and Wu, Chenshu and Zhang, Yi and Zhang, Guidong and Yang, Zheng and Liu, Yunhao},
    title = {Widar2.0: Passive Human Tracking with a Single Wi-Fi Link},
    year = {2018},
    isbn = {9781450357203},
    publisher = {Association for Computing Machinery},
    address = {New York, NY, USA},
    url = {https://doi.org/10.1145/3210240.3210314},
    doi = {10.1145/3210240.3210314},
    booktitle = {Proceedings of the 16th Annual International Conference on Mobile Systems, Applications, and Services},
    pages = {350–361},
    numpages = {12},
    location = {Munich, Germany},
    series = {MobiSys '18}
}

@inproceedings{zhang2018crosssense,
    author = {Zhang, Jie and Tang, Zhanyong and Li, Meng and Fang, Dingyi and Nurmi, Petteri and Wang, Zheng},
    title = {CrossSense: Towards Cross-Site and Large-Scale WiFi Sensing},
    year = {2018},
    isbn = {9781450359030},
    publisher = {Association for Computing Machinery},
    address = {New York, NY, USA},
    url = {https://doi.org/10.1145/3241539.3241570},
    doi = {10.1145/3241539.3241570},
    booktitle = {Proceedings of the 24th Annual International Conference on Mobile Computing and Networking},
    pages = {305–320},
    numpages = {16},
    location = {New Delhi, India},
    series = {MobiCom '18}
}

@InProceedings{zhang2020gaitid,
    author="Zhang, Yi and Zheng, Yue and Zhang, Guidong and Qian, Kun and Qian, Chen and Yang, Zheng",
    title="GaitID: Robust Wi-Fi Based Gait Recognition",
    booktitle="Wireless Algorithms, Systems, and Applications",
    year="2020",
    publisher="Springer International Publishing",
    address="Cham",
    pages="730--742",
    isbn="978-3-030-59016-1"
}

@article{zhang2021gaitsense,
  title={GaitSense: Towards ubiquitous gait-based human identification with Wi-Fi},
  author={Zhang, Yi and Zheng, Yue and Zhang, Guidong and Qian, Kun and Qian, Chen and Yang, Zheng},
  journal={ACM Transactions on Sensor Networks (TOSN)},
  volume={18},
  number={1},
  pages={1--24},
  year={2021},
  publisher={ACM New York, NY}
}

@article{wang2025muid,
  title={MUID: Multi-person Gait Identification with Commodity Wi-Fi},
  author={Wang, Luhan and Yang, Xuan and Zhou, Shuang and Lu, Zhaoming and Yang, Ying and Tian, Changzhen},
  journal={IEEE Transactions on Instrumentation and Measurement},
  year={2025},
  publisher={IEEE}
}

@article{jiang2025large,
  title={A Large-Scale Multimodal Dataset and Benchmarks for Human Activity Scene Understanding and Reasoning},
  author={Jiang, Siyang and Yuan, Mu and Ji, Xiang and Yang, Bufang and Liu, Zeyu and Xu, Lilin and Li, Yang and He, Yuting and Dong, Liran and Lu, Wenrui and others},
  journal={arXiv preprint arXiv:2512.07136},
  year={2025}
}

@article{wang2024xrf55,
  title={Xrf55: A radio frequency dataset for human indoor action analysis},
  author={Wang, Fei and Lv, Yizhe and Zhu, Mengdie and Ding, Han and Han, Jinsong},
  journal={Proceedings of the ACM on Interactive, Mobile, Wearable and Ubiquitous Technologies},
  volume={8},
  number={1},
  pages={1--34},
  year={2024},
  publisher={ACM New York, NY, USA}
}

@inproceedings{dai2025babel,
    author = {Dai, Shenghong and Jiang, Shiqi and Yang, Yifan and Cao, Ting and Li, Mo and Banerjee, Suman and Qiu, Lili},
    title = {Babel: A Scalable Pre-trained Model for Multi-Modal Sensing via Expandable Modality Alignment},
    year = {2025},
    isbn = {9798400714795},
    publisher = {Association for Computing Machinery},
    address = {New York, NY, USA},
    url = {https://doi.org/10.1145/3715014.3722068},
    doi = {10.1145/3715014.3722068},

    booktitle = {Proceedings of the 23rd ACM Conference on Embedded Networked Sensor Systems},
    pages = {240–253},
    numpages = {14},
    location = {UC Irvine Student Center., Irvine, CA, USA},
    series = {SenSys '25}
}

@article{guo2025mmpencil,
    author = {Guo, Yifan and Wang, Zhu and Qin, Qian and Lei, Yangqian and Gan, Qiwen and Sun, Zhuo and Chen, Chao and Guo, Bin and Yu, Zhiwen},
    title = {mmPencil: Toward Writing-Style-Independent In-Air Handwriting Recognition via mmWave Radar and Large Vision-Language Model},
    year = {2025},
    issue_date = {September 2025},
    publisher = {Association for Computing Machinery},
    address = {New York, NY, USA},
    volume = {9},
    number = {3},
    url = {https://doi.org/10.1145/3749504},
    doi = {10.1145/3749504},
    journal = {Proc. ACM Interact. Mob. Wearable Ubiquitous Technol.},
    month = sep,
    articleno = {81},
    numpages = {30}
}

@INPROCEEDINGS{zeng2016wiwho,
  author={Zeng, Yunze and Pathak, Parth H. and Mohapatra, Prasant},
  booktitle={2016 15th ACM/IEEE International Conference on Information Processing in Sensor Networks (IPSN)}, 
  title={WiWho: WiFi-Based Person Identification in Smart Spaces}, 
  year={2016},
  volume={},
  number={},
  pages={1-12},
  doi={10.1109/IPSN.2016.7460727},
  publisher={IEEE}
}

@inproceedings{wang2016gait,
  title={Gait recognition using wifi signals},
  author={Wang, Wei and Liu, Alex X and Shahzad, Muhammad},
  booktitle={Proceedings of the 2016 ACM international joint conference on pervasive and ubiquitous computing},
  pages={363--373},
  year={2016}
}

@article{wang2025survey,
  title={A survey on wi-fi sensing generalizability: Taxonomy, techniques, datasets, and future research prospects},
  author={Wang, Fei and Zhang, Tingting and Zhao, Bintong and Xing, Libao and Wang, Tiantian and Ding, Han and Han, Tony Xiao},
  journal={arXiv preprint arXiv:2503.08008},
  year={2025}
}

@article{WiFiSurvey,
author = {Wei, Zhongcheng and Chen, Wei and Ning, Shuli and Lin, Weidong and Li, Nan and Lian, Bin and Sun, Xiang and Zhao, Jijun},
title = {A Survey on WiFi-based Human Identification: Scenarios, Challenges, and Current Solutions},
year = {2025},
issue_date = {January 2025},
publisher = {Association for Computing Machinery},
address = {New York, NY, USA},
volume = {21},
number = {1},
issn = {1550-4859},
url = {https://doi.org/10.1145/3708323},
doi = {10.1145/3708323},
journal = {ACM Trans. Sen. Netw.},
month = jan,
articleno = {10},
numpages = {32}
}

@article{liang2024dcs,
  title={DCS-Gait: a class-level domain adaptation approach for cross-scene and cross-state gait recognition using Wi-Fi CSI},
  author={Liang, Ying and Wu, Wenjie and Li, Haobo and Chang, Xiaojun and Chen, Xiaojiang and Peng, Jinye and Xu, Pengfei},
  journal={IEEE Transactions on Information Forensics and Security},
  volume={19},
  pages={2997--3007},
  year={2024},
  publisher={IEEE}
}

@article{CSI_DIFF,
author = {Li, Wenwei and Gao, Ruiyang and Xiong, Jie and Zhou, Jiarun and Wang, Leye and Mao, Xingjian and Yi, Enze and Zhang, Daqing},
title = {WiFi-CSI Difference Paradigm: Achieving Efficient Doppler Speed Estimation for Passive Tracking},
year = {2024},
issue_date = {June 2024},
publisher = {Association for Computing Machinery},
address = {New York, NY, USA},
volume = {8},
number = {2},
url = {https://doi.org/10.1145/3659608},
doi = {10.1145/3659608},
journal = {Proc. ACM Interact. Mob. Wearable Ubiquitous Technol.},
month = may,
articleno = {63},
numpages = {29}
}

@ARTICLE{Through-wall,
  author={Ren, Zhihui and Wang, Zhu and Sun, Zhuo and Guo, Yifan and Song, Wenchao and Zhang, Hualei and Chen, Chao and Guo, Bin and Yu, Zhiwen and Zhou, Xingshe and Zhang, Daqing},
  journal={IEEE Transactions on Mobile Computing}, 
  title={Characterizing the Through-Wall Sensing Mechanism of Wi-Fi Signals With a Refraction-Aware Fresnel Zone Model}, 
  year={2024},
  volume={23},
  number={12},
  pages={13438-13454},
  doi={10.1109/TMC.2024.3425847}}

@article{Lww,
author = {Li, Wenwei and Zhou, Jiarun and Xiong, Jie and Xie, Yuhui and Wang, Leye and Zhang, Duo and Zhang, Daqing},
title = {Rethinking WiFi-based Angle Estimation for Robust Passive Indoor Localization},
year = {2025},
issue_date = {December 2025},
publisher = {Association for Computing Machinery},
address = {New York, NY, USA},
volume = {9},
number = {4},
url = {https://doi.org/10.1145/3770662},
doi = {10.1145/3770662},
journal = {Proc. ACM Interact. Mob. Wearable Ubiquitous Technol.},
month = dec,
articleno = {190},
numpages = {28}
}

@ARTICLE{tong2025stagr,
  author={Tong, Xinyu and Xu, Xiaoqiang and Yu, Aiwen and Xie, Xin and Liu, Xiulong and Qu, Wenyu},
  journal={IEEE Transactions on Mobile Computing}, 
  title={STAGR: Simultaneous Tracking and Gait Recognition With Commodity Wi-Fi}, 
  year={2025},
  volume={24},
  number={10},
  pages={10868-10885},
  doi={10.1109/TMC.2025.3570993}
}

@inproceedings{freeGait,
author = {Yan, Dawei and Yang, Panlong and Shang, Fei and Han, Feiyu and Yan, Yubo and Li, Xiang-Yang},
title = {freeGait: Liberalizing Wireless-based Gait Recognition to Mitigate Non-gait Human Behaviors},
year = {2024},
isbn = {9798400705212},
publisher = {Association for Computing Machinery},
address = {New York, NY, USA},
url = {https://doi.org/10.1145/3641512.3686362},
doi = {10.1145/3641512.3686362},
booktitle = {Proceedings of the Twenty-Fifth International Symposium on Theory, Algorithmic Foundations, and Protocol Design for Mobile Networks and Mobile Computing},
pages = {241–250},
numpages = {10},
location = {Athens, Greece},
series = {MobiHoc '24}
}

@ARTICLE{liu2025csid,
  author={Liu, Yu and Hu, Jingyang and Jiang, Hongbo and Yang, Kehua and Zhang, Wei and Qin, Zheng},
  journal={IEEE Transactions on Mobile Computing}, 
  title={CSID: Enhancing Wi-Fi Based Gait Recognition via Adversarial Learning}, 
  year={2025},
  volume={24},
  number={11},
  pages={12076-12087},
  doi={10.1109/TMC.2025.3583946}
}

@article{yan2025pushing,
  title={Pushing the Limits of WiFi-Based Gait Recognition Towards Non-Gait Human Behaviors},
  author={Yan, Dawei and Yang, Panlong and Shang, Fei and Han, Feiyu and Yan, Yubo and Li, Xiang-Yang},
  journal={IEEE Transactions on Mobile Computing},
  year={2025},
  publisher={IEEE}
}

@article{yang2025environment,
  title={Environment Independent Gait Recognition Based on Wi-Fi Signals},
  author={Yang, Wei and Li, Zhixiang and Chen, Sheng},
  journal={IEEE Transactions on Mobile Computing},
  year={2025},
  publisher={IEEE}
}

@article{wang2024rdgait,
  title={Rdgait: A mmwave based gait user recognition system for complex indoor environments using single-chip radar},
  author={Wang, Dequan and Zhang, Xinran and Wang, Kai and Wang, Lingyu and Fan, Xiaoran and Zhang, Yanyong},
  journal={Proceedings of the ACM on Interactive, Mobile, Wearable and Ubiquitous Technologies},
  volume={8},
  number={3},
  pages={1--31},
  year={2024},
  publisher={ACM New York, NY, USA}
}

@article{meng2020gait, 
    title={Gait Recognition for Co-Existing Multiple People Using Millimeter Wave Sensing},
   volume={34}, 
   url={https://ojs.aaai.org/index.php/AAAI/article/view/5430},DOI={10.1609/aaai.v34i01.5430}, 
   number={01}, 
   journal={Proceedings of the AAAI Conference on Artificial Intelligence},
   author={Meng, Zhen and Fu, Song and Yan, Jie and Liang, Hongyuan and Zhou, Anfu and Zhu, Shilin and Ma, Huadong and Liu, Jianhua and Yang, Ning},
   year={2020},
   month={Apr.},
   pages={849-856} 
}

@article{cheng2021person,
  title={Person reidentification based on automotive radar point clouds},
  author={Cheng, Yuwei and Liu, Yimin},
  journal={IEEE Transactions on Geoscience and Remote Sensing},
  volume={60},
  pages={1--13},
  year={2021},
  publisher={IEEE}
}

@article{meng2025gr,
  title={Gr-fall: A fall detection system with gait recognition for indoor environments using siso mmwave radar},
  author={Meng, Chengzhen and He, Chenming and Wang, Dequan and Xiao, Yuxuan and Wang, Lingyu and Fan, Xiaoran and Zhang, Lu and Zhang, Yanyong},
  journal={Proceedings of the ACM on Interactive, Mobile, Wearable and Ubiquitous Technologies},
  volume={9},
  number={3},
  pages={1--26},
  year={2025},
  publisher={ACM New York, NY, USA}
}

@article{li2023passive,
  title={Passive multiuser gait identification through micro-Doppler calibration using mmWave radar},
  author={Li, Jincheng and Li, Binbin and Wang, Lin and Liu, Wenyuan},
  journal={IEEE Internet of Things Journal},
  volume={11},
  number={4},
  pages={6868--6877},
  year={2023},
  publisher={IEEE}
}

@article{niu2022rethinking,
  title={Rethinking Doppler effect for accurate velocity estimation with commodity WiFi devices},
  author={Niu, Kai and Wang, Xuanzhi and Zhang, Fusang and Zheng, Rong and Yao, Zhiyun and Zhang, Daqing},
  journal={IEEE Journal on Selected Areas in Communications},
  volume={40},
  number={7},
  pages={2164--2178},
  year={2022},
  publisher={IEEE}
}

@inproceedings{qian2017widar,
    author = {Qian, Kun and Wu, Chenshu and Yang, Zheng and Liu, Yunhao and Jamieson, Kyle},
    title = {Widar: Decimeter-Level Passive Tracking via Velocity Monitoring with Commodity Wi-Fi},
    year = {2017},
    isbn = {9781450349123},
    publisher = {Association for Computing Machinery},
    address = {New York, NY, USA},
    url = {https://doi.org/10.1145/3084041.3084067},
    doi = {10.1145/3084041.3084067},
    booktitle = {Proceedings of the 18th ACM International Symposium on Mobile Ad Hoc Networking and Computing},
    articleno = {6},
    numpages = {10},
    location = {Chennai, India},
    series = {Mobihoc '17}
}

@article{zeng2019farsense,
    author = {Zeng, Youwei and Wu, Dan and Xiong, Jie and Yi, Enze and Gao, Ruiyang and Zhang, Daqing},
    title = {FarSense: Pushing the Range Limit of WiFi-based Respiration Sensing with CSI Ratio of Two Antennas},
    year = {2019},
    issue_date = {September 2019},
    publisher = {Association for Computing Machinery},
    address = {New York, NY, USA},
    volume = {3},
    number = {3},
    url = {https://doi.org/10.1145/3351279},
    doi = {10.1145/3351279},
    journal = {Proc. ACM Interact. Mob. Wearable Ubiquitous Technol.},
    month = sep,
    articleno = {121},
    numpages = {26}
}

@InProceedings{Garreau2011Gait,
  author    = {Garreau, Guillaume and Andreou, Charalambos M and Andreou, Andreas G and Georgiou, Julius and Dura-Bernal, Salvador and Wennekers, Thomas and Denham, Sue},
  title     = {Gait-based person and gender recognition using micro-doppler signatures},
  booktitle = {Biomedical Circuits \& Systems Conference},
  year      = {2011},
}

@INPROCEEDINGS{AprilTag2011,
  author={Olson, Edwin},
  booktitle={2011 IEEE International Conference on Robotics and Automation}, 
  title={AprilTag: A robust and flexible visual fiducial system}, 
  year={2011},
  volume={},
  number={},
  pages={3400-3407},
  doi={10.1109/ICRA.2011.5979561}
}

@article{xu2019acousticid,
    author = {Xu, Wei and Yu, ZhiWen and Wang, Zhu and Guo, Bin and Han, Qi},
    title = {AcousticID: Gait-based Human Identification Using Acoustic Signal},
    year = {2019},
    issue_date = {September 2019},
    publisher = {Association for Computing Machinery},
    address = {New York, NY, USA},
    volume = {3},
    number = {3},
    url = {https://doi.org/10.1145/3351273},
    doi = {10.1145/3351273},
    journal = {Proc. ACM Interact. Mob. Wearable Ubiquitous Technol.},
    month = sep,
    articleno = {115},
    numpages = {25}
}

@article{topham2022human,
    author = {Topham, Luke K. and Khan, Wasiq and Al-Jumeily, Dhiya and Hussain, Abir},
    title = {Human Body Pose Estimation for Gait Identification: A Comprehensive Survey of Datasets and Models},
    year = {2022},
    issue_date = {June 2023},
    publisher = {Association for Computing Machinery},
    address = {New York, NY, USA},
    volume = {55},
    number = {6},
    issn = {0360-0300},
    url = {https://doi.org/10.1145/3533384},
    doi = {10.1145/3533384},
    journal = {ACM Comput. Surv.},
    month = dec,
    articleno = {120},
    numpages = {42}
}

@article{wang2016recognizing,
    author = {Wang, Tianben and Wang, Zhu and Zhang, Daqing and Gu, Tao and Ni, Hongbo and Jia, Jiangbo and Zhou, Xingshe and Lv, Jing},
    title = {Recognizing Parkinsonian Gait Pattern by Exploiting Fine-Grained Movement Function Features},
    year = {2016},
    issue_date = {January 2017},
    publisher = {Association for Computing Machinery},
    address = {New York, NY, USA},
    volume = {8},
    number = {1},
    issn = {2157-6904},
    url = {https://doi.org/10.1145/2890511},
    doi = {10.1145/2890511},
    journal = {ACM Trans. Intell. Syst. Technol.},
    month = aug,
    articleno = {6},
    numpages = {22}
}

@article{mirelman2019gait,
  title={Gait impairments in Parkinson's disease},
  author={Mirelman, Anat and Bonato, Paolo and Camicioli, Richard and Ellis, Terry D and Giladi, Nir and Hamilton, Jamie L and Hass, Chris J and Hausdorff, Jeffrey M and Pelosin, Elisa and Almeida, Quincy J},
  journal={The Lancet Neurology},
  volume={18},
  number={7},
  pages={697--708},
  year={2019},
  publisher={Elsevier}
}

@InProceedings{Kalgaonkar2007Acoustic,
  author    = {Kalgaonkar, Kaustubh and Raj, Bhiksha},
  title     = {Acoustic Doppler sonar for gait recogination},
  booktitle = {IEEE Conference on Advanced Video \& Signal Based Surveillance},
  year      = {2007},
}

@book{braune2012human,
  title={The human gait},
  author={Braune, Wilhelm and Fischer, Otto},
  year={2012},
  publisher={Springer Science \& Business Media}
}
\end{document}